%% file: main.tex
\input{macro_lib.tex}

\documentclass[12pt]{thesis}
\usepackage{epsfig,amssymb,amsmath,graphicx} 

\usepackage[round,comma]{authyear}
\def\lsim{\lower.5ex\hbox{$\; \buildrel < \over \sim \;$}}
\def\gsim{\lower.5ex\hbox{$\; \buildrel > \over \sim \;$}}

\usepackage{makeidx}
\makeindex

\begin{document}
\setcounter{section}{0}
\setcounter{figure}{0}
\setcounter{table}{0}
\vskip 3cm
\thispagestyle{empty}
{\baselineskip 25pt
\centerline{\Large\bf MONTE CARLO SIMULATIONS OF THE} 
\vskip 0.5cm
\centerline{\Large\bf ADVECTIVE INFLOW AND OUTFLOW}
\vskip 0.5cm
\centerline{\Large\bf AROUND A BLACK HOLE}
\vskip 0.5cm
\vskip 0.5cm
\vskip 0.5cm
\vskip 0.5cm
\begin{center}
{\large\bf  Thesis submitted for the degree of\\
            Doctor of Philosophy (Science)\\
             of the \\
            Jadavpur University\\ }
\end{center}
\vfill
\centerline{\large\bf Himadri Ghosh}
\vskip 0.2cm
{\baselineskip 15pt
\centerline{\large S. N. Bose National Centre for Basic Sciences}
\centerline{\large Salt Lake, Sector-III, Block-JD}
\centerline{\large Kolkata-700098, West Bengal, India}}
\vskip 0.5cm
\centerline{\large\bf Supervisor:}
\vskip 0.2cm
{\baselineskip 15pt
\centerline{\large\bf Sandip K. Chakrabarti}
\centerline{\large    Senior Professor,}
\centerline{\large    S. N. Bose National Centre for Basic Sciences}
\centerline{\large    Salt Lake, Sector-III, Block-JD}
\centerline{\large    Kolkata-700098, West Bengal, India}

\newpage
\pagestyle{newheadings}
\pagenumbering{roman}
\setcounter{page}{0}
\vskip 3cm
\centerline{\bigbf CERTIFICATE FROM THE SUPERVISOR}
\vspace{0.7in}
{\baselineskip=20pt
This is to certify that the thesis entitled
{\bf ``MONTE CARLO SIMULATIONS OF THE ADVECTIVE INFLOW AND OUTFLOW 
AROUND A BLACK HOLE"}, submitted by {\bf Sri Himadri Ghosh} 
(Index No. 352/07/Phys/17), who got his name registered on 
{\bf 26.06.2007} for the award of Ph.D. (Science) degree of 
Jadavpur University, is absolutely based upon his own work 
under the supervision of {\bf Professor Sandip K. Chakrabarti}
and that neither this thesis nor any part of it has been
submitted for either any degree/diploma or any other academic
award anywhere before.
\\}
 \vspace{1 in}
  {\flushright  {\bf Sandip K. Chakrabarti}\\
	       Senior Professor,\\
	       S. N. Bose National Centre for Basic Sciences,\\
	       Salt Lake, Sector-III, Block-JD,\\
               Kolkata-700098, West Bengal, India.\\}
\today\\
\newpage
\vskip 2cm
\input{abstract.tex}
\newpage
\vskip 3cm
\centerline{\Large\bf ACKNOWLEDGMENTS}
\vskip 2cm
\input{ack.tex}
\newpage
\vskip 3cm

\input{publication.tex}
\vskip 2cm
\newpage

\tableofcontents
\newpage
\pagestyle{myheadings}
\pagenumbering{arabic}
\alpheqn
\resec
\refig
\retab
\input{CHAPTER1/chap1.tex} 
	\reseteqn
        \resetsec
	\resetfig
	\resettab

\resec
\alpheqn
\refig
\retab
\input{CHAPTER2/chap2.tex}
	\reseteqn
	\resetsec
	\resetfig
	\resettab

\alpheqn
\resec
\refig
\retab
\input{CHAPTER3/chap3.tex}
	\reseteqn
	\resetsec
	\resetfig
	\resettab
\alpheqn
\resec
\refig
\retab
\input{CHAPTER4/chap4.tex}
	\reseteqn
	\resetsec
	\resetfig
	\resettab

\resec
\alpheqn
\refig
\retab
\input{CHAPTER5/chap5.tex}
	\reseteqn
	\resetsec
	\resetfig
	\resettab
\resec
\alpheqn
\refig
\retab
\input{CHAPTER6/chap6.tex}
	\reseteqn
	\resetsec
	\resetfig
	\resettab
\resec
\alpheqn
\refig
\retab
\input{conclusion.tex}
        \resetsec
        \resettab
\vfill\eject
\pagestyle{newheadings}
{\baselineskip 15pt\input{appendx.tex}

%
\printindex 

\end{document}

%% file: macro_lib.tex
\newcounter{saveeqn}%
\newcommand{\alpheqn}{\setcounter{saveeqn}{\value{equation}}%
\stepcounter{saveeqn}\setcounter{equation}{0}%
\renewcommand{\theequation}
    {\mbox{\arabic{saveeqn}-\arabic{equation}}}}%
\newcommand{\reseteqn}{\setcounter{equation}{\value{saveeqn}}%
\renewcommand{\theequation}{\arabic{equation}}}%

\newcounter{savesec}%
\newcommand{\resec}{\setcounter{savesec}{\value{section}}%
\stepcounter{savesec}\setcounter{section}{0}%
\renewcommand{\thesection}
    {\mbox{\arabic{savesec}.\arabic{section}}}}%
\newcommand{\resetsec}{\setcounter{section}{\value{savesec}}%
\renewcommand{\thesection}{\arabic{section}}}%

\newcounter{savefig}%
\newcommand{\refig}{\setcounter{savefig}{\value{figure}}%
\stepcounter{savefig}\setcounter{figure}{0}%
\renewcommand{\thefigure}
    {\mbox{\arabic{savefig}.\arabic{figure}}}}%
\newcommand{\resetfig}{\setcounter{figure}{\value{savefig}}%
\renewcommand{\thefigure}{\arabic{section}}}%

\newcounter{savetab}%
\newcommand{\retab}{\setcounter{savetab}{\value{table}}%
\stepcounter{savetab}\setcounter{table}{0}%
\renewcommand{\thetable}
    {\mbox{\arabic{savetab}.\arabic{table}}}}%
\newcommand{\resettab}{\setcounter{table}{\value{savetab}}%
\renewcommand{\thetable}{\arabic{table}}}%

\def\ket{\vert \vert  \{ \emptyset \} \rangle}
\def\ket2{\vert \vert \otimes \{ R \} \rangle}

  \def\ket{\vert \vert	\{ \emptyset \} \rangle}
  \def\ket2{\vert \vert \otimes \{ R \} \rangle}

\def\.#1{\mathaccent 95#1}
\def\^#1{\mathaccent 94 #1}
\def\~#1{\mathaccent "7E #1}

\def\equal{\enskip =\enskip}

  \def\ket{\vert \vert	\{ \emptyset \} \rangle}
  \def\ket2{\vert \vert \otimes \{ R \} \rangle}

\def\gt{\; > \;}

\def\.#1{\mathaccent 95#1}
\def\^#1{\mathaccent 94 #1}
\def\~#1{\mathaccent "7E #1}

\def\equal{\enskip =\enskip}

  \def\ket{\vert \vert	\{ \emptyset \} \rangle}
  \def\ket2{\vert \vert \otimes \{ R \} \rangle}

\def\.#1{\mathaccent 95#1}
\def\^#1{\mathaccent 94 #1}
\def\~#1{\mathaccent "7E #1}

\def\equal{\enskip =\enskip}

  \def\ket{\vert \vert	\{ \emptyset \} \rangle}
  \def\ket2{\vert \vert \otimes \{ R \} \rangle}

\def\gt{\; > \;}

\def\be{\begin{equation}}
\def\ee{\end{equation}}

\font\bigbf=cmbx10 scaled\magstep 3

\def\equal{\enskip =\enskip}

%% file: abstract.tex
\centerline{\underline{\bf ABSTRACT}}
 
\vskip 0.4cm

In this Thesis, we describe the development of a three-dimensional radiative transfer 
code using Monte Carlo technique and its application to various astrophysical problems. 
This code is capable of simulating the radiation spectra coming out of the electron 
cloud of an accretion disk around compact objects, such as black hole X-ray binaries 
(XRBs). Physical processes included in this code are the relativistic Maxwell-Juttner 
momentum distribution of the electrons, Compton scattering with these electrons, 
and gravitational red shift of the photons.  Due to general nature of the code, 
processes like synchrotron radiation, 
bremsstrahlung radiation, Coulomb coupling and the pair production are also possible 
to incorporate. Various types of photon energy distribution (e.g., mono-energetic, 
power-law, black-body and multi-colour black body) and geometry of the photon source 
(e.g., point or disk) can be used. In this Thesis, we have mainly used multi-color black 
body photons coming out of a Keplerian disk as the source of soft radiation. This soft 
radiation is intercepted by the electron cloud. Depending on 
the optical depth of the cloud, and energy of the electrons,
the soft photons may get Comptonized and inverse-Comptonized via multiple 
scattering or may suffer no scattering at all, and they emerge out of the cloud as a relatively 
harder radiation. Our simulations provide information regarding the accretion disk and 
the central compact object. We apply the code to the Two Component Accretion Flow 
(TCAF) model of black hole XRBs to explain the cause of spectral state transitions. 
We have also applied this code in a system where both the inflow and outflow/jet are present. 
We find that the diverging outflow actually causes the down-scattering of the photons 
whereas the infalling matter upscatters them, thus the final spectrum in presence of 
jet/outflow is a complex mixture of both kinds of photons. In addition to using 
various static models (e.g., torus, sphere or sphere with a conical jet) arising 
out of the toy models or analytical solutions present in the literature, we also use realistic 
accretion flows obtained from time dependent hydrodynamic simulations. In the last part of the Thesis, 
we use the output of this simulation as the input flow configuration for our Monte Carlo 
simulation to calculate the spectral, timing and directional properties of the output 
radiation at each time step. We present results of zero angular momentum Bondi solution 
and low angular momentum accretion flows. One major conclusion is that in the presence 
of an axisymmetric disk which emits soft photons, even an originally spherically 
symmetric accreting Compton cloud losses its symmetry and becomes 
axisymmetric as there are considerably higher cooling along the axis.
This effect becomes more prominent for low angular momentum flow which produces shock 
waves close to the axis. The post-shock region cools down and the flow velocity is 
also increased in the region. The effect of the bulk velocity of the electrons on 
the spectra is highlighted. We show that in the soft states, the bulk-motion Comptonization 
leaves its mark as a power-law at high energies.

In Chapter 1, we have given a description of the important observations regarding 
the black hole candidates that have developed the subject in a historical perspective. 
Next, we have briefly discussed the major accretion flow models 
present in the literature. We have started with the Bondi flow. Spherical flows
have very low efficiency of the outcoming radiation. This is due to the fact that the 
flow has a very high infall velocity. We then discuss the standard Keplerian disk model. 
This model explains the nature of the multi-colour soft X-ray spectrum very well but it fails to
explain very high energy radiation coming from the stellar mass black holes and distant 
Quasars and AGNs. This brings the advective flows in the picture. This 
component has lower angular momentum than a Keplerian disk, and is called a sub-Keplerian flow. A 
realistic accretion flow may have both the components, a sub-Keplerian flow 
on the top of a Keplerian flow. This is the so-called two-component
advective flow or TCAF model.
In case of black hole physics, a full general relativistic
approach is essential, but it makes the time dependent 
hydrodynamic equation which includes radiative transfer so 
complex that it is almost impossible to handle in a finite time. This
problem is circumvented using a pseudo-Newtonian potential. 
This we discuss in this Chapter. This Chapter ends with a short note 
about the units and dimensions used throughout the Thesis.

In Chapter 2, we have presented the basic radiative processes in an accretion flow
around a black hole. We have discussed only the two relevant radiative processes 
namely, black body radiation and Compton scattering, that we have used in this 
Thesis. 

We then incorporated the physical processes described in Chapter 2 into
the Monte Carlo technique, into our radiative transfer code. This 
is elaborated in Chapter 3. Different methods to model random 
variables have been discussed in this Chapter. Next, we apply these 
methods to model injected multi-colored black body spectrum, Compton 
scattering, scattering cross section and momentum of the electrons. 
This Chapter ends with a basic Monte Carlo exercise. 

In Chapter 4 we have shown the application of our Monte Carlo code for a static electron cloud. 
We compute the effects of thermal Comptonization of soft photons emitted 
from a Keplerian disk around a black hole by the hot toroidal electron cloud.
We show that the spectral state transitions of black hole candidates 
could be explained either by varying the size of the Compton cloud or 
by changing the central density of the same which is governed by the 
rate of the sub-Keplerian flow. We confirm the conclusions of the 
previous theoretical studies that the interplay between the intensity 
of the soft photons emitted by the Keplerian flow and the optical depth and electron 
temperature of the Comptonizing cloud is responsible for the state transitions in a black hole.

In a black hole accretion, the Keplerian component supplies low-energy (soft) photons while the sub-Keplerian 
component supplies hot electrons which exchange their energy with the soft photons
through Comptonization or inverse Comptonization processes. In the sub-Keplerian flow,
a shock is generally produced due to the centrifugal force. The post-shock region is known as the 
CENtrifugal pressure supported BOundary Layer or CENBOL. 
In Chapter 5, we compute the effects of the thermal and the 
bulk motion Comptonization on the soft photons emitted from a 
Keplerian disk by the CENBOL, the pre-shock sub-Keplerian disk and the outflowing jet.
We study the emerging spectrum when both the converging inflow 
and the diverging outflow (generated from the CENBOL) 
are simultaneously present. From the strength of the shock, we calculate the 
percentage of matter being carried away by the outflow and determined
how the emerging spectrum depends on outflow rate. 
The pre-shock sub-Keplerian flow was also found to Comptonize the soft photons significantly.
The interplay among the up-scattering and down-scattering effects
determines the effective shape of the emerging spectrum. By simulating several cases with
various inflow parameters, we conclude that 
whether the pre-shock flow, or the post-shock CENBOL or the emerging jet 
is dominant in shaping the emerging spectrum, strongly depends on the geometry of the 
flow and the strength of the shock in the sub-Keplerian flow.

In Chapter 6, we carry out the time dependent numerical simulation where hydrodynamics and 
the radiative transfer are coupled together. We consider two component accretion 
flow in which the Keplerian disk is imersed inside an accreting low angular momentum 
flow (halo) around a black hole. The injected soft photons from the Keplerian 
disk are reprocessed by the electrons in the halo. We show that in presence of 
an axisymmetric soft-photon source, spherically symmetric Bondi flow losses its 
symmetry and becomes axisymmetric. In our simulation, the low angular momentum 
flow slows down and forms a centrifugal barrier which adds new features into the 
spectrum. We generated the radiated spectra from a two  
component system as functions of the accretion rates. We find that the transition from hard state 
to soft state is determined by the mass accretion rates of the disk and the halo. 
We separate out the signature of bulk motion Comptonization and discuss its significance in identifying
a black hole candidate. We study how the net spectrum is contributed by photons suffering multiple scatterings 
and spending different amounts of time inside the Compton cloud. We study the 
directional properties of the spectrum as well. In the last part of the Thesis we 
have taken a moderate angular momentum flow where a shock is present. We find that 
the shock oscillates with time and the nature of the oscillation changes due to Compton 
cooling. 

Finally, in Chapter 7, we draw concluding remarks and briefly mention our future plans.

%% file: ack.tex
\hskip -0.6cm 

In the first place, I owe my deepest gratitude to Prof. Sandip K. Chakrabarti for 
giving me the opportunity to work with him. He has demonstrated how to think like 
a scientist and to produce clear scientific writings. His truly scientist intuition 
has made him as a constant oasis of ideas and passion in science, which exceptionally 
inspired and enriched my growth as a student, a researcher and a scientist want to be. 

I also want to thank Prof. Philippe Laurent for helping me learning Monte Carlo techniques. 
In particular my radiative transfer code was based on one that Philippe developed.
I want to thank my collaborators Sudip Garain and Kinsuk Giri: it is a pleasure to 
collaborate with you! I would like to thank all the members of Indian Centre for 
Space Physics (ICSP) for providing a broad spectrum of Astrophysics. 
I would like to show my gratitude to all my seniors in S. N. Bose National Centre 
for Basic Sciences and ICSP. They have made available their support in various number of ways.
I would also like to thank the members of my thesis committee, Prof. Parthasarathi Majumdar, 
Prof. Archan S. Majumdar and Prof. Debashis Gangopadhyay for their time and insightful questions.

I gratefully acknowledge the funding sources that made my Ph.D. work possible. I
was funded by a RESPOND project of Indian Space Research Organization for the first 3 years
and was honored to be a Senior Research Fellow at SNBNCBS for the years 4 \& 5.

My heartiest thanks goes to my friends and colleagues in SNBNCBS, with whom
I have shared my stay here. Particularly, I must mention the names of Bibhas, 
Kinsuk, Sujay, Sudip, Wasim and Tamal, with whom I have shared moments of
various shades. 

Words fail me to express my gratitude to my parents and Pusun for supporting my interest 
in carrying out research. Finally, I want to thank my wife, Manisree and my brother, 
Sushovan, for supporting me in so many ways throughout the whole process. They have 
made many sacrifices to help make this happen. I thank them for their patience, support 
and love. This thesis is for Manisree and Sushovan.

%% file: publication.tex
\centerline{{\Large \bf LIST OF PUBLICATIONS}}
\vskip 1.50cm

\hskip -0.65cm 1. S. K. Garain, \underline{H. Ghosh} and S. K. Chakrabarti, {\bf Effects of Compton Cooling on Outflow in a Two Component Accretion Flow around a Black Hole: Results of a Coupled Monte Carlo-TVD Simulation}, {\it The Astrophysical Journal} (to appear) (2012)
\\
\\
2. \underline{H. Ghosh}, S. K. Garain, K. Giri and S. K. Chakrabarti, {\bf Effects of Compton Cooling on the Hydrodynamic and the Spectral Properties of a Two-Component Accretion Flow around a Black Hole}, {\it Mon. Not. R. Astron. Soc.}, 416 (pp 959-971) (2011)
\\
\\
3. \underline{H. Ghosh}, S. K. Garain, K. Giri, S. K. Chakrabarti, {\bf Monte Carlo Simulations of Comptonization Process in a Two Component Accretion Flow Around a Black Hole in Presence of an Outflow}, {\it Proc. 12th Marcel Grossman Meeting on General Relativity, Eds. T. Damour, R. T. Jantzen \& R. Ruffini}, World Scientific (pp 985-989) (2011)
\\
\\
4. \underline{H. Ghosh} , S. K. Garain, S. K. Chakrabarti and P. Laurent, {\bf Monte Carlo Simulations of the Thermal Comptonization Process in a Two-Component Accretion Flow Around a Black Hole in the presence of an Outflow}, {\it International Journal of Modern Physics D}, 19 (pp 607-620) (2010)
\\
\\
5. \underline{H. Ghosh} , S. K. Chakrabarti and P. Laurent, {\bf Monte Carlo Simulations of the Thermal Comptonization Process in a Two-Component Accretion Flow Around a Black Hole}, {\it International Journal of Modern Physics D}, 18 (pp 1693-1706) (2009)
\\
\\
6. \underline{H. Ghosh}, S. K. Chakrabarti and Philippe Laurent, {\bf Inverse Comptonization in a Two Component Advective Flow: Results of a Monte Carlo simulation}, {\it Proc. 2nd Kolkata Conference on Observational Evidence for Black Holes in the Universe, Eds. S. K. Chakrabarti \& A. S. Majumder}, AIP Conference Proceedings, 1053 (pp 373-376) (2008)
\\
\\
7. S. K. Chakrabarti, D. Debnath, P.S. Pal, A. Nandi, R. Sarkar, M.M. Samanta, P.J. Wiita, \underline{H. Ghosh} and D. Som, {\bf Quasi periodic oscillations due to axisymmetric and non-axisymmetric shock oscillations in black hole accretion}, {\it Proc. 11th Marcel Grossman Meeting on General Relativity, Eds. H. Kleinert, R. T. Jantzen \& R. Ruffini}, World Scientific (pp 569-588) (2007)
\\
\\
8. S. K. Chakrabarti, \underline{H. Ghosh} and D. Som, {\bf Astrophysical black holes - do they have boundary layers?},  {\it Proc. 11th Marcel Grossman Meeting on General Relativity, Eds. H. Kleinert, R. T. Jantzen \& R. Ruffini}, World Scientific (pp 1085-1097) (2007)



%% file: CHAPTER1/chap1.tex

\newpage
\markboth{\it Introduction}
{\it Introduction}
\chapter{Introduction}

Astronomy and Astrophysics is a discipline which conducts scientific study of celestial objects. It includes
the observation and interpretation of radiations that comes from various compact objects such as, black holes, 
neutron stars, white dwarfs, Gamma Ray Bursts (GRBs), Active Galactic Nuclei (AGNs), Quasars etc. In this Thesis 
we will concentrate only on black hole astrophysics.

\section{Stars and Compact Objects}

The life story of the stars has an important massage to give. The outer 
layers of a star sprayed out in space during a supernova explosion. They 
come in contact with fresh hydrogen and form a gaseous mixture, which contains 
all the chemical elements. This gaseous mixture is enriched by the products 
of many such explosions and forms a gas cloud. New stars are formed out of these 
clouds. The Sun is also one such star which is made up of countless supernova 
remnants dating back to the earliest years of the Galaxy. All the planets including 
the Earth is also composed almost entirely of these debris. Thus, we owe our 
existence to events that took places billions of years ago, in stars 
that lived and died long before the solar system came into being.

Compact objects, namely, white dwarfs, neutron stars and black holes, are `born' when 
most of the nuclear fuel of normal stars has been consumed, that is, when the normal 
stars `die'. These compact objects are different from the normal stars in two 
fundamental ways. First, they are unable to stop the gravitational collapse by generating thermal 
pressure due to the lack of nuclear fuel. To survive the complete collapse, White dwarfs 
use the degenerate electron pressure, while the neutron stars are supported by the 
pressure of degenerate neutrons. Black holes are stars that find no 
ways to prevent this inward pull due to gravity and completely collapse to 
singularities. All these three objects are essentially static over the lifetime of the 
Universe and represent the final stage of stellar evolution. Second, these 
compact objects are smaller in size compared to the normal stars. 
Due to the smaller size the surface gravitational fields of the compact objects are much 
stronger compared to that of a normal star of similar mass. In Table 1.1, we show the 
comparison (Shapiro \& Teukolsky, 1983).
\vskip 0.4cm
\begin {tabular}[h]{ccccc}
\multicolumn{5}{c}{Table 1.1}\\
\hline Object & Mass & Radius & Mean Density & Surface Potential \\
 & (M) & (R) & (g $cm^{-3}$) & ($GM/Rc^2$) \\
\hline Sun & $M_{\odot}$  & $R_{\odot}$ & $1$  & $10^{-6}$  \\
White dwarf & $\lsim M_{\odot}$ & $\sim 10^{-2}R_{\odot}$ & $\lsim 10^7$ & $\sim 10^{-4}$  \\
Neutron star & $\sim 1-3 M_{\odot}$ & $\sim 10^{-5} R_{\odot}$ & $\lsim 10^{15}$  & $\sim 10^{-1}$ \\
Black hole & Arbitrary & $2GM/c^2$ & $\sim M/R^3$ & $\sim 1$ \\
\hline
\end{tabular}
\vskip 0.4cm
Because of the presence of very high density of matter inside the compact objects, the hydrodynamics
 of matter and the nature of interparticle forces in and around them are very different from that of 
the normal stars. The large surface potential (Table 1.1) encountered in compact objects implies the 
 importance of general relativity in determining their structure.

For a blackbody of temperature $T$ and radius $R$, the flux varies as $T^4$, so that the 
luminosity varies as $RT^4$. White dwarfs are characterized by their higher 
effective temperatures than the normal stars even though they have lower luminosities. 
The reason for this mysterious behaviour is their smaller $R$ value. Thus, white dwarfs 
are much `whiter' than normal stars, hence their name. The reason behind the name `neutron' star is 
the dominance of neutrons in their interior, following the mutual elimination 
of electrons and protons by inverse $\beta$-decay. White dwarfs can be observed 
directly using optical telescopes during their long cooling period. Neutron stars 
can be observed directly as pulsating radio sources (`pulsars') and indirectly as 
gas accreting, periodic X-ray sources (`X-ray pulsars'). Black holes are literally 
`black' as it does not emit radiations that is observable with present day 
instruments. It can only be `observed' indirectly by observing their environment. 
For example, a black hole can be observed as gas-accreting aperiodic or 
quasi-periodic X-ray sources under appropriate circumstances. The objective of this Thesis 
is to find out what could be the observable signatures of a black hole.

\section{Black Holes}

A large number of the `stars' in the sky are actually multiple systems, 
having two (or more) stars orbiting around each other about their common centre 
of mass. These are called binary star system.
X-ray binaries belong to a class of binary stars that are luminous in X-rays. The X-rays 
are produced by matter falling from one component, called the donor (usually a 
relatively normal star) to the other component, called the accretor, which is compact: 
a white dwarf, neutron star, or black hole. The infalling matter releases gravitational 
potential energy as X-rays.

\subsection{Classification of X-ray binaries}

X-ray binaries are further subdivided into several subclasses. Though these classifications 
are sometimes overlapping, but they reflect the underlying physics better. The classification by 
mass (high, intermediate, low) refers to the optically visible donor ($M_2$), not to the 
compact X-ray emitting accretor ($M_X$).

{\bf a) Low-Mass X-ray Binaries}

In the low-mass X-ray binaries (LMXB), one of the stars (primary one) is normally a 
black hole or a neutron star. The donor star transfers mass to the compact object 
by Roche-lobe overflow. The donor (or, secondary) of LMXB systems is less massive 
than the primary ($M_2 \sim 1 M_{\odot}$). White dwarfs, late-type main-sequence 
stars, A-type stars and F-G-type subgiants are found among the low-mass companion 
stars. A typical LMXB emits almost all of its radiation in X-rays 
which is originated from the accretion disk around the 
compact object. These objects are very bright in X-rays but are optically 
faint. The orbital periods of these binaries vary from few minutes to hundred of days. 
XTE J1550-564 is a well studied LMXB system having a low mass donor 
($M_2 \sim 1.5 M_{\odot}$) with a black hole of mass $10 \pm 1.5 M_{\odot} $ and 
orbital periods $1.5$ days. GRO J1655-40 ($M_X \sim 7.02 M_{\odot}$, 
$M_2 \sim 2.3 M_{\odot}$ and orbital period $\sim 2.6$ days) 
and GRS 1915+105 ($M_X \sim 14 M_{\odot}$, $M_2 \sim 1.2 M_{\odot}$ 
and orbital period $\sim 33.5$ days) are other two most well studied LMXBs 
(Liu et al. 2007).

$\bullet$ {\bf Soft X-ray Transients (SXTs)}

SXTs are characterized by long periods of quiescence interrupted by intense outbursts where 
their luminosity can increase by more than $100$ times. These outbursts emit radiations 
throughout the entire electromagnetic spectrum, including the radio (not always sufficient 
to detect), optical and X-ray wave lengths (van Paradjis, McClintock, 1995). SXTs contain 
primary object of mass $\gt 3M_{\odot}$. It is most likely that the majority of SXTs 
contain black holes (Reilly, 2002).

{\bf b) High-Mass X-ray Binaries}

The HMXB systems are made up of a compact primary object 
orbiting a high mass OB class star (secondary). The primary component is a strong 
X-ray emitting neutron star or a black hole. Conventionally, HMXBs can be further 
divided into two subgroups: first, where the donor is a Be star and the 
compact object is a neutron star (Be/X-ray binary) (Zhang, Li \& Wang 2004). 
X-ray outbursts occur when the compact object passes through the Be-star disk, 
accreting from the low-velocity and high-density wind around Be
stars. These X-ray spectra are generally hard in nature. 
The hard X-ray spectrum, along with transience, is the most interesting 
characteristic of the Be/X-ray binaries. In the second group of HMXB 
systems, the donor is a supergiant early-type star 
(SG/X-ray binary). The compact object orbits deep inside the highly supersonic
wind. For these HMXB systems the X-ray luminosity is powered by either the strong 
stellar wind of the optical companion or Roche-lobe overflow. The X-ray 
luminosity of $10^{35}-10^{36} \mathrm erg s^{-1}$ is produced by 
the stellar wind. Whereas, a much higher ($\sim 10^{38} erg {s^{-1}}$) X-ray luminosity 
is produced in a Roche-lobe overflow system where matter flows via the inner 
Lagrangian point to an accretion disk (Liu et al. 2006).

To summarize, the donor mass ($\gsim 10 M_{\odot}$) of HMXB system 
is much higher compared to LMXB ($\lsim 1 M_{\odot}$). In HMXB the 
emission dominates in optical wave length and therefore easy to detect. One of the 
most well known HMXB is Cyg X-1, which was the first detected stellar-mass 
black hole (Giacconi et al. 1962). This binary system consists of high mass donor 
of mass $20-30 M_{\odot}$ with a compact object of mass $6-10 M_{\odot}$.

{\bf c) Intermediate-mass X-ray binaries}

Above we have described the HMXB and LMXB systems with companion stars 
having $M_2 \gsim 10 M_{\odot}$ and $\lsim 1 M_{\odot}$, respectively. 
There are also a large number of Galactic compact binaries with 
companion star masses in the interval $1 - 10 M_{\odot}$ (van den Heuvel, 1975). 
These are the so-called intermediate mass X-ray binaries (IMXBs). In this Thesis, 
we have mainly worked with HMXB and LMXB systems.

\subsection{Classification of Black Holes}

Depending on their physical masses, there are four classes of black holes: 
stellar, intermediate, supermassive and primordial (or mini). A stellar black 
hole is a region of space into which a star 
has collapsed. This type of black holes are created when a star whose remnant 
core can be more than $2 - 3 M_{\odot}$ (the Tolman-Oppenheimer-Volkoff 
limit for neutron stars) reaches the end of its thermonuclear life. The 
self-gravity of the star then overcomes both electron and neutron degeneracy 
pressure and the star collapses to a critical size, here gravity dominates 
all other forces. Black holes found at the center of 
galaxies have a mass up to 100 million solar masses and are called supermassive 
black holes. Between these two scales, there are believed to be intermediate 
black holes with a mass of several thousand solar masses. Primordial black holes, 
proposed by Stephen Hawking, could have been created at the time of the BIG-BANG 
due to perturbations in the homogeneous background density field (Zel'dovich 
and Novikov, 1966; Hawking, 1971), when some regions might have got so 
compressed that they underwent gravitational collapse. With original masses 
comparable to that of earth or less, these mini black holes could be of the 
order of $1 cm$ or smaller. Their existence is, as yet, not confirmed.

We can also classify black holes according to their physical properties. 
The simplest massive black hole has neither charge nor angular momentum. 
These non-rotating black holes are the Schwarzschild black 
holes after Karl Schwarzschild who discovered this solution 
in 1915. An electrically charged but non-rotating black hole is 
called Reissner-Nordstrom black hole. Rotating black holes can also be 
classified in two types. One type of this class does not have
any charge: they are called Kerr (after Roy Kerr) black holes. Another type 
are not electrically neutral and are called the Kerr-Newman black holes. 
These rotating black holes obey exact black 
hole solutions of Einstein’s equations of General Relativity. 
When a massive spinning star suffers gravitational collapse rotating 
black holes are formed. The collapse of a collection of stars 
or gas with an average non-zero angular momentum can also form a rotating 
black hole. As most stars rotate it is expected that most black holes 
in nature are rotating black holes. 
In the present Thesis, we work only with non-rotating Schwarzschild black hole.

\begin{center}
\begin {tabular}[h]{ccc}
\multicolumn{3}{c}{Table 1.2}\\
\hline Class & Mass & Size \\
\hline Supermassive BH & $\sim 10^6 - 10^9 M_{\odot}$ & $0.001 - 10 AU$  \\
Intermediate-mass BH & $\sim 10^3 M_{\odot}$ & $10^3 km = R_{\oplus}$ \\
Stellar-mass BH & $\sim 10 M_{\odot}$ & $30 km$ \\
Primordial BH & $\gsim M_{Moon}$ & $\lsim 1 cm$ \\
\hline
\end{tabular}
\end{center}

To conclude this Section, a black hole is defined as a region 
of spacetime that can not communicate with the external universe. 
The boundary of this region is called the surface of the black hole, 
or the {\it event horizon}. The ultimate fate of the collapsing matter, 
once it has crossed the event horizon, is not known. Black Holes are 
the simplest of the celestial objects to describe mathematically. 
To describe the astrophysical processes around a black hole one requires 
only two constant parameters about the black hole, namely, the mass 
$M$ and the spin $a$ which can vary from $0$ to $1$. Stars would 
require more parameters such as the temperature, pressure, ionization, 
boundary layer etc. on the star surface.

\section{X-ray Observations from Black Holes: Historical Perspective}

No observable radiation can come out of a black hole horizon. Thus, black holes 
must be detected by monitoring the radiations emitted by matter 
accreting on them. There are many indirect ways that tell us if there 
is any black hole at galactic centers or in binary systems. For instance, 
one could detect the evidence of rotation of the disk matter by observing 
the Doppler shifts of the lines emitted from the accretion disk. 
Alternatively, one could look for the powerful jets which are ejected from 
the centre of the galactic nuclei. Generally the accretion flows are 
targeted for identification of black holes as the radiations coming out of 
an accretion flow carry information about the matter that are being accreted 
onto the black holes. The radiation is necessarily dependent on the nature 
of the density and temperature distribution of the accreting matter, 
which in turn depend on the hydrodynamics of the flow. The hydrodynamics 
of the flow again depends on the angular momentum of the flow and the nature 
of the heating and cooling processes in the matter. Thus, the whole 
process of detecting black holes involves a thorough understanding of the 
radiative hydrodynamics of the flow in a curved geometry.

Fortunately, there are reasonable assumptions which may be used. For instance,
one can replace the general relativistic computation procedure by a pseudo-Newtonian
computation procedure discussed below in \S 1.8. The basic features of the 
solutions remain almost identical to those obtained in a curved space-time 
around a black hole. In our study, we shall assume one such simpler model of 
the accretion flows.


One could study the spectral properties in three distinct ways: the first is through 
direct observation of the radiation. Different ground based (IR, optical, 
very high energy gamma rays) or space based (optical, UV, X-rays, gamma-rays) 
instruments are used in this method. Here, one tries to find out if there are fast 
variabilities or spectral signatures in the observed radiation. These 
observations may point to massive compact objects at the core or some 
spectral signatures especially predicted for black holes. A second method, which is 
used in this Thesis, is to solve equations which govern the motion of matter around 
a black hole and produce the most likely solutions which explain both
the steady and time-dependent behaviour of the radiation. The third approach is intermediate between 
the other two approaches: the theoretical results are assimilated to give 
models and then one fits the observational results with these models using 
several free parameters. 

To understand the production of X-rays and gamma-rays, one needs to 
realize that the radiation energy is generally due to the thermal energy or the 
bulk energy of matter. Since $1$ eV radiation comes out of a gas of typical 
temperature $T\sim 10^4$ K, for an X-ray emission of about $10$ keV the electron cloud
need to have $T\sim 10^8$ K. X-rays from accretion onto compact objects could not be studied 
from ground because of severe absorption by the atmosphere. 

In 1962 Giacconi and his team analyzed the data from a rocket carrying a payload consisting of 
three large Geiger counters. The analysis revealed a considerable 
flux of radiation in the night sky that was identified as 
consisting of soft X-rays. Thus Cosmic X-Ray sources were discovered.
Cyg X-1, the first black hole candidate, was discovered in these early attempts in 1962 
(Giacconi et al. 1962; Giacconi et al. 1965; Gursky et al. 1966).

In 1970, the first Astrophysics related satellite UHURU was 
launched. Black hole candidate Cyg X-1 was among the first few objects 
which was observed by this satellite. In the first few papers 
(Giacconi et al. 1971) it was shown that this source emits 
X-rays of energy larger than 100 keV. At similar times, analytical 
solutions of the disk variables such as temperature, density etc. 
as a function of the radial distance from the axis were presented by 
Shakura and Sunyaev, 1973 (hereafter SS73). This model is known as 
standard disk model. SS73 also computed the radiation spectrum by adding 
black body contribution from successive annulus of the disk to 
get the multi-colour black body spectrum. The hall mark of the 
standard disk model was this so-called multi-colour black body spectrum. 
It was apparent from X-Ray observations 
(Agrawal et al. 1972; Tananbaum et al. 1972) that Cyg X-1 actually has a 
`soft state' (when X-ray power is emitted mostly in soft X-rays) and a `hard state' 
(when X-ray power is mostly emitted in hard X-rays) along with the two distinct 
components of the spectrum. Subsequently, it was argued by Thorne and Price (1975) that SS73 
disk model is unable to explain the high energy X-rays found in 
Cyg X-1 since the temperature of standard disk is too low. 
Observed spectra of black hole candidates, such as Cyg X-1 
(Sunyaev \& Truemper, 1979) clearly indicated the presence of 
two distinct components, a low energy black-body component 
and a high energy power-law component. 
Observations similar to Cyg X-1 were made for active galaxies and 
Quasars (Katz, 1976). It was shown that the multi-colour black 
body emission from a Keplerian disk could explain the soft X-ray bump
in these spectra. In 1979, Galeev, Rosner and Vaiana used 
Comptonization of softer photons by hot electrons from a magnetic 
Corona on an accretion disk to explain the power-law component of 
the spectra. 


The nature of the emitted spectrum from an optically thick converging inflow 
in Newtonian geometry was studied by Payne and Blandford (1981). They found 
that the energy spectral index $\alpha$ ($I(\nu) \sim \nu^{-\alpha}$) would 
be $\approx 2$. Considering the general relativistic calculations, $\alpha$ 
became $1.5$ and turns out to be the most convincing signature of a black hole 
candidate (Chakrabarti \& Titarchuk, 1995, hereafter CT95). 
Wandel, Yahil and Milgrom  (1984) and Colpi, Maraschi and Treves (1984) 
gave a more general solution in Newtonian geometry. They studied single 
temperature solution of non-adiabatic flows. It was shown  
that the Quasar luminosity increases with the accretion rate of the 
converging inflow and the luminosity is a few percent of the Eddington luminosity. 
However, spherical or converging flows failed to explain the spectrum of 
Seyfert galaxies and Quasars, where clear excess of radiation in ultraviolet 
was seen (Malkan \& Sergent, 1982). Several ultraviolet-optical-infrared 
spectra of quasars and AGNs were fitted using improved multi-colour blackbody 
radiation emitted by Shakura-Sunyaev Keplerian disk (SS73) and satisfactory 
results were found (Sun \& Malkan, 1989). The mass of the black hole was 
found to be $\sim 10^8M_\odot$. It was shown by Wandel and Petrosian (1988) 
that basically two parameters, namely the mass of the black hole and the accretion rate
can describe the spectrum of a thin accretion disk very well and the 
black hole mass was in the range $10^8-10^{9.5} M_\odot$ for 
the Quasars and $10^{7.5}-10^{8.5}M_\odot$ for the Seyferts. 
The disk model was improved by Ross, Fabian and Mineshige (1992) 
by properly treating the inner edge of the accretion disk. They showed 
that even for disks around a massive black hole, a significant radiation 
could be in soft X-rays. Generally, line emissions along with continuum 
emissions are also shown 
by active galactic nuclei. Rapidly moving clouds on both sides of an 
accretion disk are believed to emit these lines. The mass of the 
central object can be estimated by the measurement of the Doppler shift 
due to the motion of the cloud and the distances of the cloud from 
reverberation mapping (variation in strengths of the central
photoionizing source in a quasar or Seyfert galaxy generates variations 
in the strengths and profiles of the emission lines) (Blandford \& McKee, 1982). 

\begin{figure}[h]
\begin{center}
\vskip 6.5cm
\includegraphics{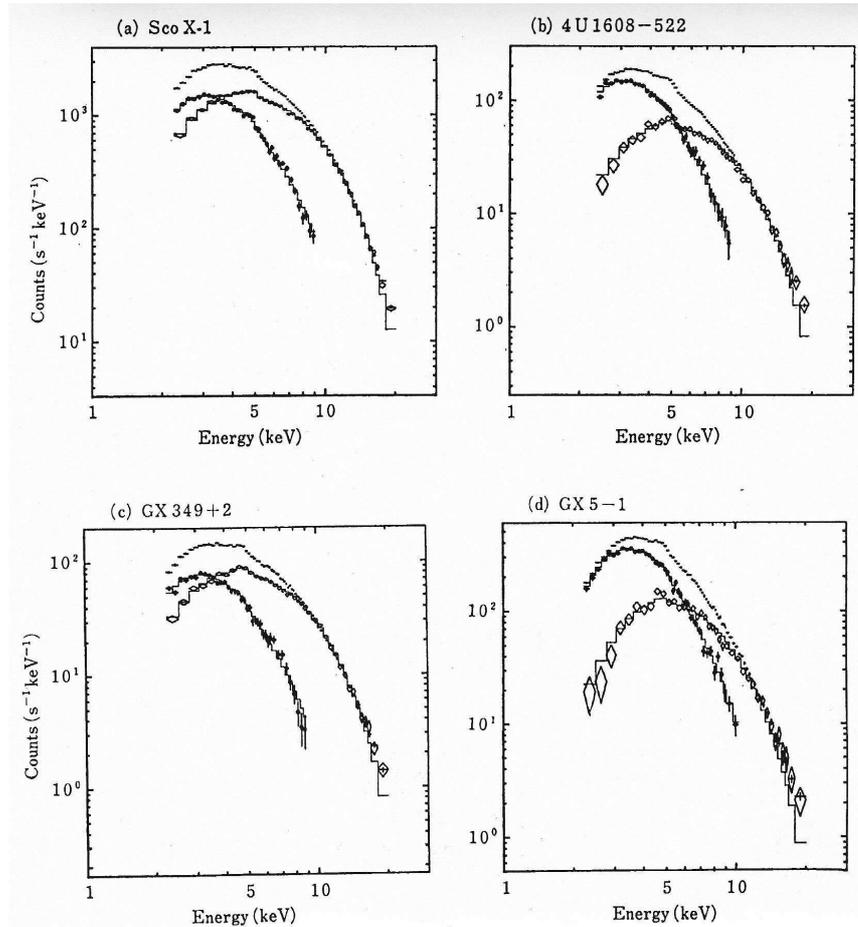}
\vskip 5.0cm
\caption{Examples of early observation of the net spectra (cross) for Sco X-1, 4U 1608-522, GX 349+2, GX 5-1 and the 
hard (rhomb) and soft (crossed circle) components of the net spectra. 
The best-fit 2-keV blackbody and the `multicolor' 
spectra are shown by the histograms (Mitsuda et. al. 1984).}
\label{fig.1.4}
\end{center}
\end{figure}

The launching of EXOSAT in 1983 and GINGA in 1987 made the subject of 
high energy astrophysics more interesting. For Low Mass X-ray 
binaries (LMXRB), a large amount of data in X-Rays were becoming available. 
Important discovery, like the Quasi-periodic Oscillations (QPOs) 
in LMXRBs and X-ray pulsars were made from these data. It was very 
clear from the observations that a source of hot electrons is 
required for Compton scattering as indicated by Sunyaev \& Titarchuk (1980, 1985; 
hereafter ST80 \& ST85). 
Sources such as Cyg X-1, LMC X-3, A0620-00 were identified 
to be  black hole candidates (McClintock and Remillard, 1986) by 
dynamical considerations. Observations of X-ray spectra for four low-mass 
binary X-ray sources, Sco X-1, 4U 1608-522, GX 5-1 and GX 349+2 performed 
with the gas scintillation proportional counters on board Tenma, revealed that 
every observed spectrum can be expressed by a sum of two spectral components 
(Fig. \ref{fig.1.4}): a hard power-law component and a soft multi-color 
blackbody component, whose spectral shapes are fixed for each 
individual source (Mitsuda et al. 1984).

It is very important to identify the source of the high energy X-rays 
to understand the observed X-ray properties. Earlier, it was believed 
that Comptonization of soft photons by `Compton Clouds' floating around 
the disk or hot corona resulted the hard radiation. Two phase accretion 
in AGNs were considered by Haardt and Maraschi (1991, 1993), one with 
cold component along the equatorial plane, and the other with hot 
component above the cooler component. CT95 introduced the two component 
advective disk model. Though this model requires more improvement, 
this picture is generally adopted in most of the models of accretion flows. 

\section{Spectral Properties of Galactic and Extragalactic Black Holes in X-rays}

\subsection{The Eddington Limit}

Before discussing the spectral properties of an accretion disk, let us define an useful 
term called the Eddington limit (Shapiro \& Teukolsky, 1983). `Eddington luminosity' is the maximum  theoretical
luminosity for which the inward gravitational pull on protons in a 
fluid element balances the outward force of radiation on electrons 
(i.e., the maximum luminosity at which matter can be accreted),
$$
L_{Edd}=\frac{4\pi G M m_p c}{\sigma_T} = 1.3 \times 10^{47} M_9 
{\rm erg\ s^{-1}},
\eqno{(1.1)}
$$
where, $m_p$ is the mass of the proton, $\sigma_T$ is the Thomson
scattering cross-section and $M_9$ is the mass of the central black 
hole in units of $10^9 M_\odot$. The luminosity of the quasars and active galaxies
has been observed to be sometimes as high as $10^{47}$ erg s$^{-1}$.
The usual explanation for such a high energy output is that the energy 
is mostly coming from the gravitational binding energy of matter accreting
onto a massive black hole with $M\sim 10^9 M_\odot$. The corresponding 
Eddington rate is defined as ${\dot M_{Edd}}=L_{Edd}/c^2$.

The characteristic black body temperature of the radiation that is emitted
from an accretion disk with luminosity $L_{Edd}$ at $r=r_g$ is given by (Chakrabarti, 1996a),
$$
T_E \sim 2.8 \times 10^5 M_9^{-1/4} \ K.
\eqno{(1.2)}
$$

Eddington rate is an upper limit of the accretion rate. It is an 
indicator of the rate at which matter is falling onto a black hole. 
Usually, the actual accretion rate is much lower than the Eddington rate,
there are evidences of the presence of higher accretion rates as well. 
However, higher accretion rates are also allowed for black holes as some of the radiations could
be trapped and advected  inside the horizon.



\subsection{Examples of different spectral states}

\begin{figure}[]
\begin{center}
\vskip 3.0cm
\includegraphics{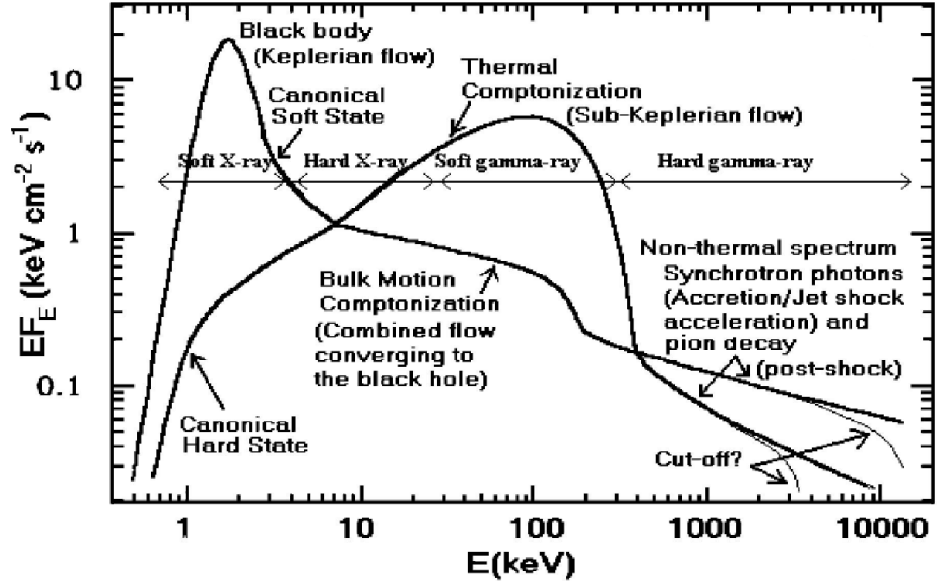}
\vskip 5.9cm
\caption{Power ($E F_E$) as a function of energy $E$ in the 
soft and hard states of a stellar mass black hole. Soft states
are dominated by high power at lower energy X-rays, while the hard states are
dominated by high power at intermediate energies ($\sim 40-100$ keV)
while the $2-20$ keV region is well described by a power-law.
There could be a non-thermal power-law  component at high energies (low power) 
in both the cases.}
\label{fig.1.5}
\end{center}
\end{figure}

We now provide an example of the spectral state variation of stellar 
mass black holes in in Fig.  \ref{fig.1.5}. We plot the power ($EF_E$) 
in the y-axis and the energy ($E$) in the X-axis. In the so-called 
soft state, the power is high in the low energy X-rays which is dominated 
by a multi-colour black body spectrum. At high energies, the power 
drops off rapidly. Occasionally a power-law is seen extending to 
$\sim 0.5$ MeV and could be due to bulk motion of the matter entering 
into the black hole. An extended power law till $\sim 20$ Mev which is also 
seen sometimes, could be due to non-thermal Comptonization of the 
non-thermal synchrotron radiation. In the so-called hard states, the 
power is very low at low energy X-rays due to the low accretion rate 
of the Keplerian disk. However the power is high typically at $\sim 40-100$ keV. 
In the $2-20$ keV range the spectrum is a power-law (straight line in a 
log-log scale). After the exponential cut-off, a power-law spectrum 
is again seen in the $\sim 200-1000$ keV range which could be due to 
Comptonization of the non-thermal synchrotron  photons. The non-thermal 
electron population may come from shock acceleration or other 
non-thermal means. This spectrum is taken from Chakrabarti (2005) and 
references therein (McConnell et al. 2002; Ling \& Wheaton, 2003, 2005; Case et al. 2005).

In case of supermassive black holes, from a purely theoretical point of view, 
it is expected that a similar spectral states should be present, though the 
blackbody spectral bump is at UV band, rather than at soft X-rays. 
In stellar mass black holes, the state transition may occur in a few tens 
of seconds to a few days. In supermassive black holes, where the mass could 
be  up to $10^9$ times larger, this time scale would be proportionally higher. 
It is unlikely that the transition could be seen  at all.

\begin{figure}[]
\begin{center}
\vskip 4.0cm
\includegraphics{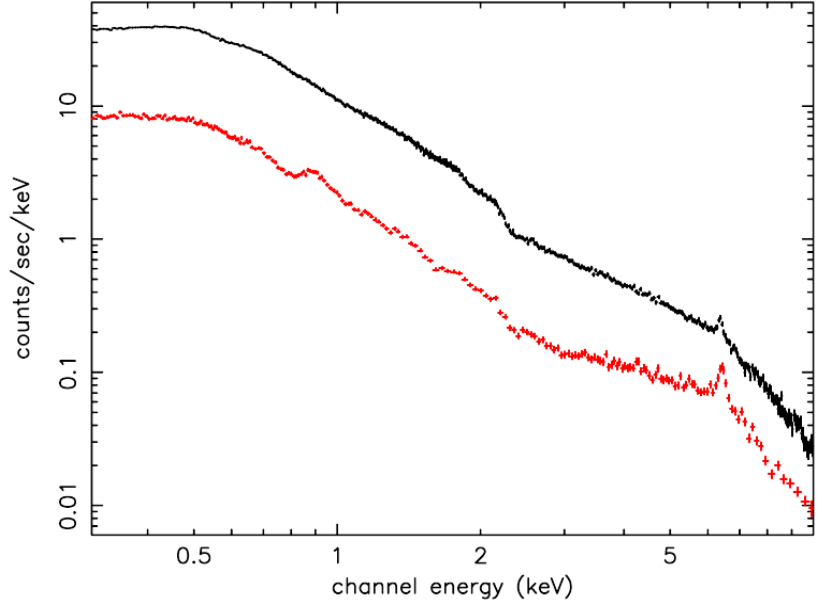}
\vskip 5.9cm
\caption{Background subtracted {\it XMM-Newton} X-ray spectra of the nearby Seyfert 
galaxy NGC 4051, taken in relatively high (2001 May, black) and low-flux 
(2002 November, red) states. See, Pounds et al. (2004) and the references therein.}
\label{fig.agn}
\end{center}
\end{figure}

In Fig. \ref{fig.agn}, the broad-band X-ray spectra of NGC 4051, integrated over 
the separate {\it XMM-Newton} observations, are shown 
(Figure taken from Pounds et al. 2004). We notice here that the mean 
flux levels of these two spectra are very much different. From $\sim 0.3$ to $3$ keV 
the spectral shape is almost unchanged, only difference is that the flux in 2001 is about 
$5$ times higher than that in 2002. From $\sim 3$ keV up to the emission line 
at $\sim 6.4$ keV, the flux ration decreases. In this energy band the low-state 
(red plot) spectrum indicates a flatter continuum slope. The emission line at 
$\sim 6.4$ keV appears to be unchanged for these two spectra.

\section{An Overview of the Accretion Flow Models}

In astrophysical context, the accretion is a process by which 
diffused matter in orbital motion is collected around a central 
object and an accretion disk is a structure where the gravitational 
energy is converted into kinetic and thermal energy. 
Accretion occurs in systems like X-ray binaries, active galactic 
nuclei (AGNs) etc. where white dwarfs, neutron stars and even super massive black 
holes may be located at the center. In low mass X-ray binary system, 
there are two ways by which matter can be supplied from the companion 
to the compact object. Matter with Keplerian angular momentum 
is accreted through the Lagrange point when the Roche lobe of the 
companion overflows. If the companion has a significant wind with it, and the 
compact object moves through the wind, then some winds with a very little 
angular momentum is also accreted. 
Another way to accrete matter from the companion is through tidal disruption. 
In this case the companion star is tidally deformed and matter flows out 
from it to the compact object. Matter can also be accreted from the 
interstellar medium at a very low rate in case of an isolated object. 
In many of the galactic centers there are evidences 
of the presence of supermassive black holes. In absence of companions, these 
black holes accrete matter from winds of surrounding stars. Those surrounding stars
may also get tidally disrupted if they come very close to the black 
hole and the matter would be accreted from the disrupted star 
to the central black hole.
 
There are several accretion and wind solutions in the literature, 
but their applicability is restricted. To describe a black hole 
accretion, several quantities are important: specific energy 
($\epsilon$), specific angular momentum ($\lambda$), viscous 
stress etc. 
A mass accretion rate ($\dot{M}$) is also needed. Some models use 
two mass accretion rates, one for the Keplerian disk and the other for the sub-Keplerian disk.

(a) {\bf Steady, spherically symmetric accretion: Bondi Flow}

The steady state behaviour of spherically accreting matter which has no angular 
momentum, was first studied by H. Bondi (1952). A detailed description of this 
particular type of flow has been given in Theory of Transonic Astrophysical Flows 
(Chakrabarti, 1990; hereafter C90). Let us assume a compact object of mass $M$, accretes matter from a large gas cloud.
The object is at rest with respect to the cloud. The matter flow is spherically
symmetric, and adiabatic in nature, i.e., there is no exchange of energy between the
gas cloud and outside world. In case of Bondi flow, the mass accretion rate is given by, 
$$
\dot{M} = 4 \pi r^2 \rho v,
\eqno{(1.3)}
$$
which is constant throughout the flow. To conserve the accretion
rate, the density ($\rho$) and velocity ($v$) will go
up simultaneously, as matter accretes radially towards the compact object. 
In case of a Bondi flow onto a Schwarzschild black hole, the matter 
velocity must reach the velocity of light (c) at the horizon making the density close to
zero as most of the matter vanishes through the horizon. On the other-hand, the 
flow at infinity where the matter is almost at rest, is characterized by the velocity $v_{\infty}$
and density $\rho_{\infty}$. Apart from these two boundary conditions, for a fluid 
of highly relativistic point particles, maximum attainable sound speed is $c/\sqrt{3}$ 
(Weinberg, 1972). Thus the black hole accretion is transonic in nature,
and the flow velocity becomes equal to the velocity of sound at a particular distance
r, away from the compact object. The distance is called sonic radius ($r_s$) at which
the flow changes from subsonic to supersonic. 
We can calculate the mass accretion rate in terms of $\rho_{\infty}$ 
and $T_{\infty}$ , where $T_{\infty}$ is the temperature of the flow at infinity. 
It is given by (Shapiro \& Teukolsky, 1983),
$$
\dot{M} = 1.2 \times 10^{10} \left( \frac{M}{M_{\odot}} \right)^2 \left( \frac{\rho_{\infty}}{10^{-24} \rm{g} \rm{{cm}^{-3}}}\right) \left( \frac{T_{\infty}}{10^4 \rm{K}} \right)^{-3/2} \rm{g} \hspace{0.1cm} \rm{s^{-1}}.
\eqno{(1.4)}
$$
This mass accretion rate would produce a luminosity of $\sim 10^{31}$ erg/s 
($\sim 1 \%$ of solar luminosity). Thus Bondi flow is radiatively inefficient. 
Unless magnetic fields and other dissipative mechanisms
are added, this flow radiate very little and thus cannot explain the high luminosities as
in a quasar. 

(b) {\bf Standard disks:}

The first accretion disk flow model was proposed by Shakura (1972), but the
complete work on the model was published in a subsequent article by Shakura \&
Sunyaev (SS73) and the disk is generally known as the Shakura-Sunyaev disk (SS
disk) or the standard disk. A relativistic version of the model was put forward 
by Novikov \& Thorne in 1973. In this disk model, it is assumed
that the accreted matter forms a geometrically thin disk in which matter rotates
in Keplerian orbits and that the inflow velocity is much smaller than the free-fall
velocity. The matter has angular momentum distribution same as the Keplerian 
distribution (obtained by equating gravitational force with centrifugal force). 
The accretion velocity is negligible, though it can have any reasonable accretion 
rate. The viscous torque ($t_{r\phi}$) which acts on different layers of the accreting 
matter transports the angular momentum outwards and thus makes the accretion possible. 
In this model $t_{r\phi}$ was assumed to be proportional to the total 
vertically averaged pressure of the gas in the disk, $p_{total}$, i.e.,
$$
t_{r\phi} = \alpha_{ss} p_{total},
\eqno{(1.5)}
$$
where, $\alpha_{ss} < 1$ is the so called viscosity parameter. The larger the value of $\alpha_{ss}$, 
the more efficient is the transport of angular momentum outwards. As $\alpha_{ss}$ increases, 
radial velocity (at a constant $\dot{M}$) of the flow increases and the surface density ($\Sigma$) of
the disk decreases.

For the steady-state disk, the energy flux, radiated from unit surface area of the 
disk per unit time at a radius $R$ can be written as,
$$
Q(R) = \frac{3}{8 \pi} \dot{M} \frac{GM}{R^3} \left[1 - \left(\frac{R_0}{R}\right)^{1/2}\right],
\eqno{(1.6)}
$$
and the corresponding luminosity produced by the disk in between $R_1$ and $R_2$ is given by,
$$
L(R_1,R_2) = 2 \int{Q(R) 2 \pi R dR},
$$
which yields,
$$
L(R_1,R_2) = \frac{3GM \dot{M}}{2} \left[\frac{1}{R_1} \left[1-\frac{2}{3}\left(\frac{R_0}{R_1} \right)^{1/2} \right] - \frac{1}{R_2} \left[1-\frac{2}{3}\left(\frac{R_0}{R_2} \right)^{1/2} \right]\right].
\eqno{(1.7)}
$$
For $R_1 = R_0$ (the radius of the central star or the Schwarzschild radius) and $R_2 = \infty$ 
the total luminosity radiated from the accretion disk becomes,
$$
L_{disk} =\frac{GM \dot{M}}{2 R_0}. 
\eqno{(1.8)}
$$
So, half of the gravitational energy is radiated away from the accretion disk.

In case the Shakura-Sunyaev disk is optically thick and opacity due to free-free 
absorption is more important than the opacity due to scattering, each element of 
the disk surface radiates blackbody spectrum with surface 
temperature $T(R)$ given by equating the dissipation rate to the blackbody flux,
$$
T(R) \approx 5 \times 10^7 \left(\frac{M}{M_{\odot}} \right)^{-1/2} {{\dot{M}}_{17}}^{1/4} (2R)^{-3/4} \left(1 - \sqrt{\frac{3}{R}} \right)^{1/4} K.
\eqno{(1.9)}
$$
In this equation $M$ is measured in units of $M_{\odot}$, $\dot{M}_{17}$ is in units 
of $10^{17}$ g $s^{-1}$ and $R$ is measured in units of $2 G M M_{\odot}/c^2$.

This disk is radiatively very efficient. But, the disk is terminated at the 
last stable circular orbit ($3r_g$ for Schwarzschild black hole).
The effective temperature of the radiation is around $\sim 1$ keV for stellar black holes
($M \sim 10 M_{\odot}$). However, for Quasars, the radiation emitted from 
such a disk is in the ultraviolet region and is widely known as 
the big blue bump. This disk is an ideal solution, as it assumes the angular 
momentum distribution to be pre-determined (Keplerian $\lambda_K$) and it has no 
solution below $r = 3r_g$. The gas pressure is assumed to be negligible.
This model is unable to explain the observed emission features of accreting
black holes at energies higher than 10 keV. 

(c) {\bf Thick Accretion Disks:}

In order that the disks are more efficient radiators and also 
to collimate the jets, one needs to incorporate the effects 
of pressure. Paczy\'nski and his collaborators assumed 
the disk to be radiation pressure dominated and radial velocity 
of the matter is zero (Paczy\'nski \& Wiita, 1980, hereafter PW80). 
Due to high pressure, the initial angular momentum of the disk 
is deviated from its Keplerian distribution. Thus, the disk 
puffs up and the height becomes comparable to the radial 
distance, hence the name thick disk. For low accretion rate 
and radiatively inefficient flow the thermal pressure (Rees et al. 1982) 
instead of radiation pressure can also make a thick disk. In such 
disks, the region around the vertical axis is devoid of matter 
due to centrifugal force. Just like the SS disk, this solution 
is also not transonic as the horizon is not joined to infinity 
(no radial velocity). In a General Relativistic thick disk, 
Chakrabarti-distribution of angular momentum derived using 
von-Zeipel theorem is used (Chakrabarti, 1985).

In late 80's and early 90's, improvements in space-borne (satellite) 
and ground based (radio antennas) instruments have triggered the 
challenge to explain the observed X-ray spectra of the black hole 
binaries. In low energy range ($< 10$ keV), the spectrum was modeled as
a blackbody spectra but at high energies (few 100 keV) there was 
power-law like behaviour whose slope changes from time to time. 
The Standard disk and the thick-disk both were unable to reproduce 
the hard spectra extending upto 100 keV and beyond. To explain this, 
the presence of hot ($T_e \sim 10^9$ K), optically thin plasma was required. 
At the same time a very interesting feature called the quasi-periodic 
oscillations (QPOs) was also observed in most of the galactic black hole 
cadidates. It was found that the X-ray variabilities are quasi-periodic 
in nature and their Fourier decomposition shows a very prominent peak 
in the power density spectrum (PDS). The radio observations confirmed 
that the jets and outflows that are observed in GBHs are coming
out from vicinity of the hole. To explain all the above features, 
a complete accretion disk model was needed instead of different models 
to explain different observations.
\begin{figure}[h]
\begin{center}
\vskip 5.0cm
\includegraphics{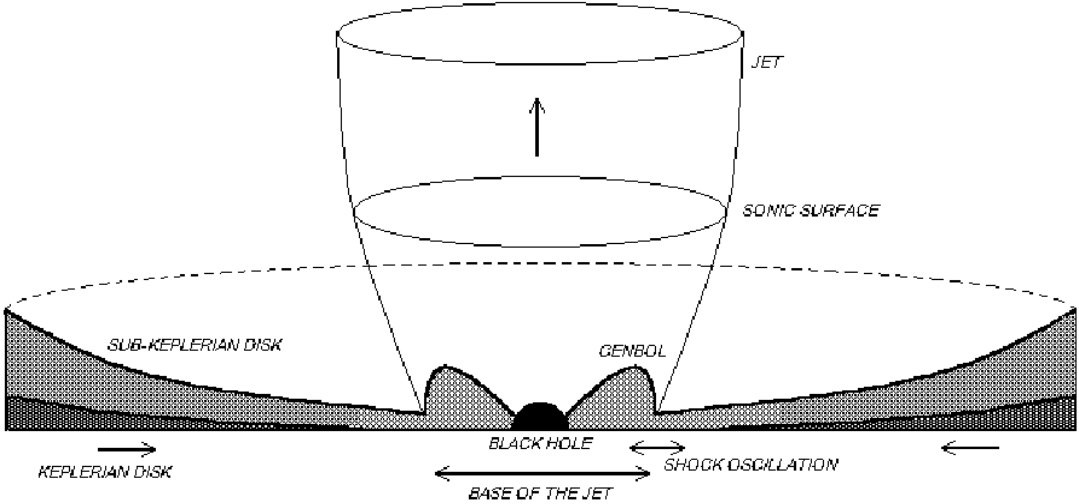}
\vskip 3.0cm
\caption{Cartoon diagram of the Two Component Advective Flow (TCAF) solution 
around a black hole, which shows the disk-jet connection  highlighting 
different components. The centrifugal pressure dominated boundary layer or 
CENBOL is the postshock region of the combined Keplerian and 
sub-Keplerian flow. There could be transient shocks close to the inner sonic point (Chakrabarti, 2003).}
\label{fig.1.8}
\end{center}
\end{figure}

(d) {\bf Two Component Advective Flows (TCAF):}

In early 90's, Chakrabarti and his collaborators (C90; Chakrabarti
\& Molteni, 1993; CT95; Chakrabarti, 1996a; 1996b; 
Molteni, Sponholz \& Chakrabarti, 1996; hereafter referred
to as MSC96) presented a accretion disk solution incorporating 
all the physical processes by solving the most general
flow equations. 
A schematic representation of the accretion disk in this model is shown in Fig. \ref{fig.1.8}.

An advective accretion disk is one which advects, or carry `something',
namely, mass, entropy, energy etc. This means, an advective disk should have a finite radial
velocity which may even reach the velocity of light on the horizon. Therefore, before
entering into a black hole, matter had to be supersonic (i.e., Mach number,
$M_a = v/a > 1$, where $v$ and $a$ are radial velocity and sound speed respectively). 
As $v \sim 0$ at infinity, the matter must pass through at least one sonic point ($M_a = 1$), 
and as a sub-Keplerian flow (i.e., a flow with specific angular momentum $\lambda < \lambda_K$, 
the Keplerian angular momentum). Close to the black hole the matter falls very rapidly 
towards the horizon, thus making $\lambda \sim constant$. As the matter approaches 
the black hole, the centrifugal force ($\sim 1/r^3$) grows much faster compared 
to the gravitational force ($\sim 1/r^2$) and creates a shock or `boundary layer' 
before entering into a black hole.  For a large region of the parameter space 
(C90), a stable solution can have a standing shock wave. Depending 
on the physical parameters, a shock may be steady or oscillating in nature, 
it may even be absent. The oscillations of the shock may give rise to the 
temporal variability in the form of QPOs which are observed in many of the 
BH candidates. In this `boundary layer', the flow kinetic energy is converted 
into the thermal energy forming a hot Compton cloud which can inverse-Comptonize 
the soft photons into hard photons and produce outflows and winds 
(Chakrabarti, 1999; hereafter C99). 
This boundary layer is called the CENtrifugal pressure supported BOundary Layer (or, CENBOL).

Two Component Advective Flow (TCAF) (Fig. \ref{fig.1.8}) is a combination of two
types of flows: a highly viscous Keplerian component which is accreted in long, 
viscous time scale and a initially sub-Keplerian component, with higher radial 
velocity and lower angular momentum. The sub-Keplerian flow is accreted in 
the short, free-fall time scale. 
The Keplerian disk, because of its low energy, resides at the equatorial
plane, while matter with lower angular momentum flows above and below it. The wind 
is predominantly produced from the CENBOL area, which is the post-shock region. 
A transient shock can also be present just outside the inner sonic point. 
The inner edge of the Keplerian disk is terminated at the shock loaction. 
The amount of matter inside the jets and outflows vary due to the shock-oscillation.

The soft radiation coming from the Keplerian disk is intercepted
by the hot sub-Keplerian flows in the CENBOL region and is re-radiated 
after multiple scattering. This radiative
 transfer between the photons and electrons changes the temperature 
of the electron cloud. Depending on the relative importance of the Keplerian disk
rate $\dot{M}_d$ and the sub-Keplerian halo rate $\dot{M}_h$, the 
electrons in the sub-Keplerian disk may loose (inverse Compton scattering) 
or gain (Compton scattering) energy. When the electron cloud gets hotter, 
the system is in the hard state and if the electron cloud cools down by 
loosing energy to the photons, the system is in the soft state. The 
hard state is thus dominated by a power-law hard photon component. 
In the soft state, the electrons in CENBOL cooled down and collapses, but
due to the inner boundary condition (matter must be supersonic when it reaches 
the horizon), the matter accelerates rapidly and up-scatters photons 
to its energy ($\sim m_ec^2$). Thus, the power-law photons can also
be seen in the soft state due to this so-called, bulk motion Comptonization (BMC) (CT95). 

\begin{figure}[]
\begin{center}
\vskip 5.0cm
\includegraphics{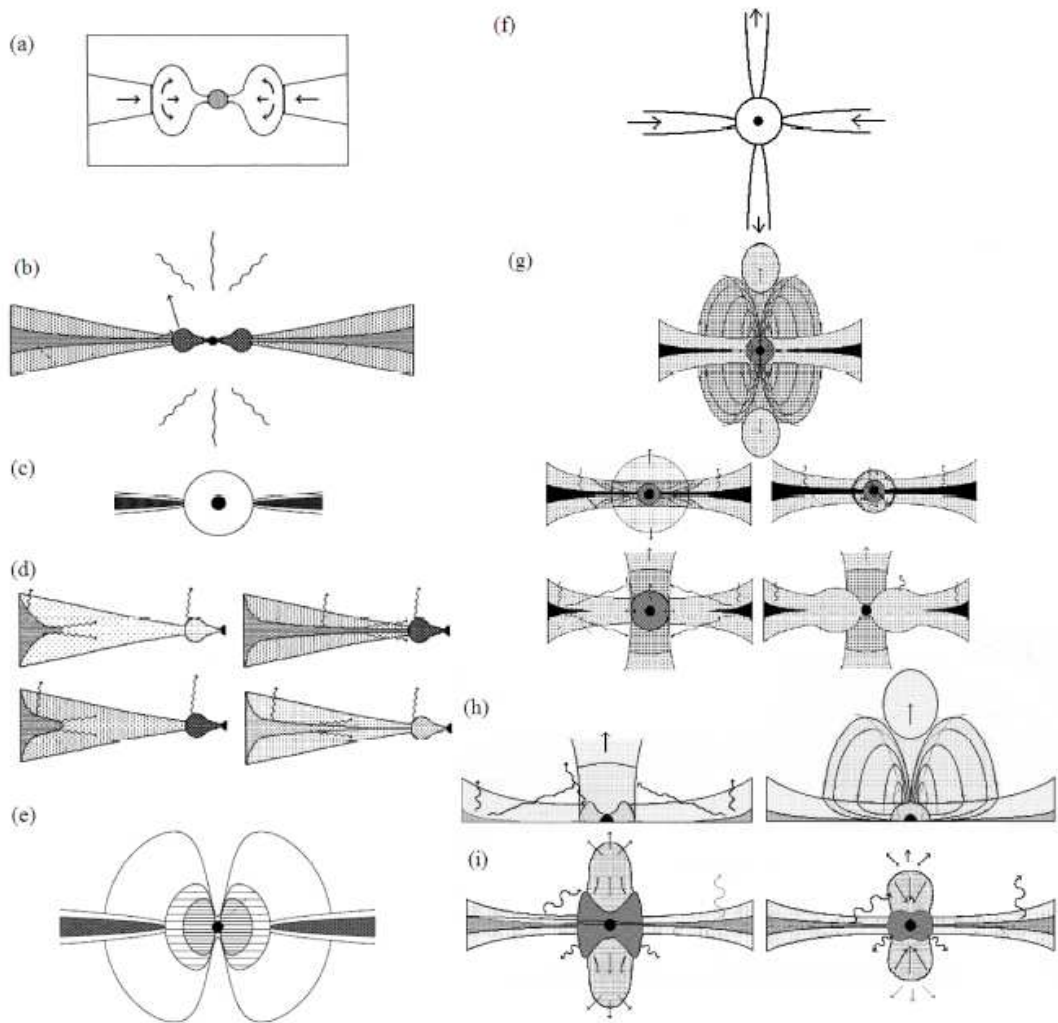}
\vskip 6.3cm
\caption{Evolution of the solution of the transonic flow to keep up with more 
and more observational data. (a) C90 ; (b) CT95;
(c) Ebisawa, Titarchuk \& Chakrabarti, 1996; (d) Chakrabarti \& Sahu, 1997; (e) Chakrabarti,
1997; (f) Chakrabarti, 1998ab; (g) Chakrabarti \& Nandi 2000; (h) Nandi et al. 2001; (i)
Chakrabarti et al. 2002 (Figure taken from CGS08). The exact configuration would depend on the 
mass accretion rates  in the two components of the flow.
}
\label{fig.1.9}
\end{center}
\end{figure}
In Fig. \ref{fig.1.9} we present the evolution of the solution 
of transonic flows since 1990 (Chakrabarti, Ghosh \& Som, 2008; 
hereafter CGS08). With more and more applications, the solution is also 
enriched. Each diagram is drawn with specific solution in mind.

\begin{figure}[]
\begin{center}
\vskip 5.0cm
\includegraphics{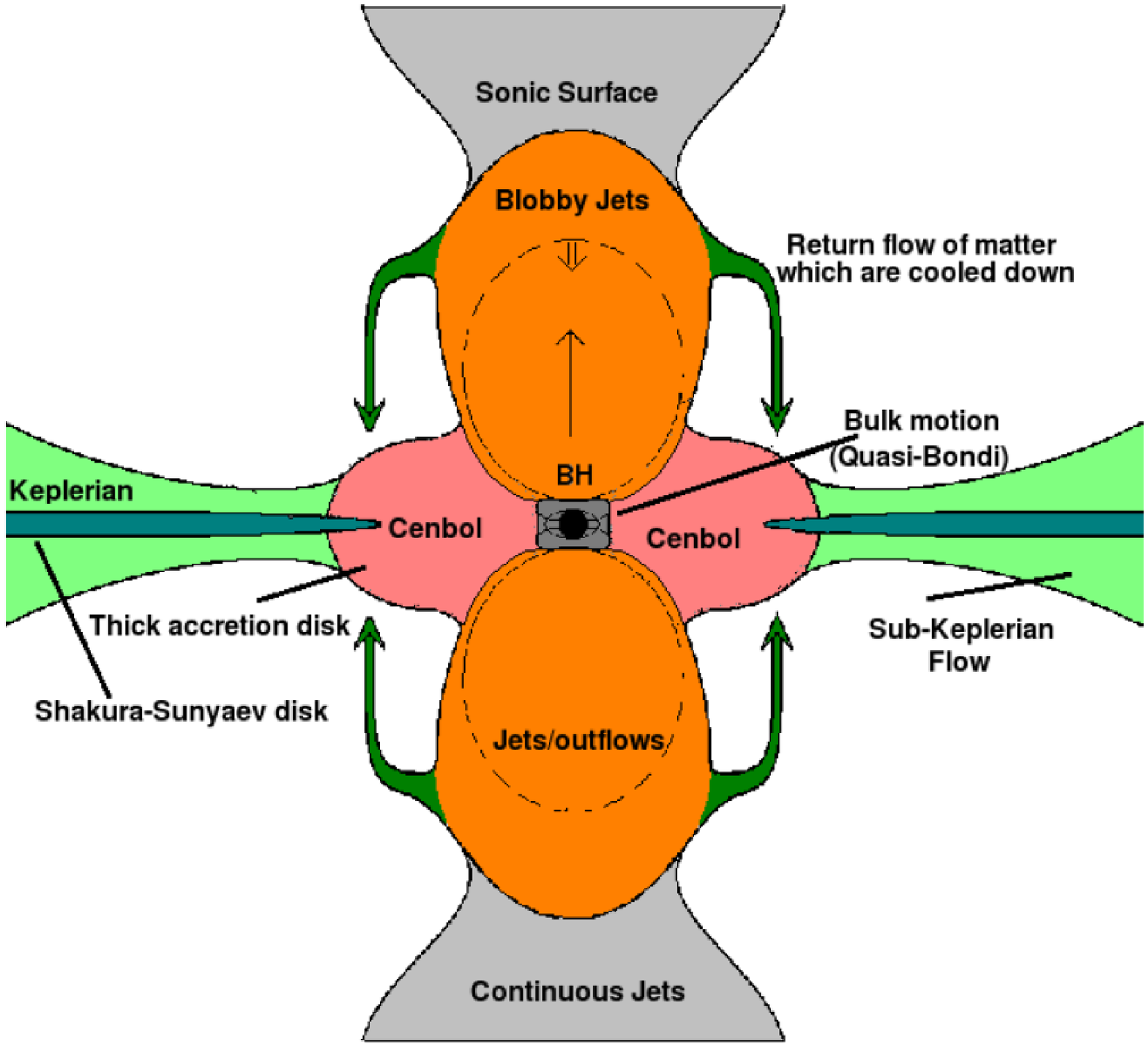}
\vskip 8.3cm
\caption{The most general type of the accretion flow which seems to encompass all the
relevant flow solutions such as Bondi flow (close to black hole horizon), standard disk (on
equatorial plane), Transonic sub-Keplerian flow (away from the plane) with shocks (CENBOL),
Thick accretion disks (post-shock flow) and the jets and outflows from CENBOL. If the CENBOL is cooled down,
then there is no outflow. If the outflows are also cooled down, some matter could fall back 
on the disk, giving rise to burst-on states (Chakrabarti, 2008).}
\label{fig.1.10}
\end{center}
\end{figure}

The general type of accretion is shown in Fig. \ref{fig.1.10} (adapted from Chakrabarti, 2008). 
We see that different parts of the flow actually behaves as individual solutions already
existing in the literature. Thus in reality, an accretion flow is a complex mixture of all
the known types of solutions: Bondi, transonic, Keplerian, thick disks, sub-Keplerian
(transonic) and outflows (transonic). In various situations, various components  may become
dominant.
\section{Spectral and Timing Studies using TCAF}

Study of spectral as well as the timing properties of the black hole candidates can only give 
us the vital clues of the understanding of the invisible central object. 
The spectrum of radiation, particularly in high energies 
give information about the thermodynamic properties of matter accreting onto a black hole. The timing 
properties give information about how these thermodynamic properties are changing with time.
The thermodynamic properties such as the mass density, temperature etc. and the dynamic 
properties such as the velocity components are the solutions of the governing equations. 
Thus, a thorough knowledge about the spectral and timing properties are essential (Chakrabarti, 1996a). 

CT95 and Chakrabarti (1997, hereafter C97) pointed out that the dynamic corona of 
a disk, namely the inner part of the sub-Keplerian flow
is indeed the Compton cloud. They described that the post-shock 
region of a rotating sub-Keplerian flow could actually be the Compton 
cloud and the Keplerian flow on the equatorial plane supplies soft photons 
to it to be inverse Comptonized. Simply put, in this so-called two component 
advective flow (TCAF) model, the state of a black hole is decided by the 
relative importance of the processing of the {\it intercepted} soft photons 
emitted from a Keplerian disk by the puffed-up post-shock region (CENBOL) formed in 
the sub-Keplerian halo. 
The TCAF model showed that the spectral properties are direct consequences 
of variation of accretion rates of the 
Keplerian (disk) and sub-Keplerian (halo) components. 
If the CENBOL remains hot (generally due to smaller number of soft-photons from a Keplerian 
disk having lower accretion rate), and emits hard X-rays, it is the 
low/hard state since more power is in the hard X-ray region. On the contrary, 
if the CENBOL is cooled down by copious number of intercepted photons, 
the black hole is in the high/soft state. CT95 also pointed out 
that even in a high/soft state, some electrons should be energized by the 
momentum deposition due to the bulk motion of the electrons rushing towards 
the horizon. These photons would have a almost constant spectral slope. 
This was later verified by Monte Carlo simulations (Laurent \& Titarchuk 1999; 2001).

Subsequenly, efforts were made to explain the timing properties 
using the TCAF model. Several numerical simulations including 
radiative processes (MSC96, Chakrabarti, Acharyya and Molteni 2004) indicate
that the resonance effects between the cooling time scale and the infall time scale causes the shocks (Chakrabarti, 
1989) to oscillate and cause the most important feature of the power density spectrum, namely, the 
quasi-periodic oscillations (QPOs) to appear. Thus is it generally established that 
the sub-Keplerian flows are responsible for both the spectral and timing properties of 
the black hole candidates. This is shown by several observations (Smith, Heindl, \& Swank, 2002; 
Wu et al. 2002, Soria et al. 2001, Pottschmidt et al. 2006, Dutta \& Chakrabarti, 2010).

\section{Relevant Radiation Mechanisms}

The observed spectrum in X-ray binaries is the superposition of 
several spectral components originating from different regions of the
 system. In this Section, we will summarize the radiation emission 
processes in X-ray binaries (Longair, 1981; Diehl, 2001). 
The details of these processes will be given in Chapter 2.
\vskip .5cm
\noindent
$\bullet$ {\bf Thermal emission}
\vskip .1cm
\noindent
High temperature gases can produce thermal emission in the form
of a black body radiation in the X-ray regime. This situation is 
very common in very hot in stellar winds, accretion disks etc.
\vskip .5cm
\noindent
$\bullet$ {\bf Compton scattering}
\vskip .1cm
\noindent
This is the interaction of photons with stationary
electrons, in such a way that the high energy photon loses energy
which is gained by the electron.
\vskip .5cm
\noindent
$\bullet$ {\bf Inverse Compton scattering}
\vskip .1cm
\noindent
In this interaction photon gains energy which is lost by the very 
energetic particle. This scattering happens in case of the accretion disk 
surrounding black holes or neutron stars in X-ray binary systems due 
to the presence of hot ($\sim 10^9$ K) plasma which becomes highly 
ionized, i.e., with a large amount of highly energetic free electrons.
\vskip .5cm
\noindent
$\bullet$ {\bf Synchrotron radiation}
\vskip .1cm
\noindent
Charged particles accelerate when interacting with a magnetic field. 
The trajectory of these particles is modified as the particle looses energy. 
The lost energy is emitted in the form of synchrotron radiation. The energy
components perpendicular to the magnetic field are quantized. Thus,
photons can be emitted or absorbed in frequencies corresponding to the
energy difference between such levels, and Cyclotron emission or absorption
can be produced.
\vskip .5cm
\noindent
$\bullet$ {\bf Bremsstrahlung}
\vskip .1cm
\noindent
Charged particles can also interact with electrostatic
fields, for example, the case of an electron passing by an
ionized atom. In this case the charged particle is 'braked' and the energy
lost is emitted in the form of bremsstrahlung emission.
\vskip .5cm
\noindent
$\bullet$ {\bf Photoelectric Absorption}
\vskip .1cm
\noindent
The atomic electrons are removed from their nuclei by the X-ray 
and $\gamma$-ray photons, thus the incident photon is absorbed 
by an electron whose binding energy was equal to the photon's energy.
 
In this Thesis, we consider only three types of radiative processes: thermal 
emission, Compton and inverse-Compton scattering.

\section{General Relativistic and Pseudo-Newtonian Approaches}

General relativity, though it is elegant and beautiful, may not be easy to use in presence
of complex physical processes. In such cases, sometimes 
it is easy to work with pseudo-potentials. In realistic cases 
when heating and cooling are important, the general relativistic 
calculations become extremely difficult and time consuming. Under 
these circumstances, one easy way out is to use the so-called 
pseudo-Newtonian geometry. This will mimic the black hole 
surrounding with very little error and we can work within the 
realms of Newtonian physics or special relativity (as the problem 
may be). We will briefly describe the environment of a 
non-rotating black hole and the pseudo-potential which will 
mimic it. In this Thesis we have used this potential to take care 
of the general relativistic effects.

\subsection{GR Approach}
The region around a non-rotating compact object is described by 
the Schwarzschild metric, and is given by,
$$
ds^2=-\left(1-2/r\right)dt^2+\left(1-2/r\right)^{-1}dr^2+r^2 d\theta^2 + r^2 sin^2 \theta d\phi^2.
\eqno{(1.10)}
$$
Here, we use the spherical polar coordinate. Also we adopt units where the
gravitational constant $G$, the central mass $M$ and the velocity of light
$c$ are all unity ($G=M=c=1$). Solving the geodesic equation one gets 
$$
-u_t = \left(1-2/r\right) u^t = E = \rm{constant\hskip 0.1cm of\hskip 0.1cm motion}
\eqno{(1.11)}
$$
and, for $\theta = \pi/2$,
$$
u_{\phi} = r^2 u^{\phi} = l = \rm{constant\hskip 0.1cm of\hskip 0.1cm motion}.
\eqno{(1.12)}
$$
Using the above equations we can get,
$$
\left( \frac{dr}{ds}\right)^2 = E^2 - \left(1 - \frac{2}{r} \right) \left(1 + \frac{l^2}{r^2} \right).
\eqno{(1.13)}
$$
The second term in the RHS of Eqn. (1.13), behaves like an effective potential, 
$$
V_{eff}^2 = \left(1 - \frac{2}{r} \right) \left(1 + \frac{l^2}{r^2} \right).
\eqno{(1.14)}
$$
\begin{figure}[]
\begin{center}
\vskip 5.0cm
\includegraphics{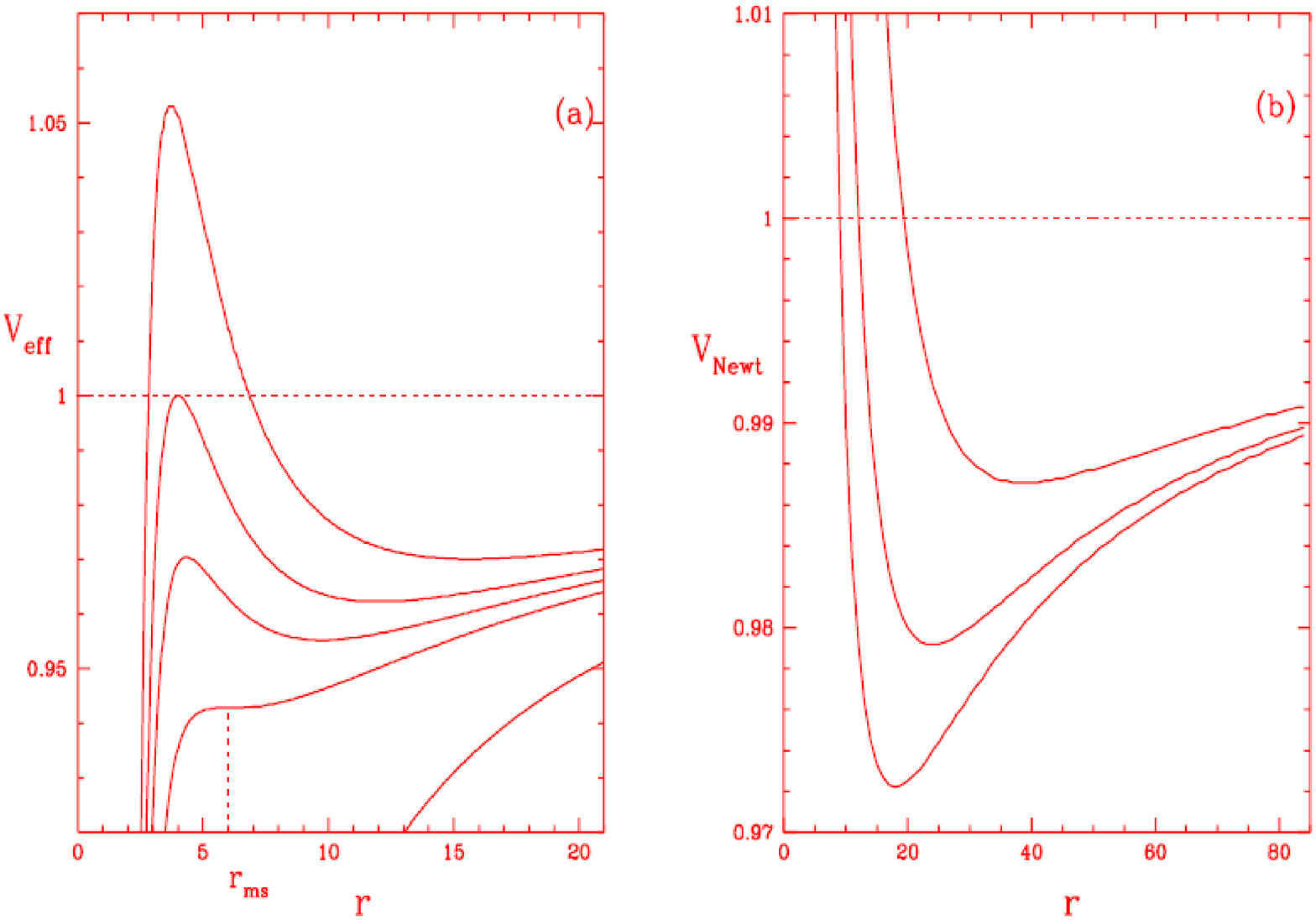}
\vskip 8.3cm
\caption{Comparison of general relativistic (a) and Newtonian (b), effective 
potentials. (a) The effective potential $V_{eff}$ is drawn for the values of 
$l = 0$, $3.464$, $3.75$, $4$ and $4.4$, from lowest curve upwards, respectively. 
$r_{ms}$ is the marginally stable radius. The dotted line for which $V_{eff} = 1$, 
denotes rest mass energy of particle falling into the black hole. 
(b) The Newtonian effective potential $V_{Newt}$, is drawn for the 
values of $\ell = 3$, $3.464$ and $4.4$ from the lowest curve upwards, 
respectively (Chattopadhyay, 2002).}


\label{pseudopot}
\end{center}
\end{figure}
The gravitational potential that a test particle around a 
Newtonian star feels is given by,
$$
\Phi_N = -\frac{1}{r}.
\eqno{(1.15)}
$$
The effective potential of a rotating gas with specific angular
momentum $l$ is obtained by the summation of the gravitational
potential and the centrifugal potential.
$$
V_{Newt} (r) = 1 + \Phi_N + \frac{1}{2} \frac{\ell^2}{r^2}.
\eqno{(1.16)}
$$

In Fig. \ref{pseudopot} we have plotted both $V_{eff}$ and $V_{Newt}$ against $r$ 
to compare the GR potential with the Newtonian one. 
We see some remarkable features. First, we notice that for small $r$, $V_{eff}$ 
dives down with decreasing $r$ and $V_{eff} = 0$ at $r = r_g = 2$. $r_g$ is 
called the event horizon or the Schwarzschild radius. Second, no matter how 
hard one throws a particle at a Newtonian star it will bounce back, 
while a black hole definitely consumes a particle if it is thrown hard enough. 
The particle then enters the black hole in trajectories known as `capture orbits'. 
We will focus on the bound particles. Conditions for the circular orbits are: 
(a) $\frac{\delta V_{eff}}{\delta r} = 0$ and, (b) $\frac{dr}{ds} = 0$.

Condition (a) gives the following equation:
$$
r^2 - l^2r + 3l^2 = 0.
\eqno{(1.17)}
$$
Thus, for $l \geq 2\sqrt3$, real values of $r$ exists, which implies $V_{eff}$ 
has an extremum for $l \geq 2\sqrt3$. For $l = 2 \sqrt3$, we have the position 
for the {\it{last stable orbit}} or the {\it{marginally stable orbit}} ($r_{ms}$; 
shown in Fig. \ref {pseudopot}). Putting $l = 2 \sqrt3$ in Eqn. (1.17), we get 
$r_{ms} = 6$. Imposing condition (b) in Eqn. (1.14), and using Eqn. (1.17) we get,
$$
E^2 = \frac{\left(r - 2\right)^2}{r\left(r - 3\right)}.
\eqno{(1.18)}
$$
If we take the definition of specific angular momentum as, $- u_\phi/u_t$ 
(Chakrabarti 1996a), the specific Keplerian angular momentum curve is the 
locus of the extrema of $V_{eff}$. Hence,
$$
\lambda_{Kep}^2 = \left(- \frac{u_\phi}{u_t}\right)^2 = \frac{r^3}{\left(r - 2\right)^2}.
\eqno{(1.19)}
$$
Putting $r = 6$ and $l = 2 \sqrt3$ in Eqn. (1.14), we have,
$$
V_{eff} (r_{ms}) = \sqrt{\frac{8}{9}}.
\eqno{(1.20)}
$$
Therefore, the binding energy at $r_{ms}$ is,
$$
E_{bind} = 1- \sqrt{\frac{8}{9}} = 5.72 \%.
\eqno{(1.21)}
$$
When a particle enters into a black hole $E_{bind}$ amount of energy will 
be liberated as radiation.

\subsection{Pseudo-Newtonian Approach}

In the case of most of the astrophysical systems involving
a rotating compact star or a black hole, it is not essential that one solves
the problem using full general relativity. Fortunately, a few tools
are now available which allow one to use the Newtonian concepts
(such as equations in flat geometry, additivity of the energy components,
etc.) at the same time retaining the salient features of a black hole
geometry. As long as one is not interested in processes very close 
(within $1-2r_g$) to the horizon, one may safely use these tools and 
obtain sufficiently accurate results. The potential we describe was proposed 
by PW80 and is called {\it{Paczy\'nski-Wiita potential}}. 
Paczy\'nski-Wiita potential or pseudo-Newtonian potential is given by,
$$
\Phi_{PW}= -\frac{1}{\left(r-2\right)}.
\eqno{(1.22)}
$$
Adding the rest mass energy to this potential and then writing the effective potential,
$$
V_{eff(PW)} = 1 + \frac{\ell^2}{2r^2} + \Phi_{PW}.
\eqno{(1.23)}
$$
Putting the condition $\delta V_{eff(PW)}/\delta r = 0$, we get,
$$
\ell_{Kep}^2 = \frac{r^3}{\left(r-2\right)^2}.
\eqno{(1.24)}
$$
We find that the specific Keplerian angular momentum produced by $\Phi_{PW}$ 
is same as that produced by exact GR calculations. We have already seen that 
$r_{ms}$ is the position of the minima of the Keplerian angular momentum. 
Taking the minima of Eqn. (1.24), we find, $r_{ms} = 6$.

Please note that we have used two different notations for specific angular 
momentum in GR ($\lambda$) and in pseudo-Newtonian ($\ell$) description, 
only to differentiate these two approaches. The binding energy at $r_{ms}$ 
in pseudo-Newtonian description is,
$$
E_{bind} = 1 - V_{eff(PW)}\left(r_{ms}\right) = 6.25 \%.
\eqno{(1.25)}
$$
Thus, we find that the pseudo-Newtonian approach is quite accurate, 
and the error is within few percent. In the next Chapters, we shall 
use $\Phi_{PW}$ to take care of the general relativistic effects.
 
\subsection{Some Remarks About Units and Dimensions}

The radius of a non-rotating black hole $r_g = 2GM/c^2$ is only $3.0$ km if $M = M_\odot$. 
A stellar mass black hole generally has $M > 3 M_\odot$. Hence the radius is
around $9$ km. For a super-massive black hole, one can scale these numbers
depending on the mass of the black hole. However, the physical processes in
accretion flows generally have length scales of the order of the Schwarzschild
radius and thus, it is convenient to choose this the unit of length.
Similarly, it is well known that the velocity of in-falling matter through the
horizon is equal to the velocity of light (Chakrabarti, 1996bc). Thus it is
expected that matter and sound velocities would be of this order and
thus the units may be chosen accordingly.

Keeping these in mind, we choose, $G=c=M =1$.
In this case the unit of velocity would be $c$,
the unit of distance would be $2GM/c^2$ (the Schwarzschild radius),
the unit of time would be $2GM/c^3$ and the unit of angular momentum
would be $2GM/c$. In this unit system, the pseudo-Newtonian potential
is written as  $-\frac {1}{2(r-1)}$.


%% file: CHAPTER2/chap2.tex

\newpage
\markboth{\it Radiation Processes in Accretion Flows}
{\it Radiation Processes in Accretion Flows}
\chapter{\bf Radiation Processes in Accretion Flows}
We discussed that black holes can be detected by 
monitoring the radiations emitted by matter 
accreting on them. These radiations are
emitted after various interactions which happen inside 
the infalling matter. Processes like bremsstrahlung 
and synchrotron radiation generate new photons. 
In Compton scattering, the energy exchange 
happens between the electrons and the photons. The flow 
parameters, density and scattering cross-section
determine the optical 
depth ($\tau$) of the medium, which determines
the number of scattering inside the medium. 
Depending on $\tau$, a photon can suffer a single 
scattering or multiple scatterings or no scattering at all
inside the accretion disk which is manifested in 
the radiation spectrum. The radiation pressure can 
also affect the hydrodynamics, but presently we
are not considering that effect. There are two 
main radiation types: thermal and non-thermal. Thermal 
radiation is the radiation emitted by matter satisfying 
Maxwell-Boltzmann thermodynamic equilibrium conditions 
for particle velocities. It is characterized by the 
temperature of the emitting gas. Black-body 
radiation is an example of thermal radiative process. 
Non-Thermal radiation is emitted by particles not in thermal 
equilibrium. Synchrotron, bremsstrahlung and Compton
processes {\it could be} non-thermal radiations. 
Below we discuss the black body, and Compton radiation. 
We will not discuss about the bremsstrahlung and synchrotron
 processes because in this Thesis we have only considered 
black body, and Compton radiation. We have followed Longair (1981), 
Rybicki \& Lightman (1979, hereafter RL79) and Poznyakov, Sobol \& Syunyaev (1983, 
hereafter, PSS83) to write this Chapter.

\section{Black Body Radiation}

If a body is irradiated with radiation of frequency $\nu$, and a 
fraction $a(\nu)$ of that radiation is absorbed and the remainder 
being either reflected or transmitted, $a(\nu)$  would be called 
the absorptance at frequency $\nu$. The fractions of the radiation 
reflected and transmitted are called the reflectance and 
the transmittance respectively. The sum of the absorptance, 
reflectance and transmittance is unity. 

A body for which $a(\nu) = 1$ for all frequencies is a black body.

Blackbody radiation is the radiation which is itself in thermal equilibrium
with the surroundings. The spectral volume density of radiation energy can be determined 
by calculating the equilibrium distribution of photons, for which 
the radiation field entropy is maximum. The energy of the photon 
with frequency $\nu$ is equal to $h \nu$, where $h$ is the Planck 
constant. Assuming the radiation field to be a gas obeying the 
Bose-Einstein statistics, then we obtain the Planck formula for 
the energy per solid angle per volume per frequency of radiation:
$$
u_{\nu} (T,\Omega) = \frac{2 h \nu^3}{c^3} \frac{1}{\left[exp(h\nu/k_BT) -1\right]},
\eqno{(2.1)}
$$
where $k_B$ is the Boltzmann constant. The intensity of the black body 
radiation $I_{\nu} (T,\Omega) = u_{\nu} (T,\Omega) \times c$, so that
$$
I_{\nu} (T,\Omega) = \frac{2 h \nu^3}{c^2} \frac{1}{\left[exp(h\nu/k_BT) -1\right]},
\eqno{(2.2)}
$$
Eqn. (2.2) expresses the {\it{Planck law}}. A plot of $I_{\nu}$ 
versus $\nu$ for a range of values of $T$ ($1$ K $< T < 10^7$ K) 
is given in Fig. 2.1.
\begin{figure}[]
\begin{center}
\vskip 4.1cm
\includegraphics{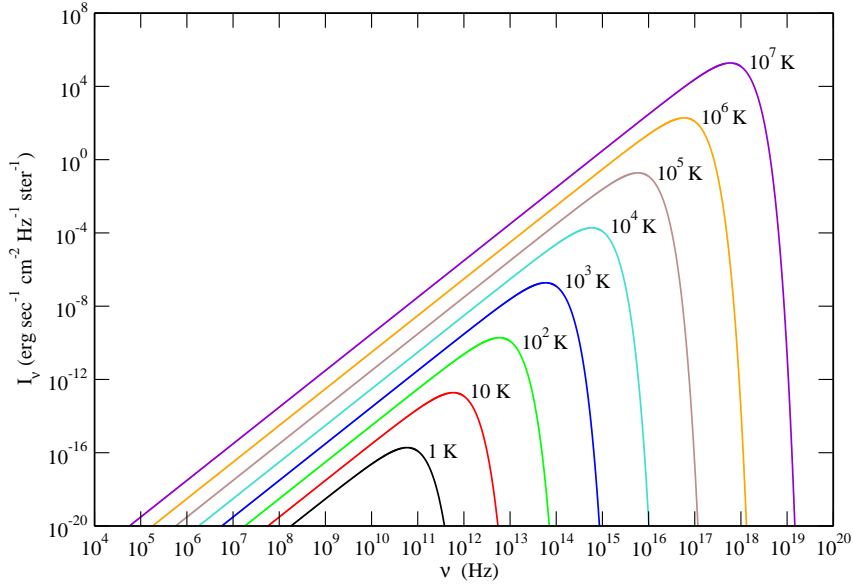}
\vskip 4.0cm
\caption{Spectrum of a blackbody radiation at various temperatures.}
\label{fig2.1}
\end{center}
\end{figure}

$\bullet$ {\bf The Rayleigh-Jeans Law ($h\nu \ll k_BT$):} In this case, 
we can expand the exponential as,
$$
exp\left( \frac{h\nu}{k_BT} \right) - 1 = \frac{h\nu}{k_BT} + ...
$$
Thus, for $h\nu \ll k_BT$ range of photon energy, we have the 
{\it{Rayleigh-Jeans Law}}:
$$
I^{RJ}_{\nu} (T,\Omega) = \frac{2\nu^2}{c^2} k_BT.
\eqno{(2.3)}
$$
The  Rayleigh-Jeans Law is valid at low frequencies. It gives the 
straight-line part of the log $I_{\nu}$ - log $\nu$ plot in Fig. 2.1. 
An important feature of Eqn. (2.3) is that, if we integrate 
$I^{RJ}_{\nu} (T,\Omega)$ over all frequencies, the total 
energy $\propto \int{\nu^2 d\nu}$ will diverge. This is known as 
the {\it{ultraviolet catastrophe}}.

$\bullet$ {\bf Wien Law ($h\nu \gg k_BT$):} This limit allows us 
to neglect the unity term in the denominator in comparison with 
$exp(h\nu / k_BT)$. Thus {\it{Wien Law}} gives:
$$
I^{W}_{\nu} (T,\Omega) = \frac{2h\nu^3}{c^2}exp\left(\frac{-h\nu}{k_BT}\right).
\eqno{(2.4)}
$$
The steep portions of curves in Fig. 2.1 are associated with the 
Wien Law.

$\bullet$ {\bf Wien Displacement Law:} The frequency $\nu_{max}$ 
at which the peak of $I_{\nu}$ occurs is given by,
$$
\left[ \frac{\partial I_\nu}{\partial \nu} \right]_{\nu=\nu_{max}} = 0,
$$
or,
$$
\frac{\nu_{max}}{T} = 5.88 \times 10^{10} \hskip 0.1cm\rm{Hz}\hskip 0.1cm \rm{deg}^{-1}.
\eqno{(2.5)}
$$
Eqn. (2.5) shows that the peak frequency shifts linearly with 
black body temperature. This is known as the {\it{Wien Displacement Law}}.

$\bullet$ {\bf Number of black body photons:} Number of black body 
photons emitted per unit volume per solid angle per sec per frequency 
of radiation is given by,
$$
N_{\nu} (T,\Omega) = \frac{u_{\nu} (T,\Omega)}{h\nu}.
$$
Integrating over the solid angle $\Omega$ and frequency we get,
$$
N (T) = \int_{\nu=0}^{\infty}{\int_{\Omega}{\frac{u_{\nu} (T,\Omega)}{h\nu}}d\Omega d\nu}.
$$
Performing the above integration we get the total number of emitted black body 
photons per unit volume per sec for temperature T, as,
$$
N (T) = 16 \pi \times 1.202057 \times \left(\frac{k_B}{hc}\right)^3 T^3.
\eqno{(2.6)}
$$
Thus, the number of photons per unit volume is proportional to 
the cube of the black body temperature.

\section{Scattering Processes}

\subsection{Thomson Scattering}
\begin{figure}[]
\begin{center}
\vskip 5.0cm
\includegraphics{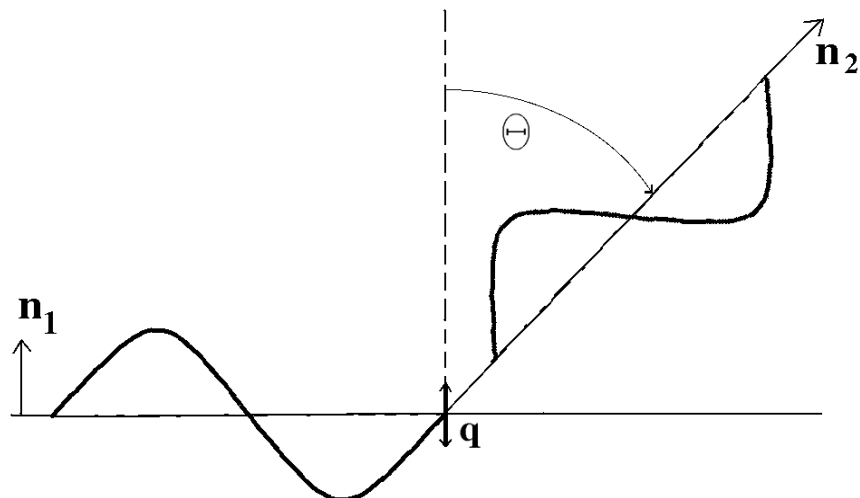}
\vskip 3.0cm
\caption{Scattering of polarized radiation by a charged particle.}
\end{center}
\end{figure}

This is a frequency-independent nonrelativistic scattering.
Suppose a lonely electron has an electromagnetic wave incident on it. The wave has a
frequency $\omega_0$.
The electric force from a linearly polarized wave is
$$
{\bf F} = q E_0 \sin{\omega_0 t} {\bf n_1} ,
$$
where $q$ is the charge and ${\bf n_1}$ is the direction of the {\bf E}-field.
From the Newton's second law and the definition of the dipole
moment ${\bf d} = q {\bf r}$, we get
$$
{\bf \ddot{d}} = \frac{q^2 E_0}{m_e} \sin{\omega_0 t} {\bf n_1},
$$
and
$$
{\bf d} = -\left(\frac{q^2 E_0}{m_e \omega_0^2} \right) \sin{\omega_0 t} {\bf n_1},
$$
which gives an oscillating dipole of amplitude
$$
{\bf d_0} = \frac{q^2 E_0}{m_e {\omega_0}^2} {\bf n_1}.
$$
Larmor's formula gives the time-averaged power per solid angle:
$$
\frac{dP}{d\Omega}=\frac{q^4 {E_0}^2}{8 \pi {m_e}^2 c^3} {\sin^2{\Theta}}.
\eqno{(2.7)}
$$
The incident flux is $\left< S \right> = (c/8\pi){E_0}^2$, so we can define a
differential cross section $d\sigma$ for scattering into the solid angle $d\Omega$:
$$
\frac{dP}{d\Omega}=\left< S \right> \frac{d\sigma}{d\Omega}=\frac{c{E_0}^2}{8\pi}\frac{d\sigma}{d\Omega}.
\eqno{(2.8)}
$$
Therefore, we have the relation
$$
\left(\frac{d\sigma}{d\Omega}\right)_{pol}=\left( \frac{q^4}{m_e c^4} \right) {\sin^2{\Theta}}.
\eqno{(2.9)}
$$
For an electron $r_0 = q^2/m_e c^2$ is known as `classical radius of the electron',
$r_0 \approx 2.8\times10^{-13}$ cm. It is the radius that gives an
electrostatic energy equal to $m_ec^2$. Integrating over all angles,
one finds the total cross section to be
$$
\sigma =\left( \frac{8\pi}{3} \right) {r_0}^2.
\eqno{(2.10)}
$$
For an electron, $\sigma = \sigma_T = 6.65\times10^{−25}$ $cm^2$.

We can now compute the differential cross section for an unpolarized radiation
by the superposition of two orthogonally polarized waves. If we define $\theta$
as the angle between the scattered radiation and the original radiation
(so that $\theta = \pi/2 - \Theta$), we have the result
$$
\left(\frac{d\sigma}{d\Omega}\right)_{unpol}=\frac{1}{2}\left[ \left(\frac{d\sigma(\Theta)}{d\Omega}\right)_{pol}+\left(\frac{d\sigma(\theta)}{d\Omega}\right)_{pol}\right]
=\frac{1}{2} {r_0}^2\left(1+{\sin^2{\Theta}}\right)
=\frac{1}{2} {r_0}^2\left(1+{\cos^2{\theta}}\right).
\eqno{(2.11)}
$$
From Eqn. (2.11), we find that the total cross section depends only on the angle
 between the incident and scattered directions, as it should be, since
 an electron at rest has no intrinsic polarization, so it has to react
 to all linear polarizations in the same way.

\subsection{Compton Scattering}

In Thomson scattering, there is no change in the frequency of the 
incoming radiation. The electron simply acts as a radiator which 
scatters the incoming radiation. This remains a good approximation 
provided the energy of the photon is much less than the rest mass 
energy of the electron i.e., $h \nu \ll m c^2$. However for high 
energy of the incident photon the collision between a free electron 
and photon leads to a change in frequency of the photon and it has 
important implications in high energy astrophysics. This is known 
as the Compton scattering. For more detailed description one may 
read PSS83.
\begin{figure}[]
\begin{center}
\vskip 5.0cm
\includegraphics{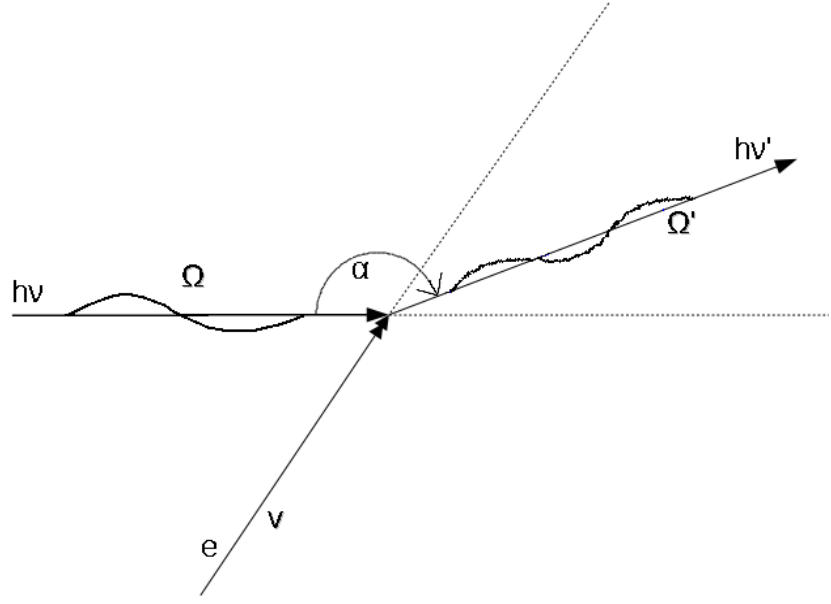}
\vskip 2.5cm
\caption{The Compton effect. A photon with energy $h\nu$ is scattered 
by an electron, moving with velocity $\bf{v}$. The scattered 
photon has energy $h\nu'$ and the scattering angle is $\alpha$.}
\end{center}
\end{figure}

Let us consider a photon of energy $h\nu$ and momentum 
$\frac{h\nu}{c}\bf{\widehat{\Omega}}$ is scattered 
by an electron of energy $\gamma mc^{2}$ and 
momentum $\overrightarrow{\bf{p}} = 
\gamma m \overrightarrow{\bf{v}}$, where the electron Lorentz factor 
$\gamma = \left( 1 - \frac{v^2}{c^2}\right)^{-1/2}$. 
Let, $h\nu'$ and $\frac{h\nu'}{c}\bf{\widehat{\Omega'}}$ 
denote the energy and momentum of the photon after 
the scattering.

\subsubsection{Shift in Photon Frequency}

If we introduce the electron and photon four-momenta before 
scattering as: $p_{el}$ and $p_{ph}$, respectively and $p_{el}'$ 
and $p_{ph}'$, after the scattering, then $p_{el}$, $p_{ph}$, 
$p_{el}'$, $p_{ph}'$ are given by:
$$
p_{el} = \left(\overrightarrow{\bf{p}}, i\gamma m c \right),
$$
$$
p_{ph} = \left(\frac{h\nu}{c}{\bf{\widehat{\Omega}}}, \frac{ih\nu}{c} \right),
$$
$$
p_{el}' = \left(\overrightarrow{\bf{p'}}, i\gamma' m c \right),
$$
$$
p_{ph}' = \left(\frac{h\nu'}{c}{\bf{\widehat{\Omega'}}}, \frac{ih\nu'}{c} \right).
$$
Conservation of energy and momentum gives,
$$
p_{el} + p_{ph} = p_{el}' + p_{ph}'.
\eqno{(2.12)}
$$
Squaring this relation and noting that:
$$
{p_{el}}^2 = {p_{el}'}^2 = -m^2c^2,
$$
and,
$$
{p_{ph}}^2 = {p_{ph}'}^2 = 0,
$$
we find,
$$
p_{el} p_{ph} = p_{el}' p_{ph}'.
\eqno{(2.13)}
$$
Now, multiplying Eqn. (2.12) by $p_{ph}'$ we find
$$
p_{el}' p_{ph}' = p_{el} p_{ph}' + p_{ph}p_{ph}'.
\eqno{(2.14)}
$$
Using Eqn. (2.13) we rewrite Eqn. (2.14) as:
$$
p_{el} p_{ph} = p_{el} p_{ph}' + p_{ph}p_{ph}'.
\eqno{(2.15)}
$$
Defining $\mu = {\bf{\widehat{\Omega}}}.\widehat{\bf{v}}$, 
$\mu' = \bf{\widehat{\Omega'}}.\widehat{\bf{v}}$, 
and the scattering angle 
$\alpha = \cos ^{-1}{\left(\bf{\widehat{\Omega}}.
\bf{\widehat{\Omega'}}\right)}$, we may write from Eqn. (2.15),
$$
\frac{\nu'}{\nu} = \frac{1 - \mu v/c}{1 - \mu' v /c + \left( h \nu / \gamma m c^2\right) \left(1 - \cos{\alpha} \right)}.
\eqno{(2.16)}
$$
If the photon is scattered by an electron at rest $\left(v = 0 \right)$, 
its frequency will change solely because of {\it{recoil}} effect of the electron:
$$
\frac{\nu'}{\nu} = \frac{1}{1 + \left( h \nu / m c^2\right) \left(1 - \cos{\alpha} \right)}.
\eqno{(2.17)}
$$
In the case of a photon with $h \nu \ll m c^2$ we will have
$$
\frac{\Delta \nu}{\nu} = - \frac{h \nu}{m c^2} \left( 1 - \cos{\alpha}\right).
$$
If instead the electrons are traveling at high speed, the {\it{Doppler effect}} 
will play the dominant role in changing the frequency of low-energy photons. 
For a reference frame comoving with the scattering electron, the photon 
frequency prior to the scattering event will be 
$$
\nu_{0} = \gamma v \left( 1 - \frac{\mu v}{c}\right), 
$$
and if $h \nu_{0} \ll m c^2$, we can assume that there is no change in 
scattered photon energy in the electron rest frame: $\nu_{0}' \approx \nu_{0}$. 
When this condition is satisfied, the scattering is closely elastic.
Reverting to the laboratory reference frame, we obtain
$$
\nu' = \frac{\nu_{0}'}{\gamma \left(1 - \mu' v /c\right)} \approx \frac{\nu_{0}}{\gamma \left(1 - \mu' v /c\right)} = \nu. \frac{\left(1 - \mu v /c\right)}{\left(1 - \mu' v /c\right)},
\eqno{(2.18)}
$$
which will agree with Eqn. (2.16) if the condition $h\nu/\gamma m c^2 \ll 1$ holds. 
In the nonrelativistic limit $v/c \ll 1$, Eqn. (2.18) gives
$$
\frac{\nu' - \nu}{\nu} = \frac{v}{c} \left(\mu' - \mu \right) + \frac{v^2}{c^2} \mu'^2.
\eqno{(2.19)}
$$

\subsubsection{Scattering Cross Section}
The differential cross section for Compton scattering can be calculated 
using quantum electrodynamics and is expressed by the formula:
$$
\frac{d\sigma}{d\Omega'} = \frac{r_{e}^{2}}{2\gamma^{2}} \frac{X}{\left( 1 - \mu v/c\right)^{2}} \left( \frac{\nu}{\nu'}\right)^{2},
\eqno{(2.20)}
$$
where the function
$$
X = \frac{x}{x'} + \frac{x'}{x} + 4\left(\frac{1}{x} - \frac{1}{x'} \right) + 4\left(\frac{1}{x} - \frac{1}{x'} \right)^{2},
\eqno{(2.21)}
$$
$$
x = \frac{2h\nu}{m c^2} \gamma \left(1 - \mu \frac{v}{c} \right),
\eqno{(2.22)}
$$
$$
x' = \frac{2h\nu'}{m c^2} \gamma \left(1 - \mu' \frac{v}{c} \right).
\eqno{(2.23)}
$$
For the case, $\nu \sim \nu'$ Eqn. (2.20) reduces to the classical expression. 
Thus, in the nonrelativistic limit,
$$
\frac{d\sigma}{d\Omega'} = \frac{r_{e}^{2}}{2} \left( 1 + cos^2 \alpha \right).
\eqno{(2.24)}
$$
Here, $r_{e} = e^2/mc^2$ is the classical electron radius. So, we find that 
the cross section reduces from its classical value for larger photon energy. 
Thus Compton scattering becomes less efficient at high energies. 
The total scattering cross section is given by Klein-Nishina formula:
$$
\sigma = \frac{2\pi r_{e}^{2}}{x} \\
\left[ \left( 1 - \frac{4}{x} - \frac{8}{x^2} \right) ln\left( 1 + x \right) + \frac{1}{2} + \frac{8}{x} - \frac{1}{2\left( 1 + x \right)^2} \right].
\eqno{(2.25)}
$$
In the nonrelativistic limit $(x \ll 1)$ we have approximately
$$
\sigma \approx \frac{8\pi}{3} r_{e}^2(1 - x) = \sigma_{T}(1 - x),
\eqno{(2.26)}
$$
while in the ultrarelativistic regime $(x \gg 1)$ we have
$$
\sigma = 2\pi r_{e}^{2}x^{-1}\left( ln x + \frac{1}{2}\right).
\eqno{(2.27)}
$$
Here, $\sigma_T$ is the Thomson scattering cross section for an 
electron, given by Eqn. (2.10).

\subsubsection{Photon Free Path}

The probability that a photon will be scattered within a path length $dl$
by a directed beam of electrons having a density $N_e(\bf{\vec{v}})$ and moving
at velocity $\vec{\bf{v}}$ is expressed by (Landau and Lifshitz, 1976)
$$
d\pi = \left(1 - \mu \frac{v}{c} \right)\sigma(x) N_e({\bf{\vec{v}}}) dl.
\eqno{(2.28)}
$$
From the scattering probability we can calculate the mean free path 
$\overline{\lambda}$ of a photon in plasma whose electrons have any 
desired isotropic distribution $N_e({\bf{\vec{v}}})$ with respect to momentum:
$$
\overline{\lambda} = \frac{4\pi\int^\infty_0{N_e(p)p^2dp}}{2\pi N_e \int^\infty_0{N_e(p)p^2} \int^{-1}_{1}{\sigma(x)\left(1 - \mu \frac{v}{c}\right) d\mu dp}} .
\eqno{(2.29)}
$$
Now we evaluate the expression (2.29) for a Maxwellian gas.

$\bullet$ {\bf{Nonrelativistic limit ($h\nu \ll mc^2$, $k_BT_e \ll mc^2$):}}
The Maxwellian momentum distribution gives:
$$
N_e(p) = N_e \left(\frac{1}{2\pi m k_BT_e}\right)^{3/2} exp\left(-\frac{p^2}{2mk_BT_e} \right).
\eqno{(2.30)}
$$
Substituting Eqns. (2.30) and (2.26) into Eqn. (2.29), we find that
$$
\frac{1}{\overline{\lambda}} = \sigma_T N_e \left[1 - 2\frac{h\nu}{mc^2} - 5\left(\frac{h\nu}{mc^2}\right)\left(\frac{k_BT_e}{mc^2}\right) \right].
\eqno{(2.31)}
$$

$\bullet$ {\bf{Ultrarelativistic limit ($h\nu \gg mc^2$, $k_BT_e \gg mc^2$):}}
The probability distribution for the momentum in this relativistic Maxwellian electron gas 
is given by the Maxwell-Juttner distribution:
$$
N_e(p) = exp\left(-\frac{\left(p^2c^2+m^2c^4\right)^{1/2}}{k_BT_e} \right).
\eqno{(2.32)}
$$
Now, if we substitute Eqn. (2.32) into Eqn. (2.29) and write the cross section from Eqn. (2.27), 
we find that
$$
\frac{1}{\overline{\lambda}} = \frac{3}{16}\sigma_T N_e \frac{mc^2}{h\nu}\frac{mc^2}{k_BT_e}\left[\ln\left(4\frac{h\nu}{mc^2}\frac{k_BT_e}{mc^2} \right) + 0.077 \right].
\eqno{(2.33)}
$$

$\bullet$ {\bf{Limit $h\nu \gg mc^2$ but $k_BT_e \ll mc^2$:}}
In this limit we will have,
$$
\frac{1}{\overline{\lambda}} = \frac{3}{8}\sigma_T N_e \frac{mc^2}{h\nu}\left(\ln\frac{2h\nu}{mc^2} + 0.5 + \frac{k_BT_e}{mc^2} \right)
\left(1 - \frac{3}{2} \frac{k_BT_e}{mc^2}\right).
\eqno{(2.34)}
$$
We find that the mean free path ${\overline{\lambda}}$ with respect to Compton scattering 
will lengthen as the Maxwellian plasma temperature ($k_BT_e$).

\subsubsection{Energy Exchange between Plasma and Radiation during Scattering}

When photons are scattered by electrons with a Maxwellian momentum distribution, 
the average relative change in the photon frequency per scattering will be
$$
\frac{\overline{\Delta \nu}}{\nu} = \overline{\lambda}\int{\left( \frac{\nu'}{\nu} - 1\right)\left(1 - \mu \frac{v}{c} \right) \frac{d\sigma}{d\Omega'} N_{e}(p)p^2dp d\Omega d\Omega' },
\eqno{(2.35)}
$$
where, $\frac{\nu'}{\nu}$ is expressed by Eqn. (2.16) and the photon mean 
free path $\overline{\lambda}$ by Eqn. (2.29).

For nonrelativistic electrons in thermal equilibrium, Eqn. (2.35) gives:
$$
\left[\frac{<\nu' - \nu>}{\nu}\right]_{NR} = \frac{4k_BT_e - h\nu}{mc^2}.
\eqno{(2.36)}
$$
If $k_BT_e > h\nu$, the photons may gain energy from the electrons. This is called 
{\it{inverse Compton scattering}}. On the other hand, for $h\nu > 4k_BT_e$, energy 
is transferred from photons to electrons via Compton scattering.

For ultrarelativistic limit, $\gamma \gg 1$, Eqn. (2.35) becomes,
$$
\left[\frac{<\nu' - \nu>}{\nu}\right]_{R} \sim 16\left(\frac{k_BT_e}{mc^2}\right)^2.
\eqno{(2.37)}
$$

\subsubsection{Compton Y Parameter}

The spectrum resulting from repeated scatterings is usually calculated numerically using
Monte Carlo techniques. Qualitatively, however, we can expect that the more scatterings that
occur, the more the seed photon distribution becomes distorted. A useful parameter that
measures the importance of scattering in a medium is the Compton Y parameter:
\vskip .3cm
\noindent
$Y \equiv $(average fractional energy change per scattering) $\times$ (mean number of scattering).
\vskip .3cm
The Y parameter in a finite media determines whether a photon will significantly change 
its energy in traversing the medium. 

The mean number of scatterings is determined by the optical depth, $\tau = \sigma N_e D$, where D is
the dimension of the scattering region. A value of $\tau \sim 1$ means that on average, 
a photon will scatter once before escaping the region. It can be shown that, the mean number 
of scatterings $N_s$ is given by,
$$
N_s \approx \tau^2, \hskip 2 cm (\tau \gg 1),
$$
and for a optically thin medium,
$$ 
N_s \approx \tau, \hskip 2 cm (\tau \ll 1).
$$
For an order of magnitude estimate in a pure scattering medium it is sufficient to use:
$$
N_s \approx \rm{Max}(\tau,\tau^2).
\eqno{(2.38)}
$$
Combining Eqns. (2.36), (2.37) and (2.38) we obtain expressions for the Compton Y parameter 
for relativistic and nonrelativistic thermal distribution of electrons:
$$
Y_{NR} = \left[\frac{<\nu' - \nu>}{\nu}\right]_{NR} \times N_s = \frac{4k_BT_e - h\nu}{mc^2} \rm{Max}(\tau,\tau^2),
\eqno{(2.39a)}
$$
$$
Y_{R} = \left[\frac{<\nu' - \nu>}{\nu}\right]_{R} \times N_s = 16\left(\frac{k_BT_e}{mc^2}\right)^2 \rm{Max}(\tau,\tau^2).
\eqno{(2.39b)}
$$

\newpage

%% file: CHAPTER3/chap3.tex

\newpage
\markboth{\it Monte Carlo Techniques}
{\it Monte Carlo Techniques}
\chapter{Monte Carlo Techniques}

\section{Introduction}
Since the discovery of X-ray and $\gamma$-ray  sources of cosmic origin, 
astrophysicists have tried to develop models that would explain the 
observed radiation spectra. The simplest model of a compact X-ray 
source is a cloud of hot plasma with a low-frequency $\nu$ photon 
source within it. The photons energy $h\nu$ is increased due to 
multiple scattering by hot electrons and it emerges from the cloud 
as hard X-ray or even $\gamma$-ray radiation. This change in the 
photon spectrum due to multiple scattering of photons by thermal 
electrons is called the {\it{Comptonization}} of radiation. This 
process is one of the chief mechanisms for producing hard radiation 
spectra in high-energy astrophysics. Radiation spectra of various 
compact sources (neutron stars, accretion disks around black holes, 
quasars, galactic nuclei, etc.) may be considered to be produced by 
Comptonization. The most efficient method for modeling such 
spectra is the Monte Carlo method.

The Monte Carlo method is a numerical method of solving mathematical 
problems by simulation of random variables. The Monte-Carlo method 
was conceived at the Los Alamos National Laboratory during the 
Manhattan Project by Nicholas Metropolis and S. Ulam, as a numerical 
method for solving the Boltzmann equation governing the 
neutron distribution function in fissile material. 
Since then it has found numerous other uses across many fields of
science. To write this Chapter we have followed the books by: 
I. M. Sobol (1994) and Pozdnyakov, Sobol \& Syunyaev (1983) (PSS83).

\section{Modeling Random Variables}
Before attempting to cope with specific problems by the Monte 
Carlo method one must describe how to model various random 
variables on a computer. Usually, three means for obtaining 
random variables are considered: tables of random numbers, 
random number generators and the pseudo-random number method. 
In developing our Monte Carlo code we have used the 
pseudo-random number method to model the random 
variables. Here we give a brief description of the method.
\subsection{Pseudo-random Numbers}
A sequence of independent values $y_1,y_2,...$ of the random variable 
$y$ distributed uniformly on the interval $(0,1)$ are called ordinary 
{\it{random numbers}}. Since the ``quality" of random numbers used for 
computations is checked by special tests, one can ignore the means by 
which random numbers are produced, as long as they satisfy the tests.

Numbers obtained by a formula that simulate the values of the random 
variable $y$ are called {\it{pseudo-random numbers}}. The word ``simulate" 
means that these numbers satisfy a set of tests just as if they were 
independent values of $y$.

The advantages of the pseudo-random number method are evident. First, 
obtaining each number requires only a few simple operations, so the 
speed of generating numbers is of the same order as the computers 
work speed. Second, the program occupies only a few addresses in  the
RAM. Third, any of the numbers $y_k$ can be reproduced easily. 
And finally, the ``quality" of this sequence need to be checked
 only once; after that, it can be used many times in calculations 
of similar problems without taking any risk.

The single shortcoming of the method is the limited supply of 
pseudo-random numbers that it gives, since if the sequence of 
numbers $y_1,y_2,...y_k,...$ is computed by an algorithm of 
the form $y_k = F(y_{k-1)}$ it must be periodic. In any 
address in RAM only a finite number of different numbers can 
be written. Thus, sooner or later one of the numbers, say $y_L$, 
will coincide with one of the preceding numbers, say $y_l$. Then 
clearly, $y_{L+1} = y_{l+1}, y_{L+2} = y_{l+2},...., $ so that 
there is a period $P = L - l$. The non periodic part of the 
sequence is $y_1,y_2,...y_{L-1}$.

The most widespread  procedure for obtaining pseudo-random numbers 
is the method of residues (also called the congruential method or 
the multiplicative method) proposed by D. H. Lehmer.
\vskip 0.1cm
$\bullet$ {\bf{Method of residues}}
\vskip 0.1cm
A sequence of integers $m_k$ is defined 
in which the initial number $m_0$ is fixed, and all subsequent 
numbers $m_1,m_2,...$ are computed by one formula:
$$
m_k \equiv g m_{k-1} \left(mod M \right),
\eqno{(3.1)}
$$
for $k = 1,2,.....;$ from the numbers $m_k$ we calculate the pseudo-random numbers
$$
y_k = \frac{m_k}{M}.
\eqno{(3.2)}
$$
The integer g is called {\it{multiplier}}, and the integer M is referred 
to as the {\it{modulus}}.

Ordinary congruential generators are not suitable in our case: if the modulus 
is small, the period is small. It was proposed by Wichman and Hill to use 
in parallel three very short number generators,
$$
m_k \equiv 171 m_{k-1} \left(mod 30269 \right),
$$
$$
m_k' \equiv 172 m_{k-1}' \left(mod 30307 \right),
$$
$$
m_k'' \equiv 170 m_{k-1}'' \left(mod 30323 \right),
\eqno{(3.3)}
$$
and to consider as pseudo-random numbers the fractional parts
$$
y_k = \left[ \frac{m_k}{30269} + \frac{m_k'}{30307} + \frac{m_k''}{30323}\right].
\eqno{(3.4)}
$$
This is how the pseudo-random numbers have been computed in our investigations. 
The period of the sequence $y_k$ is $P \approx 6.9 \times 10^{12}$. The 
initial values $m_0$, $m_0'$, $m_0''$ should not be selected at random. 
A good sequence can be obtained with $m_0 = 11$, $m_0' = 23$, $m_0'' = 101$.
\subsection{Method of Inverse Functions}
This method is one of the basic procedures for modeling 
random variables. Here we want to find the values of a random 
variable whose distribution function is $F(x)$. 
Let $x = G(y)$ be the function inverse to the function $y = F(x)$; 
then the expression $\eta = G(\xi)$ will define a random variable 
with the distribution function $F(x)$.
\vskip 0.1 cm
{\bf {Example 1:}} The free path $\lambda$ of a photon in a homogeneous 
medium is a random variable conforming to the exponential law
\begin{eqnarray*}
P\{\lambda < x \} = 1 - e^{-x/\overline{\lambda}},
\end{eqnarray*}
where $\overline{\lambda}$ is the mean free path. The method of inverse functions gives:
\begin{eqnarray*}
\xi = 1 - e^{-x/\overline{\lambda}},
\end{eqnarray*}
\begin{eqnarray*}
\Rightarrow x = -\overline{\lambda} \ln (1 - \xi).
\end{eqnarray*}
Since the random variable $\xi$ and $1-\xi$ has the same distribution, 
the expression for modeling the free path of an individual photon is given by:
\begin{eqnarray*}
\lambda = -\overline{\lambda} \ln (\xi).
\end{eqnarray*}
\vskip 0.1 cm
{\bf{Example 2:}} {\it{Energy of Power-law photons or electrons:}} Let us have a 
Power law energy distribution $F(E) = E^{-\alpha}$ between $E_1$ to $E_2$. 
The random variable whose values will be between $E_1$ and $E_2$ and 
satisfies the power law distribution function is given by:
$$
\frac{\int^{E}_{E_1}{F(E)dE}}{\int^{E_2}_{E_1}{F(E)dE}} = \xi,
\eqno{(3.5)}
$$
where, $\xi$ is the random number distributed uniformly on the interval $(0,1)$.
From this equation we obtain the modeling expression:
$$
E = \left[ \xi \left(E_2^{1- \alpha} - E_1^{1- \alpha} \right) + E_1^{1- \alpha}\right]^{\frac{1}{1- \alpha}}.
\eqno{(3.6)}
$$
\subsection{Rejection Technique}
Consider a random variable $\eta$ defined by the condition
$$
\eta = g(y_1,y_2,....,y_m) \hspace{1cm}\rm{if} \hspace{1cm} (y_1,y_2,....,y_m)\in B, 
\eqno{(3.7)}
$$
where, $B$ is some fixed region in $m-$dimensional space. To compute $\eta$ from 
expressions of the type (3.7) we choose $m$ random numbers $y_1,y_2,....,y_m$ 
and test the selection criterion $(y_1,y_2,....,y_m)\in B$. If it is satisfied, 
we compute $\eta = g(y_1,y_2,....,y_m)$; otherwise we take a new group of 
random numbers $y_1,y_2,....,y_m$ and test the criterion again.
\vskip 0.1 cm
{\bf{Example:}} Let us consider a function $f(x)$ with two important properties: 
$(i)$ $f(x)$ is single valued, and $(ii)$ for each value of x we should 
have finite value of  $f(x)$. To compute $f(x)$ we choose two random 
numbers $\xi$ and $\mu$ uniform between $(0,1)$. Any general point $x,y$ can be written as,
\begin{eqnarray*}
x = x_1 + (x_2 - x_1)\xi,
\end{eqnarray*}
\begin{eqnarray*}
y = y_1 + (y_2 - y_1)\mu,
\end{eqnarray*}
\begin{figure}[h]
\begin{center}
\vskip 5.0cm
\includegraphics{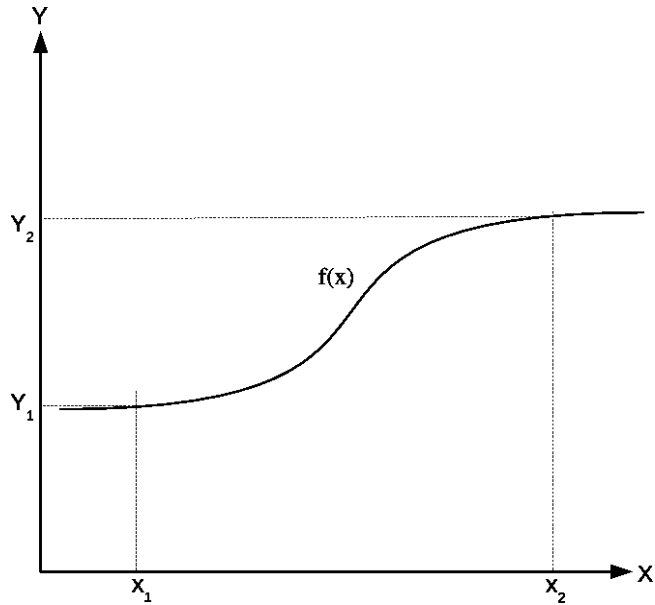}
\vskip 3.0cm
\caption{Schematic diagram to understand the rejection technique. 
Using this technique, one can choose random values of $x$ between ($x_1,x_2$) according to 
the function $f(x)$.}
\label{fig.3.1}
\end{center}
\end{figure}
\hspace{-0.18cm}where, $y_1$ and $y_2$ are respectively the minimum and maximum values of $f(x)$
in the region $x_1$ and $x_2$ (Fig. \ref{fig.3.1}). Now if $ y \leq f(x)$ we choose that $x$, 
otherwise if $ y > f(x)$ we reject that value of $x$ and again take new set of 
$\xi$ and $\mu$ to repeat the process.
\subsection{Method of Superposition}
Let us obtain values of a random variable $\eta$ whose distribution 
function $F(x)$ can be represented as a superposition of distribution functions $F_m(x)$:
\begin{eqnarray*}
F(x) = \sum c_m F_m(x),
\end{eqnarray*}
where all the $c_m > 0$, and $\sum c_m = 1$ (either finite or infinite number of terms). 
Let, $G_m(y)$ is the function inverse to $y = F_m(x)$. We introduce the random 
number $\alpha$, which may take the values $\alpha = 1,2,...$ with probabilities 
$P\{\alpha = m\} = c_m$. To draw the values of $\eta$ we use the following procedure: 

We select two random numbers $\xi_1$, $\xi_2$, and using $\xi_1$ we draw a number 
$\alpha$. From $\xi_2$ we now find the value of $\eta = G_\alpha(\xi_2)$. 
The distribution function of $\eta$ will then be $F(x)$.
\begin{figure}[h]
\begin{center}
\vskip 5.0cm
\includegraphics{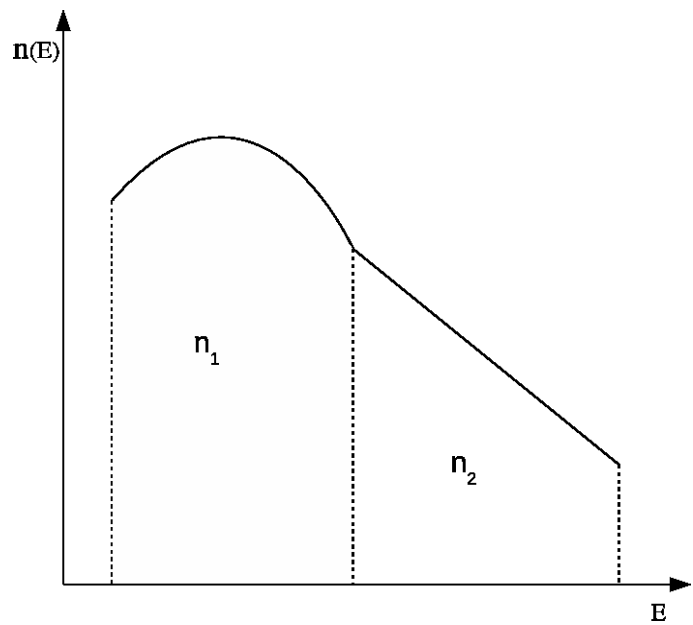}
\vskip 3.0cm
\caption{Schematic diagram to understand the method of superposition. See text for details.}
\label{fig.3.2}
\end{center}
\end{figure}
\vskip 0.1cm
{\bf{Example:}} Let us assume that the injected spectrum has two components:
the Planck spectrum and the Power-Law spectrum. Also consider that $n_1$ and $n_2$ 
are the number of photons coming out of the source obeying Planck law and 
Power-law respectively (Fig. \ref{fig.3.2}). We define the coefficient 
$\epsilon = \frac{n_1}{n_1 + n_2}$. Now we choose a uniform random number 
$\xi$ between $(0,1)$.

\hspace{3cm}If, $\xi \leq \epsilon$  \hspace{1cm}Planck spectrum, \\

\hspace{3cm}If, $\xi > \epsilon$     \hspace{1cm}Power Law.\\ 

In a similar way described above, we can take into account more than one 
physical process (e.g., pair production + Compton scattering). Here we 
have to consider the scattering cross sections of the processes. 
We have to define $ \epsilon = \frac{\sigma_1}{\sigma_1 + \sigma_2}$.

\hspace{3cm}If, $\xi \leq \epsilon$  \hspace{1cm}Process 1, \\

\hspace{3cm}If, $\xi > \epsilon$     \hspace{1cm}Process 2.\\ 
\subsection{Multidimensional Modeling Functions}
Rich opportunities are available in the simulation of random variables if we 
take advantage of a more general type of modeling function: 
$\eta = g(\xi_1,\xi_2), \eta = g(\xi_1,\xi_2,\xi_3), .... $. 
In general flows, the time intervals are independent random variables 
$\eta$ with density
$$
p(x) = \frac{a^m}{(m-1)!}x^{m-1}e^{-ax}, \hspace{3cm} 0<x<\infty.
\eqno{(3.8)}
$$
Since the integral
\begin{eqnarray*}
\Gamma (m) = \int^{\infty}_{0}x^{m-1}e^{-x}dx = (m-1)!
\end{eqnarray*}
is called the {\it{gamma-function}}, the density (3.8) is called the {\it{gamma-distribution}}. 
It can be proved that for integers $m = n$ the random variable $\eta$, with density (3.8) 
can be modeled by the equation
$$
\eta = -a^{-1}\ln(\xi_1,....,\xi_n),
\eqno{(3.9)}
$$
while for half-integers $m = n + \frac{1}{2}$ the modeling formula becomes
$$
\eta = -a^{-1}\left[\ln(\xi_1,....,\xi_n) + (\ln\xi_{n+1})\sin^2(2\pi\xi_{n+2})\right].
\eqno{(3.10)}
$$
\section{Application of the Monte Carlo Technique to the Comptonization Problem}

\subsection{Energy of Planck Photons} 
At radiation temperature $T_r$ the number density of 
photons having an energy $E = h\nu$ is expressed by
$$
p(E) = \frac{1}{2 \zeta(3)} b^{3} E^{2}(e^{bE} -1 )^{-1},
\eqno{(3.11)}
$$
where $b = 1/k_BT_r$; $\zeta(3) = \sum^\infty_1{m}^{-3} = 1.202$ is the Riemann zeta function. 
The function $(e^{bE} -1)^{-1}$ can be expanded in powers of $e^{-bE}$. Using the normalized 
densities
$$
p_{m}(E) = \frac{1}{2}m^3b^3E^2e^{-bmE},
\eqno{(3.12)}
$$
we can rewrite $p(E)$ as,
$$
p(E) = (1.202)^{-1} \sum^{\infty}_{m = 1} m^{-3} p_{m}(E).
\eqno{(3.13)}
$$
This expression shows that $p(E)$ may conveniently be modeled by the superposition 
method: each of the densities $p_{m}(E)$ represents a gamma distribution 
(given in Eqn. (3.8)) and would be modeled by the formula (3.16) (as from 
Eqn. (3.11) we see $m = 3$, an integer).

To draw a value for the energy $E = h\nu$ we select four random numbers 
$\xi_1,\xi_2,...,\xi_4$. From $\xi_1$ we define an auxiliary random number 
$\alpha$ such that
$$
\alpha = 1 \hspace{2cm}\rm{if} \hspace{1cm}1.202\xi_1 < 1,
\eqno{(3.14)}
$$
$$
\alpha = m \hspace{2cm}\rm{if} \hspace{1cm}\sum^{m-1}_{1}j^{-3}\leq1.202\xi_1 < \sum^{m}_{1}j^{-3},
\eqno{(3.15)}
$$
where $m = 2,3,....$; then from Eqn. (3.9) we set 
$$
h\nu = -\frac{k_BT_r}{\alpha}\ln(\xi_2 \xi_3 \xi_4).
\eqno{(3.16)}
$$

\subsection{Momentum of Relativistic Electrons}
The number of Maxwellian electrons having momentum $\vec{p}$ is expressed by Eqn. (2.65) 
$$
N(\vec{p})d\vec{p} = exp[-(p^2c^2 + m^2c^4)^{1/2}/k_BT_e]d\vec{p}.
\eqno{(3.17)}
$$
If all directions $\vec{p}$ are equally probable, the density of p will be proportional to 
\begin{eqnarray*}
p^2exp[-(p^2c^2 + m^2c^4)^{1/2}/k_BT_e].
\end{eqnarray*}
Introducing the dimensionless energy $ n= k_BT_e/mc^2 $ and momentum $\eta = p/mc$, 
we have for density of $\eta$:
$$
p(x) = B_2x^2 exp\left( -\frac{1}{n}\sqrt{1+x^2}\right),
\eqno{(3.18)}
$$
where $0<x<\infty$ and the normalizing constant $B_2$ is given in terms of the Macdonald function by 
\begin{eqnarray*}
B_2 = {\left[ nK_2\left(\frac{1}{n}\right) \right]}^{-1}.
\end{eqnarray*}
\vskip 0.1 cm
$\bullet$ {\bf{Low-temperature case}}
\vskip 0.1 cm
If $n \leq 0.29$ (equivalently, if $k_BT_e \leq 150$ keV), we select two random 
numbers $\xi_1,\xi_2$ and compute the auxiliary quantity 
\begin{eqnarray*}
\xi' = -\frac{3}{2} \ln \xi_1.
\end{eqnarray*}
Then if the selection criterion 
\begin{eqnarray*}
\xi_2^2 < 0.151 \left( 1 + n\xi'\right)^2 \xi' \left( 2 + n\xi'\right) \xi_1,
\end{eqnarray*}
is satisfied, we set 
\begin{eqnarray*}
\eta = \left[ n\xi' \left( 2 + n\xi'\right)\right]^{1/2};
\end{eqnarray*}
otherwise, we select new $\xi_1, \xi_2$.
\vskip 0.1 cm
$\bullet$ {\bf{High-temperature case}}
\vskip 0.1 cm
For $n > 0.29$, or equivalently, for $k_BT_e \geq 150 $ keV we select four random numbers $\xi_1,...,\xi_4$ and compute the two quantities 
\begin{eqnarray*}
\eta' = -n \ln(\xi_1 \xi_2 \xi_3),
\end{eqnarray*}
\begin{eqnarray*}
\eta'' = -n \ln(\xi_1 \xi_2 \xi_3 \xi_4).
\end{eqnarray*}
If $(\eta'')^2 - (\eta')^2 > 1$ we set $\eta = \eta'$; otherwise we draw new numbers $\xi_1,...,\xi_4$.
\subsection{Scattering Cross Section}
The cross section 
$$
\sigma(x) = 2 \pi r_e^2 \widehat{\sigma}(x),
\eqno{(3.19)}
$$
for scattering of a photon by an electron is well established, but the conventional equation (Eqn. 2.61),
$$
x\widehat{\sigma}(x) = \left( 1 - \frac{4}{x} -\frac{8}{x^2}\right) \hspace{0.1cm}\ln(1+x) + \frac{1}{2} + \frac{8}{x} - \frac{1}{2(1 + x)^2},
\eqno{(3.20)}
$$
is inconvenient for calculations when $x \ll 1$ and is rather cumbersome to integrate. 
We therefore approximate $\widehat{\sigma}(x)$ to high accuracy by means of the simpler functions
$$
\widehat{\sigma}(x)=\frac{1}{3} + 0.141x - 0.12x^2 + (1+0.5x)(1+x)^{-2},\hspace{1cm} x \leq 0.5;
$$
$$
=[\ln (1 + x) + 0.06]x^{-1},\hspace{4.5cm} 0.5 \leq x \leq 3.5;
$$
$$
=[\ln (1 + x) + 0.5 - (2 + 0.076x)^{-1}]x^{-1},\hspace{2cm} 3.5 \leq x.
\eqno{(3.21)}
$$
The error of this fit is no more than $1$ percent. We need to evaluate the following function:
$$
\Phi(x) = \int^{x}_{0} x \widehat{\sigma}(x) dx.
\eqno{(3.22)}
$$
By integrating the approximations to $x \widehat{\sigma}(x)$ we obtain the computation formulas
$$
\Phi(x)=\frac{1}{6}x^2 + 0.047x^3 - 0.03x^4 + \frac{1}{2}x^2(1+x)^{-1}, \hspace{1.7cm}0 \leq x \leq 0.5;
$$
$$
=(1+x) \ln (1 + x) - 0.94x - 0.00925, \hspace{2.5cm} 0.5 \leq x \leq 3.5;
$$
$$
=(1+x) \ln (1 + x) + 0.5x - 13.16 \ln(2 + 0.076x) + 9.214, \hspace{1cm} 3.5 \leq x.
\eqno{(3.23)}
$$

\subsection{Photon Mean Free Path}

A photon of energy $h \nu$ and with its momentum in direction $\Omega$
will have a mean free path given by Eqn. (2.65),
where the scattering cross section $\sigma(x)$ and the function $N(\vec{p})$ have been 
defined above (Eqn. (3.20) and (3.17), respectively). Here, 
$$
x = H\gamma (1 - \hat{\Omega}.{\vec{v}}/c),
\eqno{(3.24)}
$$
with $H = 2h\nu /mc^2$. Since all directions of $\vec{p}$ are equally probable, we align 
the polar axis with $\hat{\Omega}$ and introduce spherical coordinates $p, \theta, \phi$ 
to obtain the expression:
$$
\bar{\lambda} = \frac{4\pi \int^{\infty}_{0}{e^{-\gamma/n} p^2 dp}}
{2\pi N_e \int^{\infty}_{0} e^{-\gamma/n} p^2 dp 
\int^{1}_{-1}{\sigma(x) \left[ 1 - (v/c) \cos \theta \right]} d(\cos{\theta}) }.
\eqno{(3.25)}
$$
The integration over $\cos{\theta}$ is readily converted to an integration over $x$, 
because if $\gamma$ is fixed, $dx = -H \gamma (v/c) d \cos{\theta}$. Then we find that
$$
\int^{1}_{-1}{\sigma(x) \left[ 1 - (v/c) \cos \theta \right] d(\cos{\theta})}
=\frac{2\pi r_e^2}{{(H\gamma)}^2 v/c} \int^{H\gamma(1+v/c)}_{H\gamma(1-v/c)} {x\hat{\sigma}(x) dx},
$$
$$
=\frac{2\pi r_e^2}{H^2 \gamma \sqrt{\gamma^2 - 1}}\left[ \Phi(x) \right]^{H\gamma^{+}}_{H\gamma^{-}},
\eqno{(3.26)}
$$
where $\gamma^{\pm} = \gamma \pm \sqrt{\gamma^2 -1}$. We substitute this expression 
into Eqn. (3.25) with
$$
\pi r_e^2 N_e = \frac{3}{8} \sigma_T N_e = 0.375 \sigma_T N_e,
\eqno{(3.27)}
$$
and change from an integration over $p$ to an integration over $\gamma$ 
(using, $p = mc\sqrt{\gamma^2 - 1})$. As a result we finally obtain the mean free path:
$$
\bar{\lambda} = \frac{H^2 \int^{\infty}_{1}{e^{-\gamma/n}\gamma \sqrt{\gamma^2 -1}d\gamma}}
{0.375 \sigma_T N_e \int^{\infty}_{1}{e^{-\gamma/n} \left[ \Phi(x) \right]^{H\gamma^{+}}_{H\gamma^{-}} d\gamma} }.
\eqno{(3.28)}
$$
Both the integrals (numerator and denominator) are of the form 
\begin{eqnarray*}
\int^{\infty}_{1}{e^{-\gamma/n} \Psi(\gamma) d\gamma}.
\end{eqnarray*}
The change of variables 
\begin{eqnarray*}
u = exp[(1 - \gamma)/n]
\end{eqnarray*}
transforms the semi axis $(1 , \infty)$ into the interval $(0 , 1)$, 
and the resulting integral can be calculated from the rectangle formula:
$$
\int^{\infty}_{1}{e^{-\gamma/n} \Psi(\gamma) d\gamma}= n e^{-1/n}\int^{1}_{0}{\Psi (1 - n \ln u) du}
$$
$$
\approx \frac{n e^{-1/n}}{N_1} \sum^{N_1}_{m = 1} \Psi \left( 1 - n \ln \frac{m -\frac{1}{2}}{N_1}\right).
\eqno{(3.29)}
$$
The factor $n e^{-1/n}$ occurs in both integrals and cancels out.

The final computation formulas are as follows:
a) Abscissas of integration 
$$
\gamma_m = 1 - n \ln \left[ (m - \frac{1}{2})/N_1\right], m = 1,2,....,N_1.
\eqno{(3.30)}
$$
b) The constant quantity
$$
g = \frac{1}{0.375\sigma_T N_e} \sum^{N_1}_{m = 1} {\gamma_m \sqrt{{\gamma_m^2 - 1}}}.
\eqno{(3.31)}
$$
c) The general equation
$$
\bar{\lambda} = \frac{gH^2}{\sum^{N_1}_{m = 1}{\left[ \Phi(x) \right]^{H\gamma_m^{+}}_{H\gamma_m^{-}}}}.
\eqno{(3.32)}
$$
d) Case $H \ll 1$: As $H \rightarrow 0$ the Eqn. (3.32) reduces to an indeterminate 
form, with $\lim^{}_{H\rightarrow 0}{\bar{\lambda}} = (\sigma_T N_e )^{-1}$. 
As $\sigma(x) = \frac{1}{\bar{\lambda} \sigma_T N_e}$ so for  $H \ll 1$, $\sigma(x) = 1$. 
Therefore, if the largest abscissa used in the numerical integration, 
$\gamma_1 = 1 + n \ln(2 N_1)$, and the energy $H = 2h\nu /mc^2$ satisfy the condition
\begin{eqnarray*}
2\gamma_1 H \leq 0.01,
\end{eqnarray*}
or equivalently,
$$
h\nu \leq \frac{0.01mc^2}{4 \gamma_1},
\eqno{(3.33)}
$$
then we set $\sigma(x) = 1$. In our calculations we have adopted the value $N_1 = 300$.
 
\subsection{Modeling Compton Scattering}
\vskip 0.1cm
$\bullet$ {\bf{Selection of scattering electron}}
\vskip 0.1cm
The probability density of the momenta $\vec{p}$ of the scattering electrons is proportional 
to the quantity $\sigma(x) (1 - \hat{\Omega}.\vec{v}/c) N(\vec{p})$ appearing in the 
Eqn. (3.25). In Sec. 3.3 we have modeled the momentum $\vec{p}$ for a Maxwellian density. 
After computing $\vec{v}$, $x$, and $\hat{\sigma}(x)$ (Sec. 3.3), we take one random 
number $\xi$ and test the selection criterion:
$$
\xi < 0.375 \hat{\sigma}(x) (1 - \hat{\Omega}.\vec{v}/c).
\eqno{(3.34)}
$$
If it is satisfied, $\vec{p}$ will be accepted; otherwise we choose a new $\vec{p}$.
\vskip 0.1cm
$\bullet$ {\bf{Choice of coordinate system}}
\vskip 0.1cm
Let us choose the velocity vector 
\begin{eqnarray*}
\vec{v^0} = v_1^0 \hat{i} + v_2^0 \hat{j} + v_3^0 \hat{k}.
\end{eqnarray*}
Assuming
\begin{eqnarray*}
\rho^2 = (v_1^0)^2 + (v_2^0)^2 > 0,
\end{eqnarray*}
and introducing the unit vectors 
\begin{eqnarray*}
\hat{{w}^0} = (v_2^0 \hat{i} - v_1^0 \hat{j})/\rho,
\end{eqnarray*}
and
\begin{eqnarray*}
\hat{{t}^0} = (v_1^0 v_3^0 \hat{i} + v_2^0 v_3^0 \hat{j} - \rho^2 \hat{k})/\rho,
\end{eqnarray*}
which together with $\hat{{v}^0}$ form an orthonormal triad, the direction of the 
vector $\vec{\Omega'}$ may conveniently be expressed in the coordinate system 
$\hat{{v}^0}, \hat{{w}^0}$ and $\hat{{t}^0}$: 
\begin{eqnarray*}
\vec{\Omega'} = \mu' \hat{v^0} + (1 - \mu'^2)^{1/2} (\cos \phi' \hat{w}^0 + \sin \phi' \hat{t}^0),
\end{eqnarray*}
where $\phi'$ denotes the azimuthal scattering angle, measured from the direction $\hat{{w}^0}$ 
in a plane perpendicular to $\hat{v^0}$. The components of $\vec{\Omega'}$ in the stationary 
coordinate system $\hat{i}, \hat{j}, \hat{k}$ can be written as:
$$
\Omega'_1 = \mu' v_1^0 + (1 - \mu'^2)^{1/2} \rho^{-1}(v_2^0 \cos \phi' + v_1^0 v_3^0 \sin \phi'),
\eqno{(3.35)}
$$
$$
\Omega'_2 = \mu' v_2^0 + (1 - \mu'^2)^{1/2} \rho^{-1}(-v_1^0 \cos \phi' + v_2^0 v_3^0 \sin \phi'),
\eqno{(3.36)}
$$
$$
\Omega'_3 = \mu' v_3^0 - (1 - \mu'^2)^{1/2} \rho \sin \phi'.
\eqno{(3.37)}
$$
The scattering angle can then be computed from the relation
\begin{eqnarray*}
\vec{\Omega}.\vec{\Omega'} = \Omega_1\Omega'_1 + \Omega_2\Omega'_2 + \Omega_3\Omega'_3.
\end{eqnarray*}
The joint distribution density of the random variables $\mu'$ and
$\phi'$ is given by
\begin{eqnarray*}
p(\mu',\phi') = \frac{1}{\sigma} \frac{d\sigma}{d\Omega}.
\end{eqnarray*}
Expression for the differential scattering cross section is given by Eqn. (2.56), 
and the change in frequency of a photon when it is scattered is given by Eqn. (2.52). 
The joint density can be written in product form:
$$
p(\mu',\phi') = \frac{1}{\hat\sigma} p_1(\mu',\phi') Y,
\eqno{(3.38)}
$$
where the normalized density  
$$
p_1(\mu',\phi') = \frac{1}{4\pi\gamma^2 (1 - \mu' v/c)^2},
\eqno{(3.39)}
$$
and the functional factor $Y$ is bounded:
$$
Y = \left( \frac{x}{x'} \right)^2 X \leq 2.
\eqno{(3.40)}
$$
The factor $1/\hat{\sigma}(x)$ is independent of $\mu'$ and $\phi'$, and plays the role of a constant. 
X is given by Eqn. (2.57).
\vskip 0.1cm
$\bullet${\bf{Modeling algorithm}}
\vskip 0.1cm
The representation (3.38) enables us to device a rejection technique for modeling 
random variables $\mu'$ and $\phi'$. We assume that the quantities  $\mu$  
and $x$ have already been computed.

a) We take two random numbers $\xi_1,\xi_2$ to compute a possible direction 
of scattering [for the density $p_1(\mu',\phi')$]:
$$
\mu' = \frac{v/c + 2\xi_1 - 1}{1 + (v/c)(2\xi_1 -1)},
\eqno{(3.41)}
$$
$$
\phi' = 2 \pi \xi_2.
\eqno{(3.42)}
$$
(keeping in mind that these are not yet the final quantities).

b) We compute the vector $\vec{\Omega'}$, the scattering angle 
$\vec{\Omega}.\vec{\Omega'}$, and then the ratio:
$$
\frac{x'}{x} = \left[ 1 + \frac{h\nu (1 - \vec{\Omega}.\vec{\Omega'})}{\gamma m c^2 (1 - \mu' v/c)}\right]^{-1}.
\eqno{(3.43)}
$$
We also compute the factor Y from Eqn. (3.40).

c) We draw a random number $\xi_3$ and test the selection condition
$$
2\xi_3 < Y.
\eqno{(3.44)}
$$
If it is satisfied, the direction $\vec{\Omega'}$ is accepted, and the new photon energy will be
$$
h\nu' = \frac{x'}{\left[ 2 \gamma (1 - \mu' v/c)\right]} m c^2;
\eqno{(3.45)}
$$
If instead 
\begin{eqnarray*}
2\xi_3 \geq Y,
\end{eqnarray*}
we return to step (a).

\section{Results of a Sample Monte Carlo Simulation}

We use the above Monte Carlo techniques to develop a 
simulation code which we will apply to solve different 
problems in situations of our interest. 

\subsection{Statement of the Problem}

Let a spherical cloud of radius R contains a uniform density $N_e$ of Maxwellian electrons 
at temperature $k_BT_e$. At the centre of the cloud there is a black hole of 
mass $M$. Within the sphere we place either a low-frequency blackbody source having 
temperature $T_r \ll T_e$, or a hard-radiation source with a power-law spectrum 
$I(E) \propto E^{-\alpha}$. The electrons are moving towards the black hole with 
a free fall velocity. We are trying to calculate the radiation emerging from 
the electron cloud.
\begin{figure}[]
\begin{center}
\vskip 5.3cm
\includegraphics{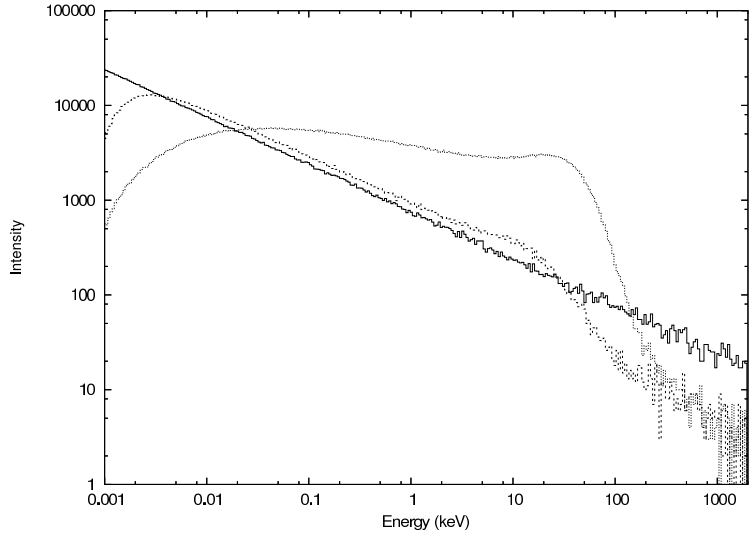}
\vskip 1.1cm
\caption{Variation of the intensity of the photons with energy. Power-law 
photons with $\alpha = 1.5$ are injected from a distributed source at the equatorial 
plane of the electron cloud with temperature $k_BT_e=5$ keV. Change in the 
injected spectrum (solid curve) in presence (dotted) and absence (dashed) of the 
bulk velocity of the electrons are shown.}
\label{fig.3.3}
\end{center}
\end{figure}
\begin{figure}[]
\begin{center}
\vskip 1.7cm
\includegraphics{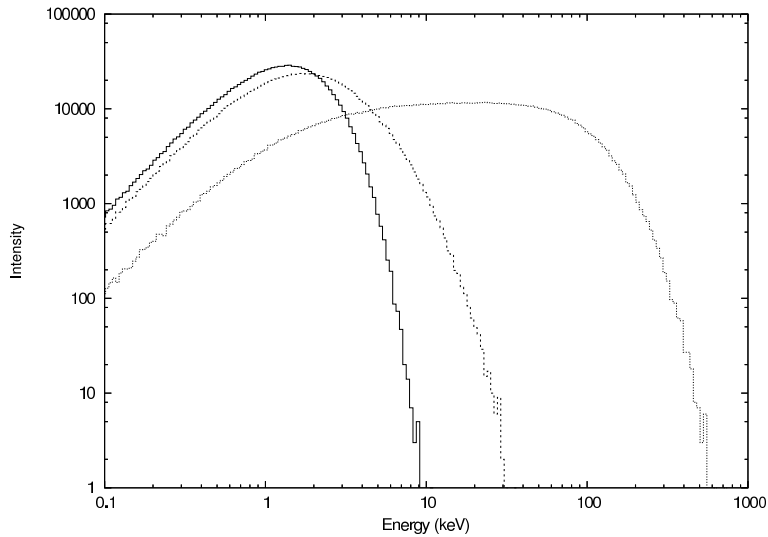}
\vskip 2.5cm
\caption{Variation of the output photon intensity with $k_BT_e$. Black Body photons 
with temperature 0.5 keV (solid curve) are injected into the cloud with temperatures 
5 keV (dashed curve) and 50 keV (dotted curve) (see text for details).}
\label{fig.3.4}
\end{center}
\end{figure}
\subsection{Results and Discussions} 

The spectral variations are shown in Fig. \ref{fig.3.3} and Fig. \ref{fig.3.4}. The mass of the black hole 
used in this simulation is $10 M_\odot$. The radius of the spherical electron cloud is 
$50 r_g$. In these two simulations we have used two different kinds of photon 
energy distributions and source geometry. For the first case (Fig. \ref{fig.3.3}), source of 
soft photons is a disk (inner radius $2 r_g$ and outer radius $50 r_g$) at the 
equatorial plane of the cloud emitting photons that follow power-law ($E^{-\alpha}$) 
energy distribution with $\alpha = 1.5$. In Fig. \ref{fig.3.3} the variation 
of the intensity of the photons coming out of the cloud is plotted against energy. 
The electron cloud has uniform temperature $k_BT_e = 5$ keV, throughout the cloud. 
The solid curve shows the injected spectrum, the dashed and the dotted curves are 
the output spectra in absence and in presence of the bulk velocity of the electrons, 
respectively. Fig. \ref{fig.3.4} shows the variation of the intensity with the 
energy when mono-energetic (temperature, $k_BT_p = 0.5$ keV) black body photons 
are injected isotropically from a point source (kept at a distance of $10 r_g$ 
from the black hole) inside the cloud. In this case the cloud temperature 
is 5 keV (dashed curve) and 50 keV (dotted curve). The solid curve shows the 
injected black body spectrum. The bulk velocity of the cloud has been kept zero 
in this case. In both the cases, electron number density at each point of 
the cloud is also kept constant at $9 \times 10^{16}$/cc. From these two plots 
we can see that an increase in the bulk velocity or temperature of the cloud 
increases the Comptonization of the photons. All these effects of thermal and 
bulk motion Comptonization will be discussed in the next Chapters.

\newpage

%% file: CHAPTER4/chap4.tex

\newpage
\markboth{\it Thermal Comptonization in an Accretion Flow}
{\it Thermal Comptonization in an Accretion Flow}
\chapter{Thermal Comptonization in an Accretion Flow}


\section{Introduction}	
In Chapter 3, we have discussed that Monte Carlo simulation has been found 
to be an essential tool to understand the formation of spectrum in compact 
bodies (PSS83). The work of ST80 showed 
that the power-law component of a black hole spectrum is due to inverse 
Comptonization. More work (ST85) firmly 
established this. Hua \& Titarchuk (1996) confirmed the conclusions drawn in 
ST80, ST85 and Titarchuk (1994) using a Monte Carlo simulation. Meanwhile, 
more efforts were made to understand the nature of the Compton cloud itself
and generally it was believed that accretion disk coronas could be responsible 
for Comptonization.

While the general results of ST80 and ST85 are of great importance, the 
computations in the literature were done with a few specific geometries 
of the cloud, such as plane slabs or spherical blobs. In reality, 
the geometry {\it must be} more complex, simply because of the angular momentum 
of matter (C90 and references therein). Indeed, time dependent numerical 
simulations of sub-Keplerian flows (Molteni, Lanzafame \& Chakrabarti, 1994; Molteni, 
Ryu \& Chakrabarti, 1996) confirm the predictions in C90 and clearly show 
that the geometry of the flow close to a black hole, especially in the post-shock 
region, is more like a torus, very similar to a thick accretion disk conceived 
much earlier (e.g., PW80; Rees et al. 1982 and references therein). 
In the latter case, the radial velocity was ignored but the angular 
momentum was assumed to be sub-Keplerian, while in the simulations of Molteni et al. (1994) 
the radial velocity was also included. In CT95 and C97, theoretical computation of the 
spectra was made by using the post-shock region as the Comptonizing cloud and by 
varying the accretion rates in the Keplerian and sub-Keplerian components. Here too somewhat 
ideal geometry (cylindrical) was chosen so as to utilize the ST80 and ST85 results 
as far as the radiative transfer properties are concerned. A result with a real toroidal 
geometry can be handled only when the Monte Carlo simulations are used.

In the present Chapter, we attempt to solve the problem of spectral 
properties using a thick accretion disk of toroidal geometry as 
the Compton cloud which is supposed to be produced by the sub-Keplerian 
inflow. The outer boundary of the thick accretion disk is treated as 
the inner edge of the Keplerian disk. One positive aspect in treating 
the CENBOL in this manner is that the distribution of electron density 
and temperature can be obtained totally analytically. In a more realistic 
case, one needs to solve the coupled transonic flow solution with 
radiative transfer. This will be described in Chapter 6. The plan of this 
Chapter is the following. In the next Section, we discuss the geometry and 
hydrodynamic properties of the Compton cloud and the source of soft 
photons used in our simulation. In \S 4.3, the results of the 
simulations are presented and in \S 4.4, we draw conclusions.

\section{The Electron Cloud and the Soft Photon Source}

\begin{figure}[h]
\begin{center}
\vskip 5.0cm
\includegraphics{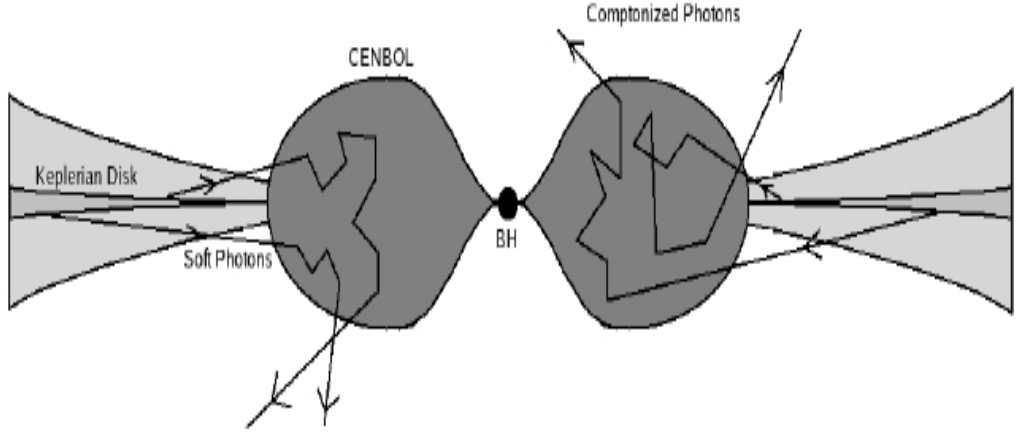}
\vskip 3.0cm
\caption{A cartoon diagram of the geometry of the Compton cloud used for the Monte Carlo Simulations
presented in this Chapter. The puffed up post-shock region surrounds the black hole and 
it is surrounded by the Keplerian disk on the equatorial plane. A tenuous sub-Keplerian flow 
above and below the Keplerian disk is also present (CT95). Typical photon scattering paths 
are shown (Ghosh et al., 2009; hereafter GCL09).}
\label{cartoon4}
\end{center}
\end{figure}

In Fig. \ref{cartoon4}, we present a cartoon diagram of our simulation set up. 
In this paradigm picture, the Compton cloud (CENBOL) is 
produced by the standing shock in the sub-Keplerian flow. 
CENBOL behaves like a boundary layer as it dissipates 
the thermal and bulk energy and produce hard X-rays 
and outflow/jets. The Keplerian disk is truncated and 
the inner edge is typically extended till the outer boundary of 
the CENBOL (shock location). However, in the soft states when 
the post-shock region is cooled down, the Keplerian disk 
can extend till the last stable circular orbit.

As the CENBOL is puffed up, its hot electrons intercept 
the soft photons and reprocess them via inverse Compton 
scattering. A photon originally emitted towards the CENBOL 
may undergo a single, multiple or no scattering with the hot electrons. The photons
which enter the black holes are absorbed.
%

\subsection{Thermodynamic Conditions Inside the Compton Cloud}

Chakrabarti (1989) showed that centrifugal pressure supported Rankine-Hugoniot shock 
solutions exist for the stationary, axially symmetric and rotating 
adiabatic accretion flows around a black hole. In order to verify these results, 
Molteni, Lanzafame and Chakrabarti (1994) carried out a two dimensional 
numerical simulation and found that the shock is indeed formed and the post-shock region (CENBOL) 
has all the properties of a thick accretion disk (PW80). 
The simulation result is more realistic than a thick disk, since the flow also has a significant 
radial motion close to the horizon. 
\begin{figure}[h]
\begin{center}
\vskip 6.0cm
\includegraphics{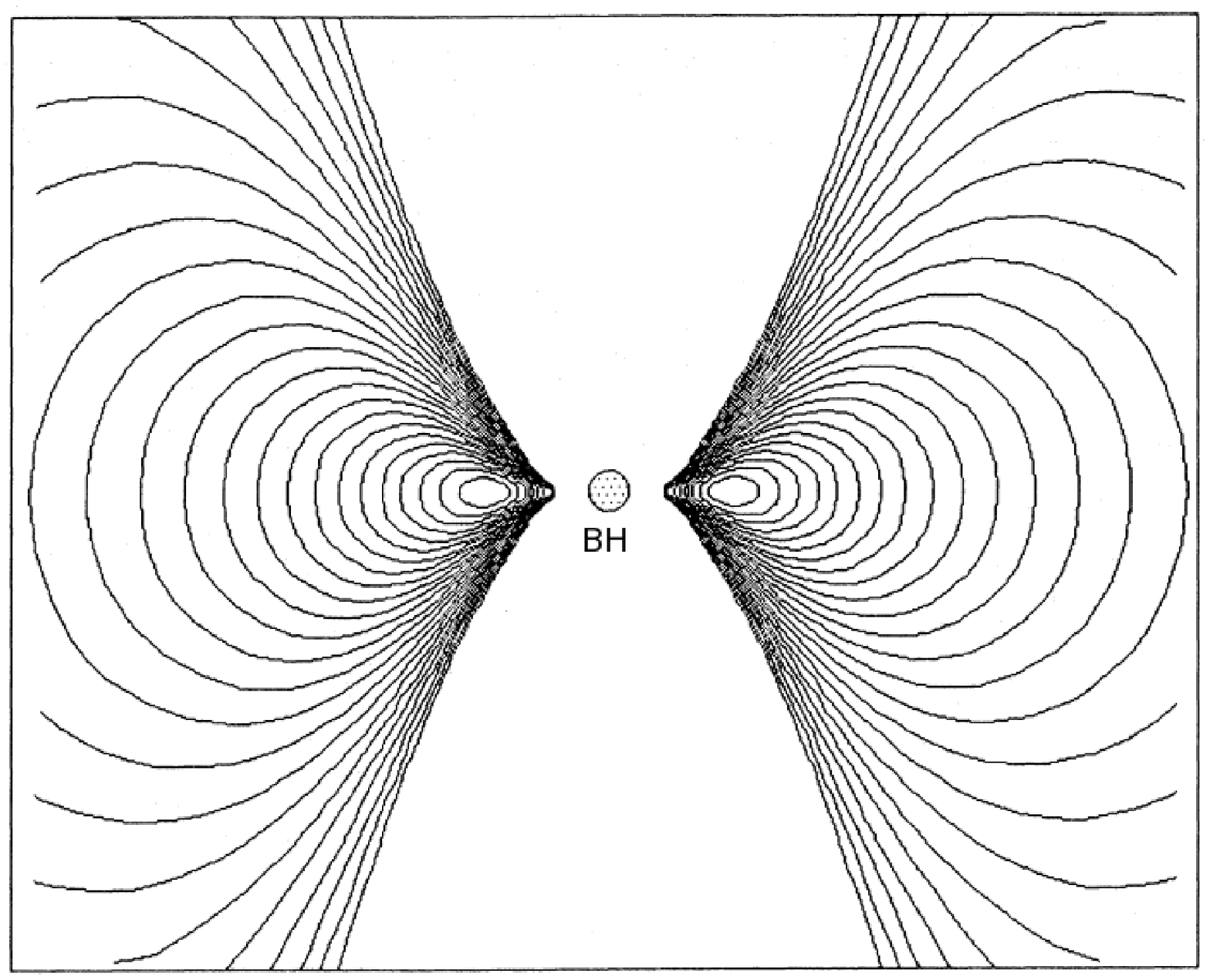}
\vskip 3.0cm
\caption{Typical structure of the equipotential contours for the thick accretion disk in 
the Schwarzschild geometry (Figure taken from Chakrabarti, Jin \& Arnett, 1987).}
\label{thickdisk}
\end{center}
\end{figure}
Fig. \ref{thickdisk} shows the equilibrium structure of the equipotential surfaces 
of a thick accretion disk for a barotropic matter in Schwarzschild geometry. Very 
far away from the black hole, the contours are almost spherical in shape since the 
effect of angular momentum is very weak as compared to gravity. The centrifugal barrier 
keeps the matter away from the axis of the disk, thus providing space for jet formation.

\begin{figure}[]
\begin{center}
\vskip 4.0cm
\includegraphics{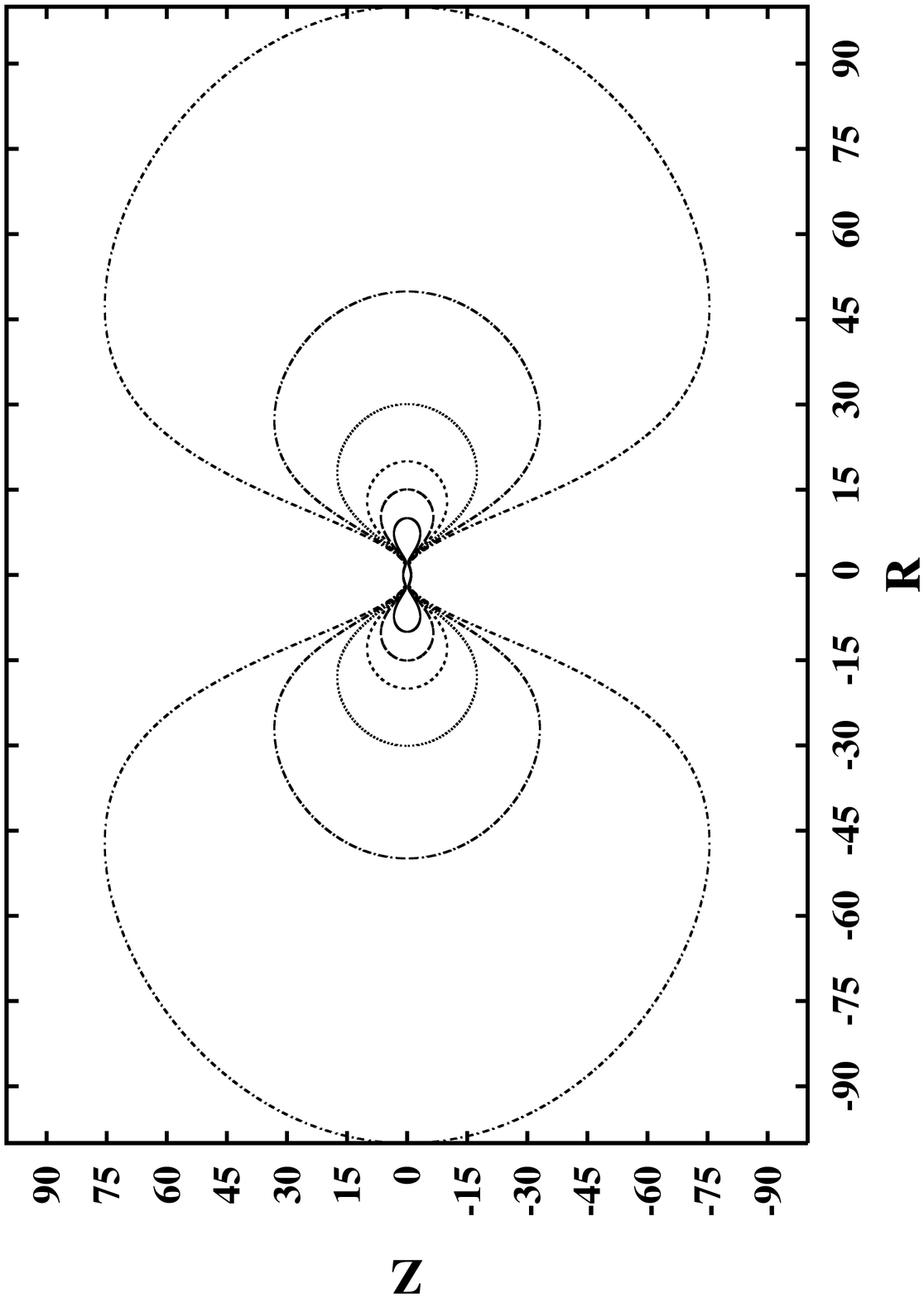}
\vskip 3.0cm
\caption{Contours of constant temperature and density inside 
the CENBOL. Each of them has been used as the outer boundary
in our simulations. $R_{out} = 10$ (solid line), $15$ (large-dashed 
line), $20$ (dashed line), $30$ (dotted line), $50$ (large dashed-dot 
line) and $100$ (dashed-dot line) (GCL09).}
\label{fig.4.2}
\end{center}
\end{figure}
For simplicity, in the present Monte Carlo simulation, we assume the CENBOL to 
have the same analytical shape as an ideal thick disk (Fig. \ref{fig.4.2}) and 
compute the matter density and temperature distribution using the prescription given in 
Chakrabarti, Jin and Arnett (1987). Here we assume that the disk is adiabatic and 
the specific angular momentum and specific entropy is constant throughout the disk. 
The equation of state is given by $p=K\rho^{\gamma}$, where $p$ and $\rho$ are the 
isotropic pressure and the matter density, respectively, $\gamma$ is the 
adiabatic index and $K$ is the entropy constant. The pressure $p$ is considered to be contributed by the 
gas pressure and the radiation pressure. Thus, 
$$
p = p_{gas} + p_{rad}.
$$
For a fully ionized ($T > 10^4$ K) nondegenerate and nonrelativistic ideal gas we have
$$
p_{gas} = \frac{T_e \rho \sqrt{r^2-z^2}}{\mu m_p},
$$
where $\mu$ is the mean molecular weight, $T_e$ is the cloud temperature and $m_p$ is 
the mass of the proton. The radiation pressure is given by
$$
p_{rad} = \frac{1}{3} a {T_e}^4,
$$
where $a$ denotes the Stefan radiation density constant. 
Using the pseudo Newtonian potential $-\frac{1}{2(r-1)}$ 
(PW80), the effective potential due to the black hole is written as: 
$$
\phi(r,z) = \frac{\lambda^2}{2(r^2 - z^2)} - \frac{1}{2(r-1)}.
\eqno{(4.1)}
$$
As the disk is static (i.e., electrons have no radial velocity) we can 
always write in terms of the sound speed $a_s$:
$$
\phi(r,z) = n {a_s}^2 = \frac{\gamma p}{\rho}.
$$
Using the above equations we calculate the distribution of temperature and 
number density of the electron cloud.
The number density of electrons within the CENBOL, $n = \frac{\rho}{\mu m_p}$ is calculated 
from the matter density $\rho$, given by: 
$$
\rho(r,z) = \left[ \frac{\phi(r,z)}{n \gamma K} \right]^{n},
\eqno{(4.2)}
$$
where, the entropy constant $K$ is given by,
$$
K(\beta,\mu) = \left[\frac{3}{a} \frac{1-\beta}{\beta^4} \frac{(k_b)^4}{(\mu m_p)^4} \right]^{\frac{1}{3}},
\eqno{(4.3)}
$$
where, $k_b$ is the Boltzmann constant and $\beta = \frac{p_{gas}}{p}$, denotes the ratio of the gas pressure 
to the total pressure. The temperature $T_e$ of the electron cloud within CENBOL is given by,
$$
T_e(r,z) = \left[\frac{\beta \mu m_p}{k_b} K \right] \rho^{\frac{1}{3}}.
\eqno{(4.4)}
$$
The velocity, mass and distance scales are measured in units of $c$, the 
velocity of light, $M_{bh}$, the mass of the black hole and 
$r_g=2GM_{bh}/c^2$, the Schwarzschild radius of the black hole. 
In this unit, the angular momentum of the disk is chosen to be 
$\lambda = 1.9$. We have chosen the values of the parameters: 
$\beta = 0.5$, polytropic index $(n) = 3$ and $\mu = 0.5$.
In the simulation we have varied the outer edge ($R_{out}$) of the CENBOL to different
values (by varying $\phi$) to change the size of the Compton cloud. 
The disk (Fig. \ref{fig.4.2}) has a centre at $4$ and the inner edge at $2.5$. 
For a better understanding of the results, we calculated an effective 
electron temperature $T_{eff}$ of the CENBOL using the prescription given 
in Sec. 2.4 of CT95. 

\subsection{Emission from a Keplerian Disk}

The soft photons are produced from a Keplerian disk whose inner edge is at the outer edge 
($R_{out}$) of the CENBOL, and the outer edge is at $500 r_g$. The source of soft 
photons have a multi-color blackbody spectrum coming from a standard (SS73) 
disk. 
As the disk is optically thick, the emission is black body type with 
the local surface temperature (Eqn. 1.9):
$$ 
T(r) \approx 5 \times 10^7 (M_{bh})^{-1/2}(\dot{M}_{17})^{1/4} (2r)^{-3/4} 
\left[1- \sqrt{\frac{3}{r}}\right]^{1/4} \rm{K},
$$
The total number of photons emitted from the disk surface is given by 
Eqn. (2.6),
$$
n_\gamma(r) = \left[16 \pi \left( \frac{k_b}{h c} \right)^3 \times 1.202057 \right].
\left[T(r)\right]^3.
$$
We divide the disk into different annuli each having an width of $\delta r$. The disk 
between radius $r$ to $r+\delta r$ injects $dN(r)$ number of soft photons 
isotropically with black body temperature $T(r)$ 
$$
dN(r) =  4 \pi r \delta r H(r) n_\gamma(r),
\eqno{(4.5)}
$$
where, $H(r)$ is the half height of the disk given by:
$$
H(r) = 10^5 \dot{M_d}_{17} \left[1- \sqrt{\frac{3}{r}}\right] {\rm cm}.
$$
\begin{figure}[h]
\begin{center}
\vskip 5.1cm
\includegraphics{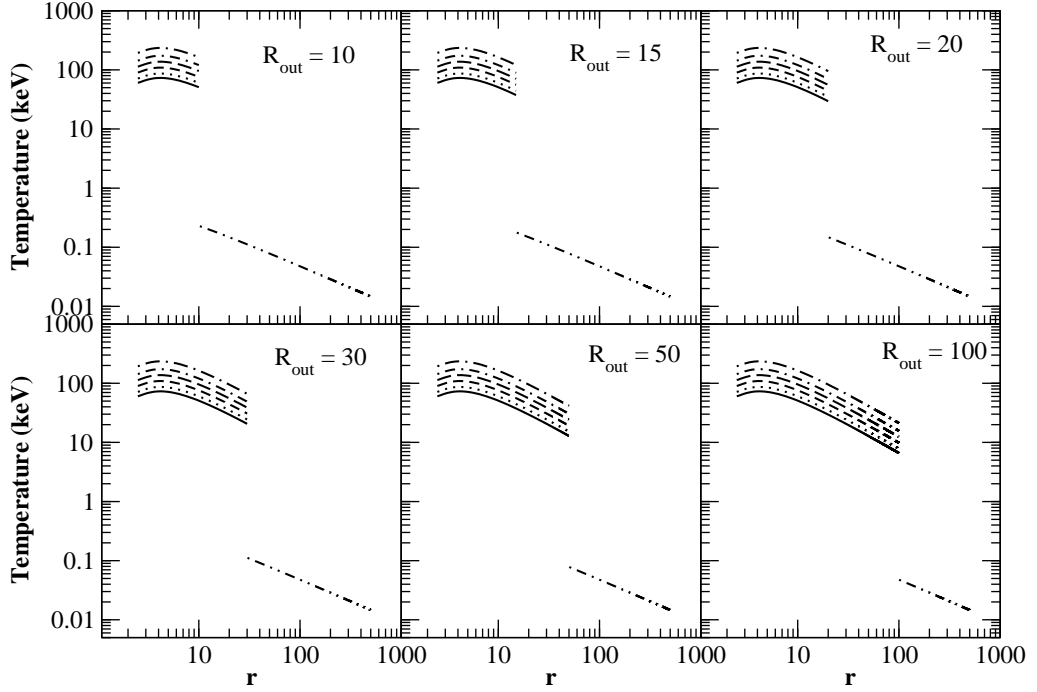}
\vskip 4.9cm
\caption{Temperature distribution (in keV) inside the disk and the CENBOL region for different 
CENBOL outer boundary ($R_{out}$) and various central densities. Table 4.1 gives the relation 
between the central densities and the line styles (GCL09).}
\label{fig.4.3}
\end{center}
\end{figure}

In the above equations, the mass of the black hole $M_{bh}$ is measured 
in units of the mass of the Sun ($M_\odot$), the accretion rate 
$\dot{M}_{17}$ is in units of $10^{17}$ gm/s. Unless otherwise stated, 
we chose $M_{bh} = 10$, accretion rate $\dot m = \frac{\dot M}{\dot M_{edd}} = 1$ 
and $\delta r = 0.5 r_g$. For the sake of completion of a simulation 
using a reasonable amount of computer time, we take a constant 
fraction of the number of photons (Eqn. 4.5) from each annulus (Table 4.2). 
Because of the number of photons we select is way below the actual number, 
the absolute value of accretion rate itself is not very meaningful. However,
the relative number of the intercepted photons and the number density of electrons inside CENBOL
appears to be more important. The result also does not depend on the choice of $\delta r$ 
as long as it is a fraction of a Schwarzschild radius. In Fig. \ref{fig.4.3}, 
we show the distribution of the temperature (in keV) in the Keplerian disk and that in the CENBOL 
which we have used in our simulations. Different panels are for different 
values of the outer edge $R_{out}$ (marked) of the CENBOL radius. 
We provide the effective temperature ($T_{eff}$) within the post-shock region in $R_{out}$ in Table 4.1.
These were obtained by changing the central density of the thick disk, which gave a temperature
distribution inside CENBOL. Subsequently, CT95 was followed to obtain an effective temperature. In 
simulations, however, the actual temperature distribution was used.

\begin{table}
\centering
\centerline {Table 4.1 (GCL09)}
\centerline {Central electron number densities ($n$ in cm$^{-3}$) and the effective temperatures (in keV)}
\centerline {for various outer edge $R_{out}$ of the CENBOL used in this Chapter.}
\vskip 0.2cm
\begin {tabular}[h]{llllllll}
\hline
$n$ (cm$^{-3}$) $\ \downarrow R_{out} \rightarrow$ & $10$ & $15$ & $20$ & $30$ &$50$ & $100$ & Line style\\
\hline
$7\times10^{19}$ & 61 & 54 & 50 & 46 & 43 & 42 & Solid\\
$1\times10^{20}$ & 73 & 64 & 59 & 54 & 51 & 50 & Dotted\\
$2\times10^{20}$ & 91 & 80 & 74 & 68 & 65 & 63 & Short dashed\\
$5\times10^{20}$ & 115 & 101 & 93 & 86 & 81 & 79 & Big dashed\\
$9\times10^{20}$ & 145 & 128 & 118 & 109 & 102 & 99 & Dash-dotted\\
$2\times10^{21}$ & 197 & 173 & 160 & 147 & 139 & 135 & Big dash-dotted\\
\hline
\end {tabular}
\end{table}
\section{Simulation Procedure}

In a simulation, we randomly generated a photon out of the Keplerian disk and using 
another set of random numbers we obtained its injected direction. With another 
random number we obtained a target optical depth $\tau_c$ at which the scattering takes place. 
The photon is followed within CENBOL till the optical depth reached $\tau_c$. 
At this point a scattering is allowed to take place and the energy exchange is computed 
through Compton or inverse Comptonization (see Chapter 3 for details). The electrons are
are assumed to obey relativistic Maxwell distribution inside CENBOL.
The photon frequencies are also gravitationally red-shifted or blue shifted depending on 
its relative location change with respect to the black hole. The process is continued till
the photon either leaves the CENBOL or is absorbed by the black hole. 
\section{Results and Discussions}

In Table 4.2, we summarize all the cases for which the simulations have been presented in this Chapter. 
In Col. 1, various cases are marked. In Col. 2, the $R_{out}$ and $T_{eff}$ in keV are listed.
In Cols. 3 we show the temperature ($T_{p}$) from the innermost annulus ($R_{out}$) 
of the Keplerian disk. Columns 4, 5, 6 and 7 show 
the total number of injected photons ($N_{inj}$), number of the photons 
intercepted by the CENBOL ($N_{int}$), 
number of photons which suffered Compton scattering ($N_{cs}$) and the 
number of photons captured ($N_{cap}$) by the black hole respectively. 
In Column 8 we calculated the percentage $p$ 
of the total injected photons that have suffered scattering through CENBOL. 
In Column 9, we present the 
energy spectral index $\alpha$ ($I(E) \sim E^{-\alpha}$) obtained from our simulations.
{\small{
\begin{center}
\begin {tabular}[]{ccccccccc}
\multicolumn{9}{c}{Table 4.2: Summary of all the simulation cases presented in this Chapter (GCL09).}\\
\hline Case & $R_{out}$, $T_{eff}$ & $T_p $ & $N_{inj}$ & $N_{int}$ & $N_{cs}$ & $N_{cap}$ & $p$ & $\alpha$  \\
\hline
1a & 10, 61 & 0.227 & 115150710 & 3042538 & 3024733 & 17805 & 2.627 & 2.10 \\
1b & 10, 73 & -do- & -do- & 3043059 & 3025416 & 17643 & 2.627 & 1.90\\
1c & 10, 91 & -do- & -do- & 3041990 & 3024452 & 17538 & 2.627 & 1.65\\
1d & 10, 115 & -do- &-do- & 3046115 & 3028743 & 17372 & 2.630 & 1.40\\
1e & 10, 145 & -do- &-do- & 3043849 & 3026646 & 17203 & 2.628 & 1.15 \\
1f & 10, 197 & -do- &-do- & 3042031 & 3025183 & 16848 & 2.627 & 0.90\\
\hline
2a & 15, 54 & 0.177 & 101283949 & 4011191 & 4005770 & 5421 & 3.955 & 1.12 \\
2b & 15, 64 & -do- & -do- & 4011473 & 4006227 & 5246 & 3.955 & 0.99 \\
2c & 15, 80 & -do- & -do- & 4012125 & 4007312 & 4813 & 3.957 & 0.82 \\
2d & 15, 101 & -do- &-do- & 4013872 & 4009153 & 4719 & 3.958 & 0.70 \\
2e & 15, 127 & -do- &-do- & 4007883 & 4003407 & 4476 & 3.953 & 0.57 \\
2f & 15, 173 & -do- &-do- & 4011584 & 4007483 & 4101 & 3.957 & 0.45\\
\hline
3a & 20, 50 & 0.147 & 91280716  & 4224551 & 4222011 & 2540 & 4.625 & 0.85\\
3b & 20, 59 & -do- & -do- & 4222520 & 4219921 & 2599 & 4.623 & 0.75\\
3c & 20, 74 & -do- & -do- & 4224950 & 4222666 & 2284 & 4.626 & 0.64\\
3d & 20, 93 & -do- & -do- & 4221994 & 4219926 & 2068 & 4.623 & 0.54\\
3e & 20, 118 & -do- &-do- & 4223468 & 4221663 & 1805 & 4.623 & 0.44\\
3f & 20, 160 & -do- &-do- & 4222218 & 4220626 & 1592 & 4.623 & 0.34 \\
\hline
4a & 30, 46 & 0.111 & 77355270  & 4369685 & 4368926 & 759 & 5.648 & 0.65 \\
4b & 30, 54 & -do- & -do- & 4366959 & 4366302 & 657 & 5.645 & 0.56\\
4c & 30, 68 & -do- & -do- & 4367505 & 4366932 & 573 & 5.645 & 0.46\\
4d & 30, 86 & -do- & -do- & 4371154 & 4370665 & 489 & 5.650 & 0.40 \\
4e & 30, 109 & -do- &-do- & 4368390 & 4367963 & 427 & 5.647 & 0.35\\
4f & 30, 147 & -do- &-do- & 4372356 & 4372014 & 342 & 5.652 & 0.29\\
\hline
5a & 50, 43 & 0.078 & 60533079  & 4194919 & 4194781 & 138 & 6.930 & 0.56 \\
5b & 50, 51 & -do- & -do- & 4195709 & 4195582 & 127 & 6.931 & 0.50\\
5c & 50, 65 & -do- & -do- & 4196096 & 4195988 & 108 & 6.931 & 0.44\\
5d & 50, 81 & -do- & -do- & 4195788 & 4195715 & 73 & 6.931 & 0.38\\
5e & 50, 102 & -do- & -do- & 4194298 & 4194244 & 54 & 6.929 & 0.32\\
5f & 50, 139 & -do- & -do- & 4194327 & 4194282 & 45 & 6.929 & 0.28\\
\hline
6a & 100, 42 & 0.048 & 39539601  & 3780685 & 3780669 & 16 & 9.562 &  0.42\\
6b & 100, 50 & -do- & -do- & 3801929 & 3801919 & 10 & 9.615 & 0.30\\
6c & 100, 63 & -do- & -do- & 3808845 & 3808843 & 2 & 9.633 & 0.24\\
6d & 100, 79 & -do- & -do- & 3812567 & 3812565 & 2 & 9.642 & 0.19\\
6e & 100, 99 & -do- & -do- & 3814646 & 3814644 & 2 & 9.648 & 0.17\\
6f & 100, 135 & -do- &-do- & 3815355 & 3815353 & 2 & 9.650 & 0.13\\
\hline
\end {tabular}
\end{center}
}}

\begin{figure}[h]
\begin{center}
\vskip 4.3cm
\includegraphics{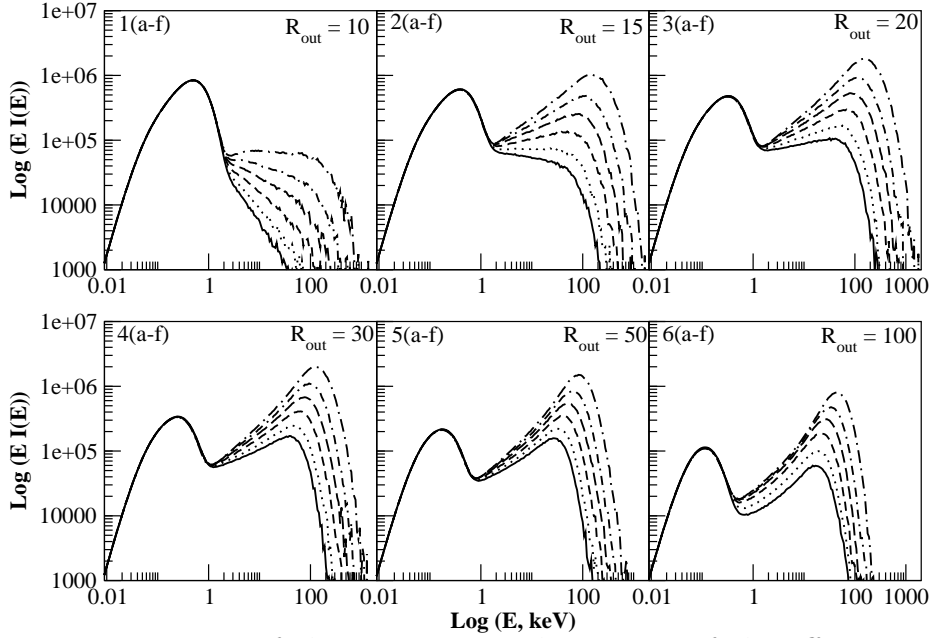}
\vskip 5.0cm
\caption{Variation of the spectrum with increase of the effective temperature 
($T_{eff}$) of the CENBOL for a fixed CENBOL size $R_{out}$.
Each panel marks the cases for which the spectrum is drawn.
Curves for (a) to (f) are from bottom to top respectively along the direction of 
increasing density and effective temperature (GCL09).}
\label{fig.4.4}
\end{center}
\end{figure}

In Fig. \ref{fig.4.4}, we summarize the results of all the cases, bunching them in 
groups with the same CENBOL size. 
Different cases are marked in each panel. Curves (a) to (f) are from bottom to top respectively. 
Note that as $T_{eff}$ is raised (a $\rightarrow $ f), the spectrum becomes harder.
Also, as $R_{out}$ is increased, the percentage $p$ of photons intercepted is 
also increased as the CENBOL 
becomes bigger. However, as the CENBOL size is increased, it becomes increasingly difficult to
soften the spectrum with the same number of injected photons. Thus the spectrum becomes 
harder. This behavior 
matches with the earlier theoretical predictions (Fig. 6 of C97). All the spectra 
are angle-averaged.

\begin{figure}[]
\begin{center}
\vskip 6.5cm
\includegraphics{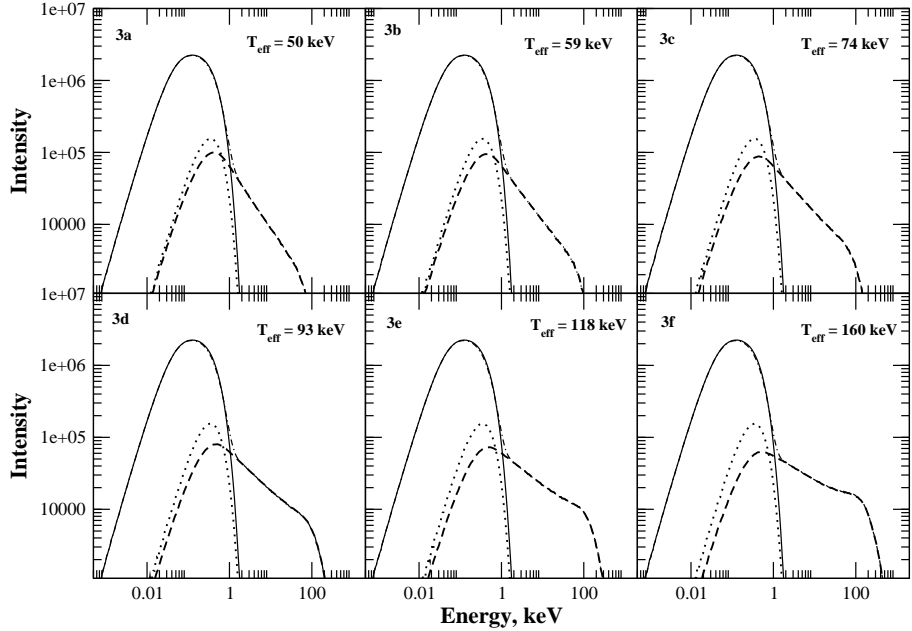}
\vskip 3.0cm
\caption{Components of the emerging spectrum with a Keplerian disk
outside $R_{out}=20$ and their variation with the effective temperature of the CENBOL.
As the electron temperature becomes hotter, the spectrum also gets harder (GCL09).}
\label{fig.4.5}
\end{center}
\end{figure}

\begin{figure}[]
\begin{center}
\vskip 5.0cm
\includegraphics{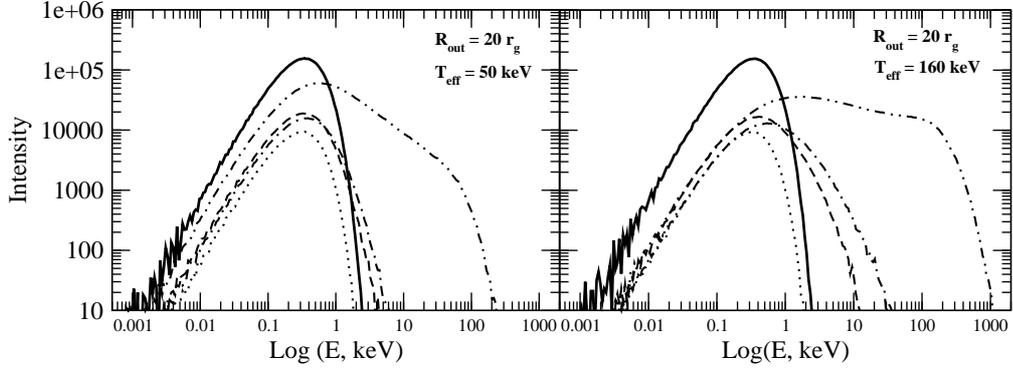}
\vskip 1.0cm
\caption{Components of the emerging spectra with a Keplerian disk outside
 $R_{out}=20$. The solid curve is for the injected photons. The dotted, dashed, 
dot–dashed and double dot-dashed curves show contributions from photons 
with number of scattering 1, 2, 3 and 4 or higher respectively. 
Cases 3a (left) and 3f (right) are shown (GCL09).}
\label{fig.4.6}
\end{center}
\end{figure}

In Fig. \ref{fig.4.5}, we take one set, namely, Cases 3(a-f), for which $R_{out}=20$ but the temperature 
distribution is varied which also changed $T_{eff}$. Here we draw each components, namely,  
the  injected component (solid), the intercepted component (dotted) and the 
Comptonized component (dashed) separately. The net spectrum is shown as the 
dash-dotted curve. As we increase the temperature of the CENBOL, it becomes increasingly harder to cool 
the electrons, and thus the spectrum becomes harder. In Fig. \ref{fig.4.6}, we  
show two typical cases (Cases 3a and 3f) in which we wish to demonstrate 
how the power-law component has been produced. The solid curve represents the injected photons. 
The dotted, dashed, dot–dashed and double dot-dashed curves show contributions 
from photons which underwent $1$, $2$, $3$ and $4$ or above number of scattering respectively. 

\begin{figure}[]
\begin{center}
\vskip 5.0cm
\includegraphics{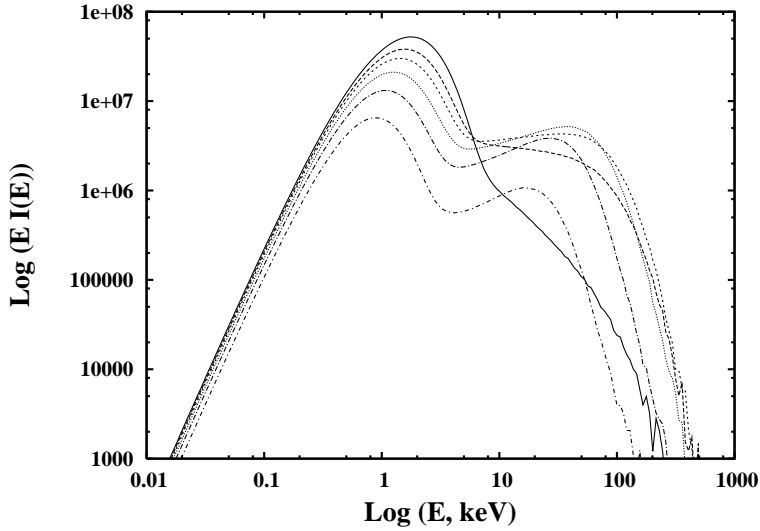}
\vskip 1.0cm
\caption{Variation of the spectrum with the CENBOL size is shown. The cases correspond to Cases (1-6)a
which are drawn in solid, dashed, small-dashed, dotted, long-dashed and small dash-dotted curves
respectively. See Table 4.2 for parameters. With increase in size of the CENBOL the spectrum
becomes harder to cool, although the optical depth weighted effective temperature becomes 
lower. The latter causes the cut-off energy to become lower (GCL09).}
\label{fig.4.7}
\end{center}
\end{figure}

\begin{figure}[]
\begin{center}
\vskip 6.0cm
\includegraphics{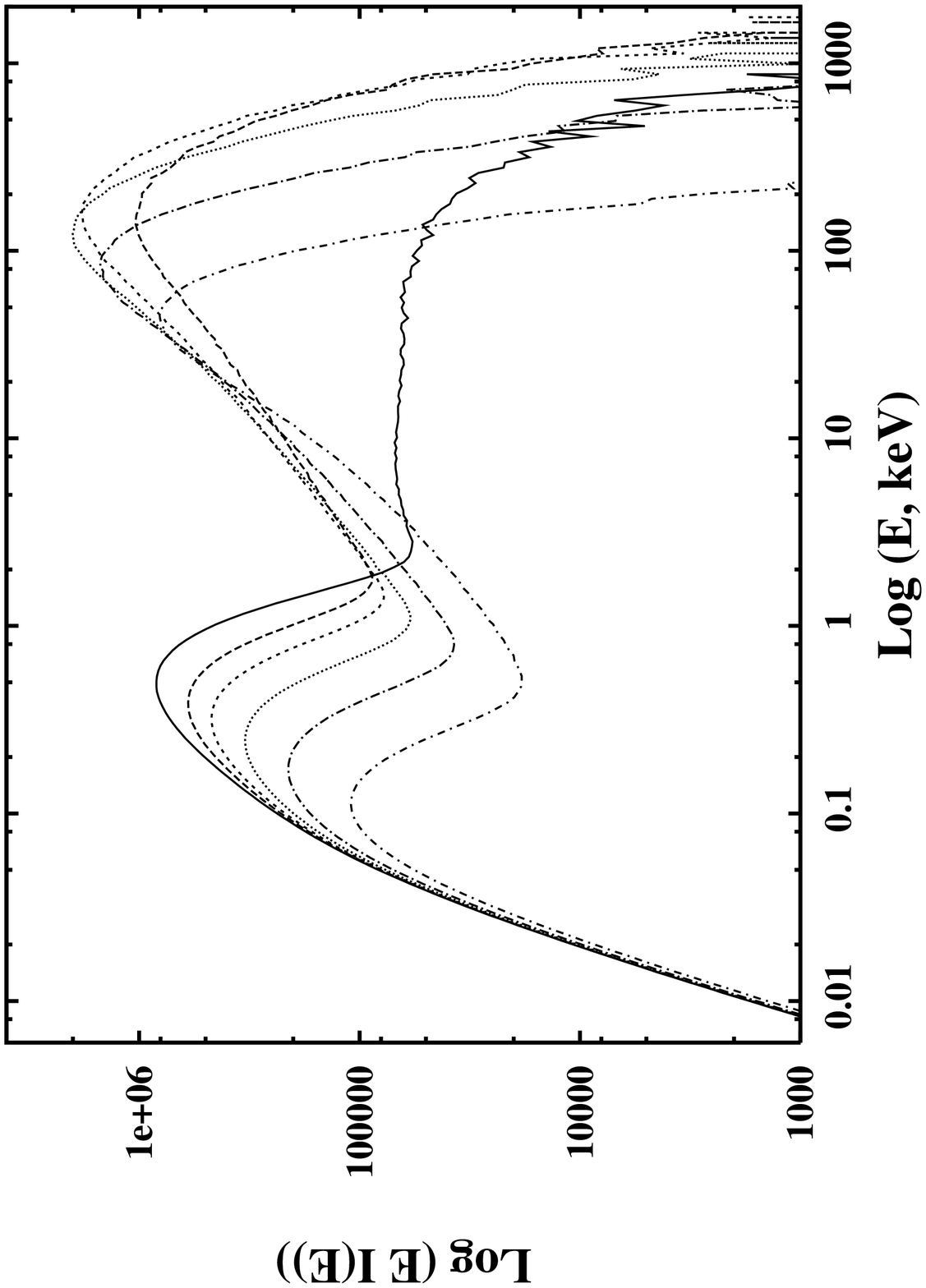}
\vskip 1.0cm
\caption{Same as Fig. \ref {fig.4.7}. The cases correspond to Cases (1-6)f
which are drawn in solid, dashed, small-dashed, dotted, long-dashed and small dash-dotted curves
respectively. Table 4.2 lists all the parameters.}
\label{fig.extra}
\end{center}
\end{figure}

The spectral variation with the CENBOL size has been plotted in Figs. \ref{fig.4.7} and \ref{fig.extra}. 
Cases (1-6)a are shown in Fig. \ref{fig.4.7} and Cases (1-6)f are shown in Fig. \ref{fig.extra}. 
Solid, dashed, small-dashed, dotted, long-dashed and small dash-dotted curves 
are drawn for Cases 1, 2, 3, 4, 5 and 6, respectively. With the increase in size of the CENBOL, the spectrum
becomes harder, although the optical depth weighted effective temperature becomes
lower. The latter causes the cut-off energy to become lower as well.

\begin{figure}[]
\begin{center}
\vskip 5.0cm
\includegraphics{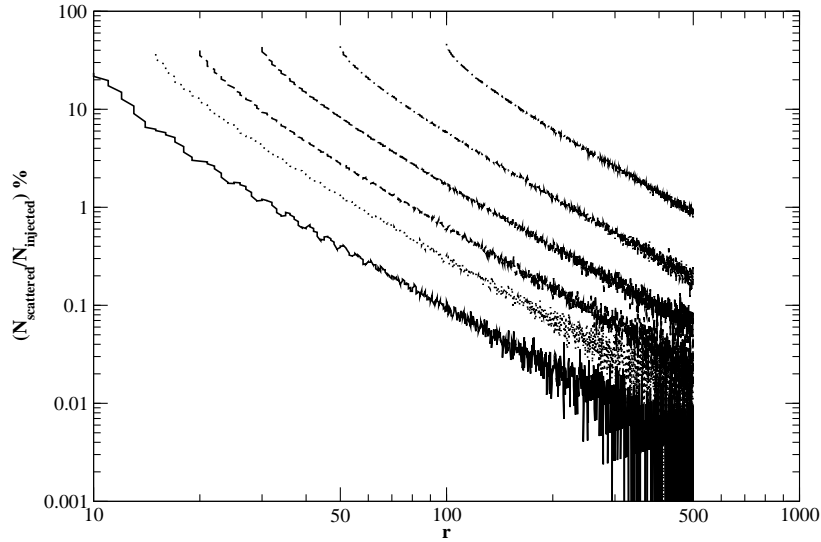}
\vskip 2.5cm
\caption{Ratio of the scattered photons and the injected photons for different annuli of the Keplerian
disk. Cases (1-6)e are drawn from bottom to top. The result is insensitive to the effective
temperature of the electrons and generally depends on the relative geometry (GCL09).}
\label{fig.4.8}
\end{center}
\end{figure}

In order to understand how the spectrum is influenced by the photons from different annuli,
we compute the fraction of injected photons from each annulus which suffer scattering.
In Fig. \ref{fig.4.8}, we show the result for various CENBOL sizes. What we find is that when the 
CENBOL size is smaller, say, $R_{out}=10$, only about $20\%$ photons are intercepted from the nearest 
annulus, but the effect of the annuli close to the periphery is negligible. 
On an average, however, only $2.6\%$ get intercepted (see Table 4.2). When the CENBOL size is bigger, say,
$R_{out}=100$, almost $50\%$ of the photons from the annulus immediately outside the CENBOL
gets intercepted and scattered. In this case, on an average, about $10\%$ photons from the whole
disk is scattered. From Table 4.2 we see that the nature of the above plot does not change for a 
particular CENBOL size even when the temperature is varied. So it is purely a geometric effect.
\begin{figure}[]
\begin{center}
\vskip 5.0cm
\includegraphics{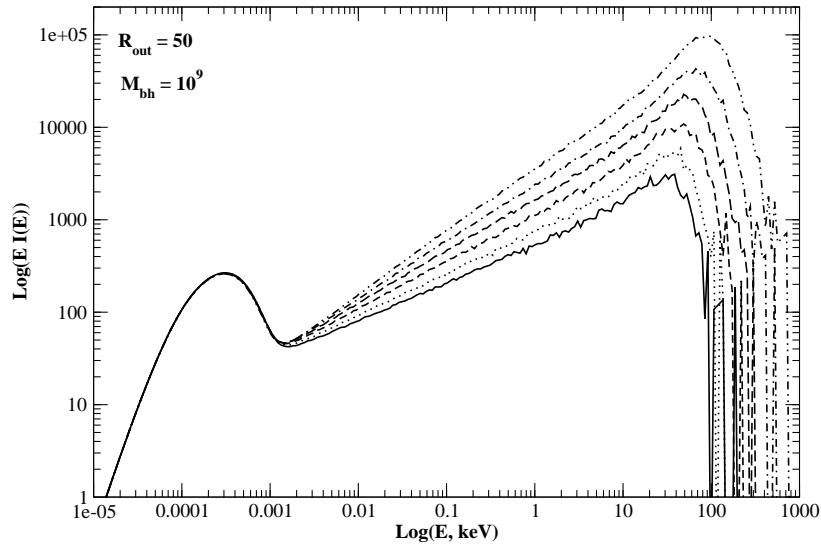}
\vskip 2.1cm
\caption{Plot of EI(E) for Cases 7(a-f) (presented in Table 4.3) are shown. 
For these simulations we considered $\dot{m}=0.001$, $M_{bh} = 10^9 M_\odot$. 
Curves (a) to (f) are from bottom to top respectively (GCL09).}
\label{fig.4.9}
\end{center}
\end{figure}

It is in general instructive to understand the behavior when the black hole mass is 
much higher. In this case, the effective temperature could be much lower and the accretion rate will also be
generally lower. In Table 4.3, we show the cases which were run for a massive black hole. We chose the
mass to be $10^9 M_\odot$ and the accretion rate ${\dot m} = 0.001$. For the sake 
of comparison with the earlier cases, we selected the CENBOL parameters exactly same as in Cases 5(a-f).
In Fig. \ref{fig.4.9}, we note that for super-massive black hole, the Keplerian photons are cooler. Nevertheless, 
the inverse Comptonization extends the spectra to very high energies. This is because the 
source of the energy is the hot electron cloud itself. The variation of $\alpha$, the spectral index
is given in the table and they are marginally softer compared to what was observed for smaller black holes
(Table 4.2). 
\begin{center}
\begin {tabular}[h]{ccccccccc}
\multicolumn{9}{c}{Table 4.3: Summary of the simulation cases for a massive black hole (GCL09).}\\
\hline Case & $R_{out}$, $T_{eff}$ & $T_p $ & $N_{inj}$ & $N_{int}$ & $N_{cs}$ & $N_{cap}$ & $p$ & $\alpha$  \\
\hline
7a & 50, 43 & 1.39E-4 & 34037468 & 2359549 & 2359463 & 86 & 6.932 & 0.59  \\
7b & 50, 51 & -do- & -do- & 2358802 & 2358725 & 78 & 6.930 & 0.55\\
7c & 50, 65 & -do- & -do- & 2359201 & 2359145 & 57 & 6.931 & 0.50\\
7d & 50, 81 & -do- & -do- & 2359354 & 2359313 & 41 & 6.931 & 0.44\\
7e & 50, 102 & -do- & -do- & 2360152 & 2360114 & 38 & 6.934 & 0.39\\
7f & 50, 139 & -do- & -do- & 2358340 & 2358314 & 26 & 6.929 & 0.32\\
\hline
\end {tabular}
\end{center}



In the present computation we assumed a stationary toroidal accretion disk. An inclusion of the radial 
component of velocity can produce an interesting effect, particularly visible when the spectral index is very high
(in the so-called soft-state of the black hole). Here the electrons become so cold that the 
thermal Comptonization is ineffective and the power-law spectrum is dominated by the bulk motion 
Comptonization (CT95). This effect for spherical cloud has been demonstrated by Laurent \& Titarchuk
(1999) and can be considered to be a signature of a black hole candidate, since the radial 
velocity is high for infalling matter around such objects. In the next Chapter, we will show 
the results of the simulations incorporating the bulk velocity of the electron.
We will  also show the effects of the presence of an outflow along with the inflow. 
In presence of a rotational motion, preliminary results (Chakrabarti, Titarchuk, Kazanas \& Ebisawa, 1996) show 
that the spectrum tends to become harder. Similarly, the outflow, which is generally believed 
to be formed out of the Compton cloud (here CENBOL) itself, can also Comptonize the injected photons
and in certain situation could be very important. 
\newpage

%% file: CHAPTER5/chap5.tex

\newpage
\markboth{\it Effects of Thermal and Bulk Motion Comptonizations}
{\it Effects of Thermal and Bulk Motion Comptonizations}
\chapter{Effects of Thermal and Bulk Motion Comptonizations in Presence of an Outflow}



\section{Introduction}	
So far, we have only discussed the effects of thermal Comptonization 
on the soft photons injected into a static Compton cloud (Chapter 4). 
In the present Chapter, we will show the effects of the bulk velocity 
of the cloud on the emitted spectrum from an accretion disk. 
As discussed in Chapter 1, the accretion flow onto a black hole 
is necessarily transonic in nature. 
In the context of the spherical flows, 
Bondi (1952) solution of accretion and Parker (1959) solution of 
winds are clear examples of transonic flows. But they have only 
one sonic point. In presence of angular momentum, the flow may have two saddle 
type sonic points with a shock in between (C90, Chakrabarti, 
1996a). The solutions with shocks have been extensively 
studied in both the accretion and the winds even when 
rotation, heating, cooling etc. are included 
(C90, Chakrabarti, 1996a). The study demonstrates that 
the accretion and the winds are inter-related, the outflows 
are generated from the post-shock region. Subsequently, 
in C99, Das \& Chakrabarti (1999) and Das et al. (2001), 
the mass outflow rate was computed as a function of the 
shock strength and other flow parameters. Meanwhile, in the 
so-called two component advective flow (TCAF) model of 
CT95 and C97, the spectral states were shown to depend on 
the location and strength of the shock. Thus, C99 for the 
first time, brought out the relationship between the jets 
and outflows with the presence or absence of 
shocks, and therefore with the spectral states of a black hole 
candidate. This paves the way to study the relative 
importance between the Compton cloud and the outflow as far 
as the emerging spectrum is concerned. 

Computation of the spectral characteristics have so far been done only 
for the advective accretion flows (CT95; Chakrabarti \& Mandal, 2006) 
and the outflow or the base of the jet was not included. 
In the Monte-Carlo simulations of Laurent \& Titarchuk (2007), 
outflows in isolation were used, but not in conjunction with inflows.
In GCL09, the results of Monte-Carlo simulations in a setup similar to 
that of CT95 was presented (Chapter 4). In the present Chapter 
(Ghosh, et al., 2010; hereafter GG10), we improve this 
and obtain the outgoing spectrum in presence of 
both inflows and outflows. We also include a Keplerian disk inside 
an advective flow which is the source of soft photons. We show how the spectrum 
depends on the flow parameters of the inflow, such as the accretion rates of the 
two components and the shock strength. The post-shock region 
being denser and hotter, it behaves like the so-called 'Compton cloud' 
in the classical model of ST80. This is 
the CENBOL region described in Chapter 1. The shock location 
(size of the Compton cloud) and its 
strength depends on the basic parameters of the flow, such as 
the specific energy, the accretion rate and the specific angular 
momentum. Thus, the basic Comptonized component of the 
spectrum is a function of the flow parameters. Since the 
intensity of soft photons determines the Compton cloud 
temperature, the result depends on the accretion rate of 
the Keplerian component also. In our result, we see the 
effects of the bulk motion Comptonization (CT95) 
because of which even a cooler CENBOL produces a harder spectrum. 
At the same time, the effect of down-scattering due the 
outflowing electrons is also seen, because of which even 
a hotter CENBOL causes the disk-jet system to emit lesser 
energetic photons. Thus, the net spectrum is a combination 
of all these effects.

In the next Section, we discuss the geometry of the soft 
photon source and the Compton cloud in our Monte-Carlo simulations. 
In \S 5.2, we present the variation of the thermodynamic 
quantities and other vital parameters inside the Keplerian 
disk and the Compton cloud which are required for the 
Monte-Carlo simulations. In \S 5.3, we describe the 
simulation procedure and in \S 5.4, we present the 
results of our simulations. Finally in \S 5.5, we make 
concluding remarks.

\section{Geometry and Properties of the Flow}

\begin{figure}[]
\begin{center}
\vskip 4.9cm
\includegraphics{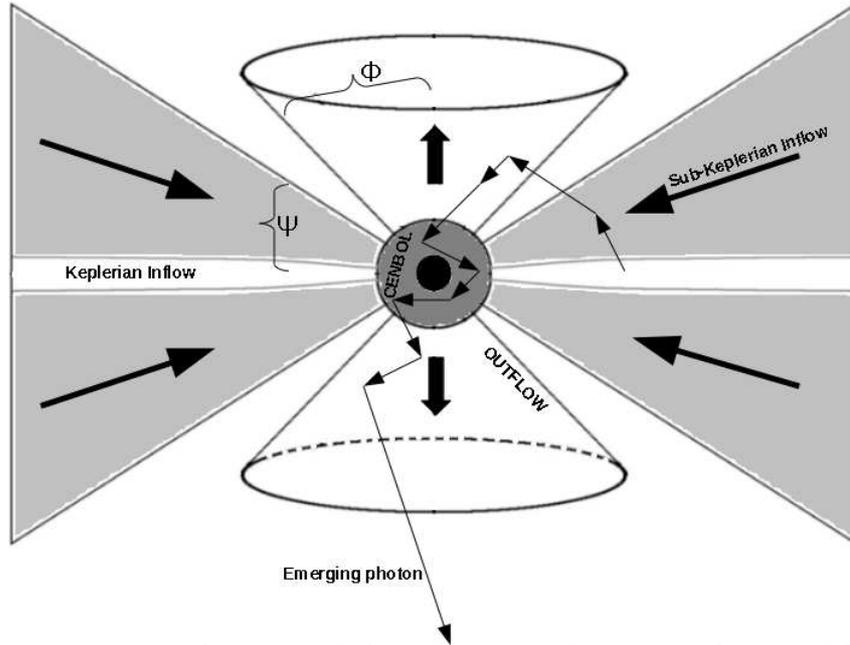}
\vskip 3.2cm
\caption{A cartoon diagram of the geometry of our initial case of Monte-Carlo 
simulation. The spherical inflowing post-shock region (CENBOL) surrounds 
the black hole. The Keplerian disk is on the equatorial plane and a 
sub-Keplerian halo is above and below. A diverging conical outflow 
is formed from the CENBOL. The zig-zag path of a typical photon is shown (GG10).}
\label{cartoon5}
\end{center}
\end{figure}

To simplify the geometry of the inflow-outflow configuration without 
sacrificing the salient features, we first assume the flow geometry as depicted in the cartoon diagram 
presented in Fig. \ref{cartoon5}. This is our simulation set up. The components of the
hot electron clouds, namely, the CENBOL, the outflow and the sub-Keplerian flow, 
intercept the soft photons emerging out of the Keplerian disk 
and reprocess them via inverse-Compton scattering. An injected 
photon may undergo a single, multiple or no scattering at all with 
the hot electrons in between its emergence from the Keplerian disk 
and its detection by the telescope at a large distance. 
The photons which enter the black holes are absorbed. The CENBOL, 
though toroidal in nature, is chosen to be of spherical shape 
for simplicity. The sub-Keplerian inflow in the pre-shock 
region is assumed to be of wedge shape of a constant angle $\Psi$. 
The outflow, which emerges from the CENBOL in this picture, is 
also assumed to be of constant conical angle $\Phi$. In reality, 
inflow and outflow both could have somewhat different shapes, depending on the 
balance of the force components. However, the final result is 
not expected to be sensitive to such assumptions.

\subsection{Compton Cloud and its Temperature, Density and Velocity}

We assume the black hole to be non-rotating and we use the pseudo-Newtonian
potential (PW80) to describe the geometry around 
the black hole. This potential is $-\frac{1}{2(r-1)}$ (Section 1.8). The velocity components and 
angular momenta are measured in units of $c$, the velocity of 
light and $r_g c$ respectively. For simplicity, we have chosen the 
Bondi accretion solution in pseudo-Newtonian geometry to describe 
both the accretion and winds. The equation of motion of the sub-Keplerian 
matter around the black hole in the steady state is assumed to be given by,
$$
u\frac{du}{dr}+\frac{1}{\rho}\frac{dP}{dr} +\frac{1}{2(r-1)^2}=0.
$$ 
Integrating this equation, we get the expression of the conserved specific energy as, 
$$
\epsilon=\frac{u^2}{2}+na^2-\frac{1}{2(r-1)}.
\eqno{(5.1)}
$$
Here $P$ is the thermal pressure and $a$ is the adiabatic
sound speed, given by $a=\sqrt{\gamma P/\rho}$, $\gamma$ being the adiabatic
index and is equal to $\frac{4}{3}$ in our case. The conserved mass flux
equation, as obtained from the continuity equation, is given by
$$
\dot{m}=\Omega\rho u r^2,
\eqno{(5.2)}
$$
where, $\rho$ is the density of the matter and $\Omega$ is the solid angle subtended 
by the flow. For an inflowing matter, $\Omega$ is given by, 
$$
\Omega_{in}=4\pi sin\Psi,
$$
where, $\Psi$ is the half-angle of the conical inflow. For the outgoing matter, the 
solid angle is given by, 
$$
\Omega_{out}=4\pi(1-cos\Phi),
$$
where $\Phi$ is the half-angle of the conical outflow. From Eqn. (5.2), we get
$$
\dot{\mu}=a^{2n}ur^2.
\eqno{(5.3)}
$$
The quantity $\dot{\mu}=\frac{\dot{m}\gamma ^n K^n}{\Omega}$ is the 
Chakrabarti rate (Chakrabarti, 1989, C90, 1996a) which includes 
the entropy, $K$ being the constant measuring the entropy of the flow,
and $n=\frac{1}{\gamma -1}$ is called the polytropic index. We take derivative
of Eqns. (5.1) and (5.3) with respect to $r$. Eliminating $\frac{da}{dr}$ from
both the equations, we get the gradient of the velocity as, 
$$
\frac{du}{dr}=\frac{\frac{1}{2(r-1)^2}-\frac{2a^2}{r}}{\frac{a^2}{u}-u}.
\eqno{(5.4)}
$$
Solving this, we obtain the Bondi accretion and wind solutions in the usual manner (C90).
Solution of Eqn. (5.4) gives the radial variation of velocity $u$. Using Eqn. (5.1),
we find the radial variation of sound speed $a$. Finally, we get the temperature  
profile of the electron cloud ($T_e$) using $T_e=\frac{\mu a^2 m_p}{\gamma k_B}$,
where $\mu=0.5$ is the mean molecular weight, $m_p$ is the proton mass and $k_B$ is
the Boltzmann constant. Using Eqn. (5.2), we calculate the mass density $\rho$,
and hence, the number density variation of electrons inside the Compton cloud. 
We ignore the electron-positron pair production inside the cloud.

\begin{figure}[]
\begin{center}
\vskip 5.2cm
\includegraphics{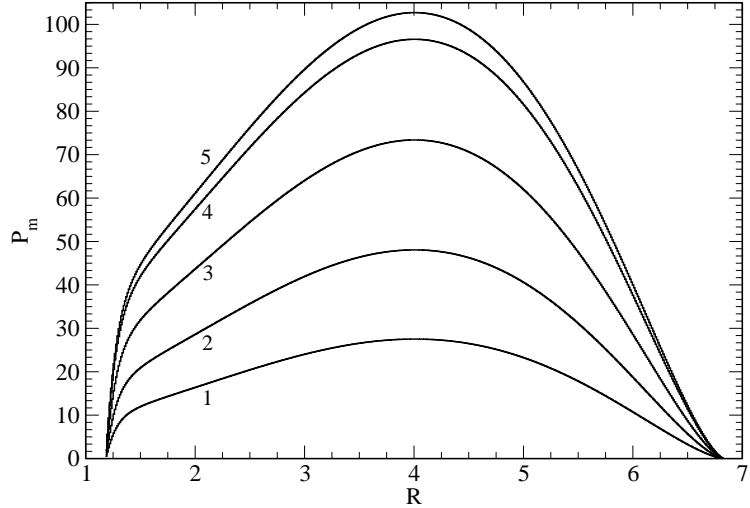}
\vskip 1.9cm
\caption{Percentage of matter $P_m = 100 \times R_{\dot{m}}$ inside 
the jet as a function of the compression ratio $R$ of the inflow 
when the outflow is adiabatic. Plots 1-5 are for different jet 
angles (30$^\circ$, 40$^\circ$, 50$^\circ$, 58$^\circ$ and 60$^\circ$, 
respectively). In simulations presented in this Chapter, we have 
used the jet angle to be 58$^\circ$ (Plot 4).}
\label{compression}
\end{center}
\end{figure}

The flow is supersonic in the pre-shock region and sub-sonic in the 
post-shock (CENBOL) region. We chose this surface at a location ($R_s$), 
where the pre-shock Mach number $M=2$. This location depends 
on the specific energy $\epsilon$ (C90). In our simulation, 
we have chosen $\epsilon = 0.015$ so that we get $R_{s} = 10$. 
We simulated a total of six cases. For Cases 1(a-c), we chose 
$\dot {m_h} = 1$, $\dot {m_d} = 0.01$ and for Cases 2(a-c), 
the values are listed in Table 5.2. The velocity variation 
of the sub-Keplerian flow is the inflowing Bondi solution 
(pre-sonic point). The density and the temperature of 
this flow have been calculated according to the above 
mentioned formulas. Inside the CENBOL, both the Keplerian and the
sub-Keplerian components are mixed together. The velocity variation of the
matter inside the CENBOL is assumed to be the same as the Bondi accretion flow solution
reduced by the compression ratio due to the shock. The compression ratio (i.e., 
the ratio between the post-shock and pre-shock densities) $R$ is also used 
to compute the density and the temperature profile of the Compton cloud and the jet. 
When the outflow is adiabatic, the ratio of the outflow to the inflow rate is 
(Das et al. 2001) given by,
$$
R_{\dot{m}} = \frac{\Omega_{out}}{\Omega_{in}} \left( \frac{f_0}{4 \gamma}\right)^3 \frac{R}{2} \left[ \frac{4}{3} \left( \frac{8(R-1)}{R^2} -1 \right) \right]^{3/2} .
\eqno{(5.5)}
$$
Here we have used $n=3$ for a relativistic flow. 
In Fig. \ref{compression} we have plotted the variation of 
$R_{\dot{m}}$ (in percentage) with R, for different opening angles $\Phi$ 
of the jet. For drawing Fig. \ref{compression}, $\Omega_{in}$ is kept constant 
($\Psi = 32^\circ$). The angle $\Phi$ is varied from 30$^\circ$, 
40$^\circ$, 50$^\circ$, 58$^\circ$ and 60$^\circ$, the corresponding 
curves are 1, 2, 3, 4 and 5, respectively. In our simulation, we have generally 
used $\Phi = 58^\circ$ (Plot 4, Fig. \ref{compression}). This is the 
highest value of $\Phi$ we can choose for our system because, for $\Phi$ 
more than $58^\circ$ the disk would be evacuated (e.g., Plot 5, where 
for $\Phi = 60^\circ$ maximum $P_m$ is more than 100\%). Thus we have chosen 
$\Phi$ to be smaller than this in order that steady jets are produced. 
The velocity variation inside the jet is obtained from the outflow branch of Bondi 
solution. Using $P_m$ and the velocity of matter inside it we compute 
the density variation inside the jet.

\subsection{Source of Soft Photons}

The soft photons are produced from a Keplerian disk whose inner 
edge coincides with the CENBOL surface, while the outer edge is 
located at $500 r_g$. The source of soft photons have the same  
multi-color blackbody spectrum as described in Section 4.2, 
coming from a standard SS73 disk. 
The soft photons are generated isotropically between the inner and outer edge 
of the Keplerian disk but their positions are randomized using the distribution 
function (Eqn. 4.5) of black body temperature $T(r)$. All the results of the simulations 
presented here have used the number of injected photons to be $6.4\times10^8$.
In the above equations, the mass of the black hole $M_{bh}$ is measured 
in units of the mass of the Sun ($M_\odot$), 
the disk accretion rate $\dot{M_d}_{17}$ is in units of $10^{17}$ gm/s. 
We chose $M_{bh} = 10$ and $\delta r = 0.5 r_g$.

\section{Simulation Procedure}

In a given simulation, we assume a Keplerian disk rate and a 
sub-Keplerian halo rate to be given. The specific energy of the halo 
determines hydrodynamic properties (such as number density of the 
electrons and the velocity variation) and the thermal properties 
of matter. Since we chose the PW80 potential, 
the radial velocity is not exactly unity at $r=1$, the horizon, 
but it becomes unity just outside. In order not to over-estimate 
the effects of bulk motion Comptonization which is due to the 
momentum transfer of the moving electrons to the horizon, we 
shift the horizon just outsize $r=1$ where the velocity is unity. 
The shock location of the CENBOL is chosen where the Mach number 
$M=2$ for simplicity and the compression ratio at the shock 
is assumed to be a free parameter. These simplifying assumptions 
are not expected to affect our conclusions. Photons are generated 
from the Keplerian disk according to the prescription in SS73 as 
mentioned before and are injected into the sub-Keplerian halo, 
the CENBOL and the  outflowing jet.

Generally, the same simulation procedure as in Chapter 4 is used, 
except that we are now counting also those photons which 
were scattered at least once by the outflow. To highlight our point, we are especially choosing the
cases when the jet could play a major role in shaping the spectrum.

\section{Results}

\subsection{When the Outflow is Present}

\begin{figure}[]
\begin{center}
\vskip 4.9cm
\includegraphics{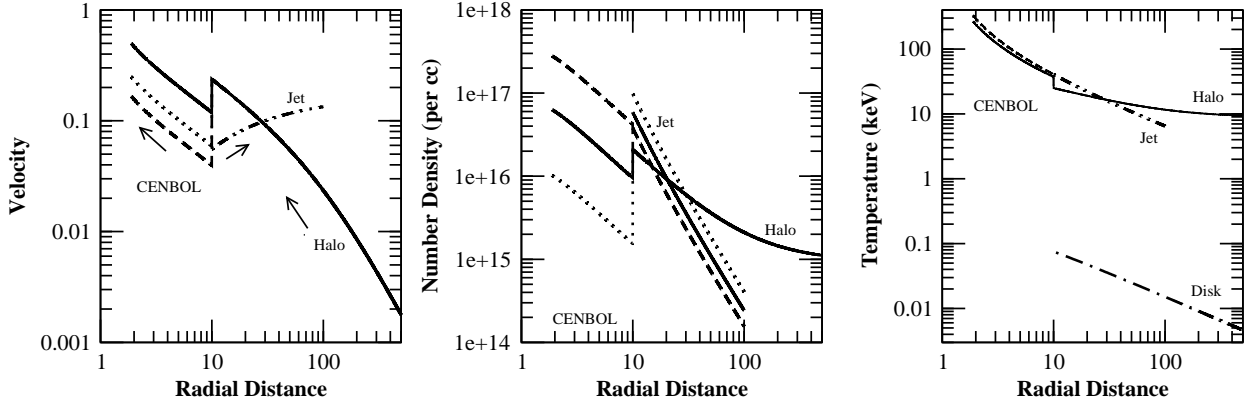}
\vskip 2.2cm
\caption{{\bf (a-c)}: Velocity (left), density (middle) and the 
temperature (right) profiles of Cases 1(a-c) as described in Table 5.1 
are shown with solid ($R=2$), dotted ($R=4$) and dashed ($R=6$) curves. 
$\dot{m_d} = 0.01$ and $\dot{m_h} = 1$ were used (GG10).}
\label{fig.5.3}
\end{center}
\end{figure}

First, we consider the cases where the outflow is present. 
In Figs. \ref{fig.5.3}(a-c) we present the velocity, electron number density 
and temperature variations as a function of the radial distance 
from the black hole for specific energy $\epsilon=0.015$. 
$\dot{m_d} = 0.01$ and $\dot{m_h} = 1$ were chosen. Three cases 
were run by varying the compression ratio $R$. These are given 
in Col. 2 of Table 5.1. The corresponding percentage of matter going in the 
outflow is also given in Col. 2. In the left panel, the bulk velocity variation is shown.
The solid, dotted and dashed curves are the same for $R = 2$ (Case 1a), $4$ 
(Case 1b) and $6$ (Case 1c) respectively. The same line style is used 
in other panels. The velocity variation within the jet does not 
change with $R$, but the density (in the unit of $cm^{-3}$) does (middle panel)
as the amount of matter inside the jet changes with $R$. 
The doubledot-dashed line gives the velocity variation of the matter within the 
jet for all the above cases. The arrows show the direction of the bulk velocity 
(radial direction in accretion, vertical direction in jets). The last panel 
gives the temperature (in keV) of the electron cloud in the CENBOL, 
jet, sub-Keplerian and Keplerian disk. Big dash-dotted line gives 
the temperature profile inside the Keplerian disk.

\begin{center}
\begin {tabular}[h]{cccccccccc}
\multicolumn{10}{c}{Table 5.1: Summary of the simulation parameters where outflow is present (GG10).}\\
\hline Case & R, $P_m$ & $N_{int}$ & $N_{cs}$ & $N_{cenbol}$ & $N_{jet}$ & $N_{subkep}$ & $N_{cap}$ & $p$ & $\alpha$\\
\hline
1a & 2, 58  & 2.7E8 & 4.0E8 & 1.4E7 & 7.5E7  & 8.4E8 & 3.4E5  & 63 & 0.43 \\
1b & 4, 97  & 2.7E8 & 4.1E8 & 2.4E6 & 1.3E8  & 8.6E8 & 3.3E5  & 65 & 1.05 \\
1c & 6, 37  & 2.7E8 & 4.0E8 & 5.4E7 & 4.8E7  & 8.3E8 & 3.1E5  & 62 & -0.4 \\
\hline
\end{tabular}
\end{center}

In Table 5.1, we summarize the parameters of all the Cases. We present the corresponding results
in Fig. \ref{fig.5.3}(a-c). In Col. 1, various Cases are 
marked. In Col. 2, the compression ratio ($R$) and percentage $P_m$ 
of the total matter that is going out as outflow (Fig. \ref{compression}) are 
listed. In Col. 3, we show the total number of photons (out of the 
total injection of $6.4 \times 10^8$) intercepted by the CENBOL 
and jet ($N_{int}$) combined. Column 4 gives the number of photons 
($N_{cs}$) that have suffered scattering inside the flow. 
Columns 5, 6 and 7 show the number of scatterings which took place 
in the CENBOL ($N_{cenbol}$), in the jet ($N_{jet}$) and in the 
pre-shock sub-Keplerian halo ($N_{subkep}$) respectively. A 
comparison of them will give the relative importance of these 
three sub-components of the sub-Keplerian disk. The number of 
photons captured ($N_{cap}$) by the black hole is given in Col. 8. 
In Col. 9, we give the percentage $p$ of the total injected photons 
that have suffered scattering through CENBOL and the jet. In 
Col. 10, we present the energy spectral index 
$\alpha$ ($I(E) \sim E^{-\alpha}$) obtained from our simulations.

\begin{figure}[]
\begin{center}
\vskip 5.6cm
\includegraphics{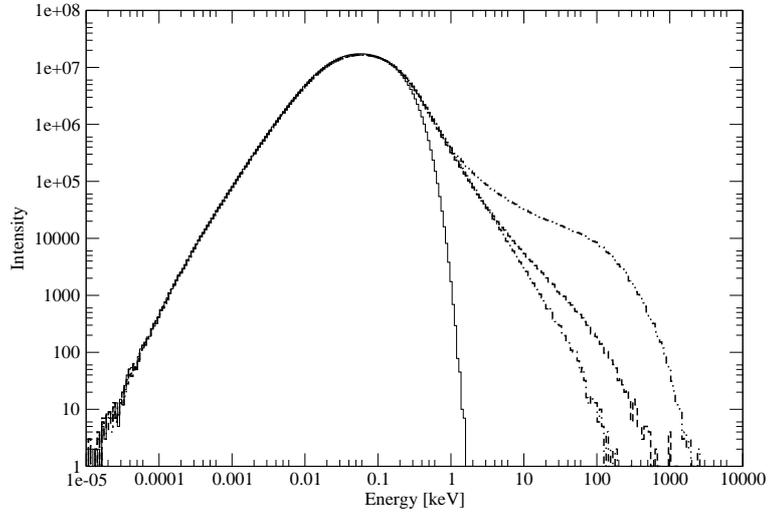}
\vskip 1.5cm
\caption{Variation of the emerging spectrum for different compression ratios.
The solid curve is the injected spectrum from the Keplerian disk. The dashed, dash-dotted 
and doubledot-dashed lines are for $R=2$ (Case 1a), $R=4$ (Case 1b) and $R=6$ (Case 1c) 
respectively. The disk and halo accretion rates used for these cases are $\dot{m_d} 
= 0.01$ and $\dot{m_h} = 1$. See, text for details (GG10).}
\label{fig.5.5}
\end{center}
\end{figure}

In Fig. \ref{fig.5.5}, we show the variation of the spectrum in 
the three simulations presented in Figs. \ref{fig.5.3}(a-c). The 
dashed, dash-dotted and doubledot-dashed lines are 
for $R=2$ (Case 1a), $R=4$ (Case 1b) and $R=6$ (Case 1c) 
respectively. The solid curve gives the spectrum of 
the injected photons. Since the density, velocity and 
temperature profiles of the pre-shock, sub-Keplerian region and the
Keplerian flow are the same in all these cases, we find that
the difference in the spectrum is mainly due to the
CENBOL and the jet. In the case of the strongest shock
(compression ratio $R=6$), only $37\%$ of the total injected
matter goes out as the jet. At the same time, due
to the shock, the density of the post-shock region increases
by a factor of $6$. Out of the three cases, the effective density of the
matter inside the CENBOL is the highest and that inside the jet is
the lowest in this case. Again, due to the shock, the temperature
increases inside the CENBOL and hence the spectrum is the hardest.
Similar effects are seen for moderate shock ($R=4$) and 
to a lesser extent, the low strength shock ($R=2$) also. 
When $R=4$, the density of the post-shock region
increases by the factor of $4$ while almost $97\%$ of total injected
matter (Fig. \ref{compression}) goes out as the jet reducing the matter density
of the CENBOL significantly. From Table 5.1 we find that the 
$N_{cenbol}$ is the lowest and $N_{jet}$ is the highest 
in this case (Case 1b). This decreases the up-scattering 
and increases the down-scattering of the photons. This 
explains why the spectrum is the softest in this case. 
In the case of low strength shock ($R=2$), $57\%$ of the 
inflowing matter goes out as jet, but due to the shock, the 
density increases by a factor of $2$ in the post-shock region. 
This makes this case similar to a non-shock case as far 
as the density is concerned, but with a little higher
temperature of the CENBOL due to the shock. So the spectrum with
the shock would be harder than when the shock is not present.
The disk and the halo accretion rates used for these cases
are $\dot{m_d} = 0.01$ and $\dot{m_h} = 1$.

\begin{figure}[h]
\begin{center}
\vskip 7.0cm
\includegraphics{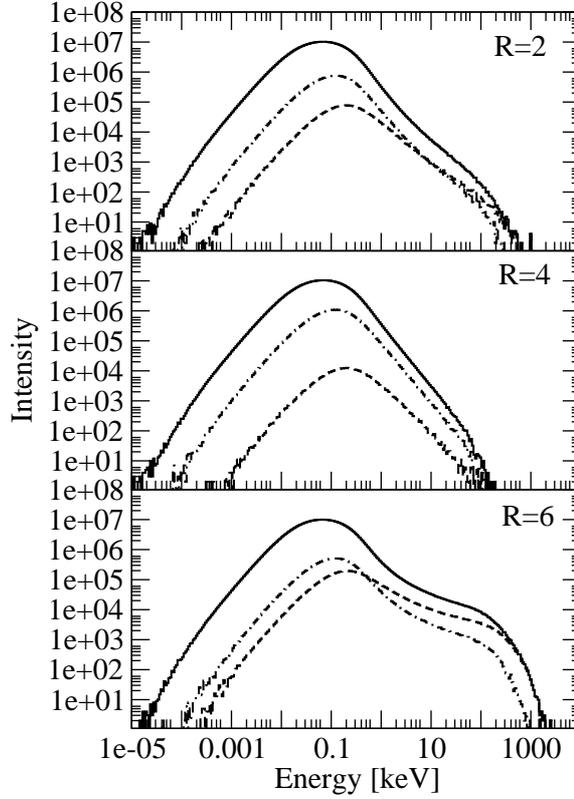}
\vskip 4.0cm
\caption{{\bf (a-c)}: Variation of the components of the emerging spectrum with 
the shock strength (R). The dashed curves correspond to the photons emerging
from the CENBOL region and the dash-dotted curves are for the photons 
coming out of the jet region. The solid curve is the spectrum for all 
the photons that have suffered scatterings. See, the text for details (GG10).}
\label{fig.5.6}
\end{center}
\end{figure}

In Figs. \ref{fig.5.6}(a-c), we show the components of the emerging spectrum for all the three cases 
presented in Fig. \ref{fig.5.5}. The solid curve is the intensity of all the photons
which suffered at least one scattering. The dashed curve corresponds to the photons 
those suffered their last scattering 
from the CENBOL region and the dash-dotted curve is for the photons
those suffered last scattering in the jet region. We find that the spectrum from
the jet region is softer than the spectrum from the CENBOL. As $N_{jet}$
increases and $N_{cenbol}$ decreases, the spectrum from the jet becomes
softer because of two reasons. First, the temperature of the jet is lesser
than that of the CENBOL, so the photons get lesser amount of energy from thermal
Comptonization making the spectrum softer. Second, the photons are
down-scattered by the outflowing jet which eventually make the spectrum
softer. We note that a larger number of photons are present in the spectrum 
from the jet than the spectrum from the CENBOL, which shows the photons
have actually been down-scattered. The effect of down-scattering is larger when $R=4$. 
For $R=2$ also there is significant amount of down scattered photons. 
But this number is very small for the case $R=6$ as $N_{cenbol}$ is 
much larger than $N_{jet}$ so most of the photons get up-scattered. 
The difference between total (solid) and the sum of the other two 
regions gives an idea of the contribution from the sub-Keplerian halo 
located in the pre-shock region. In our choice of geometry (half angles of the
disk and the jet), the contribution of the pre-shock flow is significant. In general
it could be much less. This is especially true when the CENBOL is further out.

\begin{figure}[]
\begin{center}
\vskip 4.7cm
\includegraphics{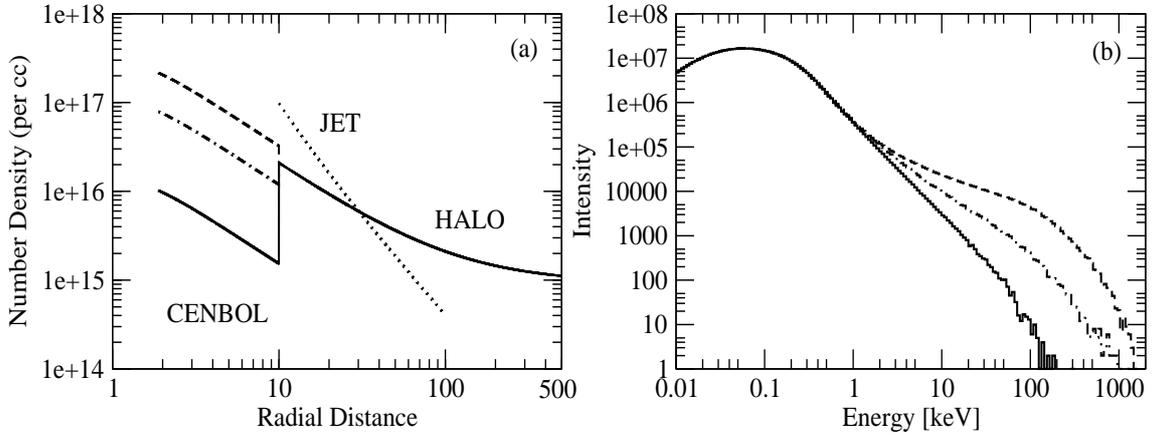}
\vskip 2.0cm
\caption{{\bf (a-b)}: (b) The variation of the output spectrum 
if we vary the Jet angle is shown. Here, the dashed, dash-dotted and solid plots 
are for $\Phi = 30^\circ$ and $50^\circ$ and  $58^\circ$, 
respectively. Compression ratio R is
$4$ in all three cases. (a) The density variation with the change in $\Phi$. 
Parameters and the line styles are same as the right panel. The dotted plot shows 
the density variation inside the outflow. See, the text for details. }
\label{jetkonvary}
\end{center}
\end{figure}

We now turn our attention to the effect of the variation of jet-angle on the output spectrum. 
In Fig. \ref{jetkonvary}b, we have plotted the spectra for three different jet angles 
[$\Phi = 30^\circ$ (dashed curve), $50^\circ$ (dash-dotted curve) and $58^\circ$ (solid curve)], keeping 
$\Psi = 32^\circ$ and $R=4$, fixed. Parameters used for the simulation: $\epsilon=0.015$. 
$\dot{m_d} = 0.01$ and $\dot{m_h} = 1$. The temperature and velocity distribution of the remain 
same for all these three cases, only difference the change in $\Phi$ makes is in the 
density of the matter inside the CENBOL. In Fig. \ref{jetkonvary}a we plot the density variation 
with radial distance. The line styles for $\Phi = 30^\circ$, $50^\circ$ and $58^\circ$ are same as 
Fig. \ref{jetkonvary}b. Here, dotted plot gives the density inside the jet for all the three 
cases. With the decrease in jet angle the density of the infalling matter inside the CENBOL increases, 
thus increasing the number of upscattered photons which makes the spectrum harder.

\subsection{When the Outflow is Absent}

\begin{figure}[]
\begin{center}
\vskip 4.5cm
\includegraphics{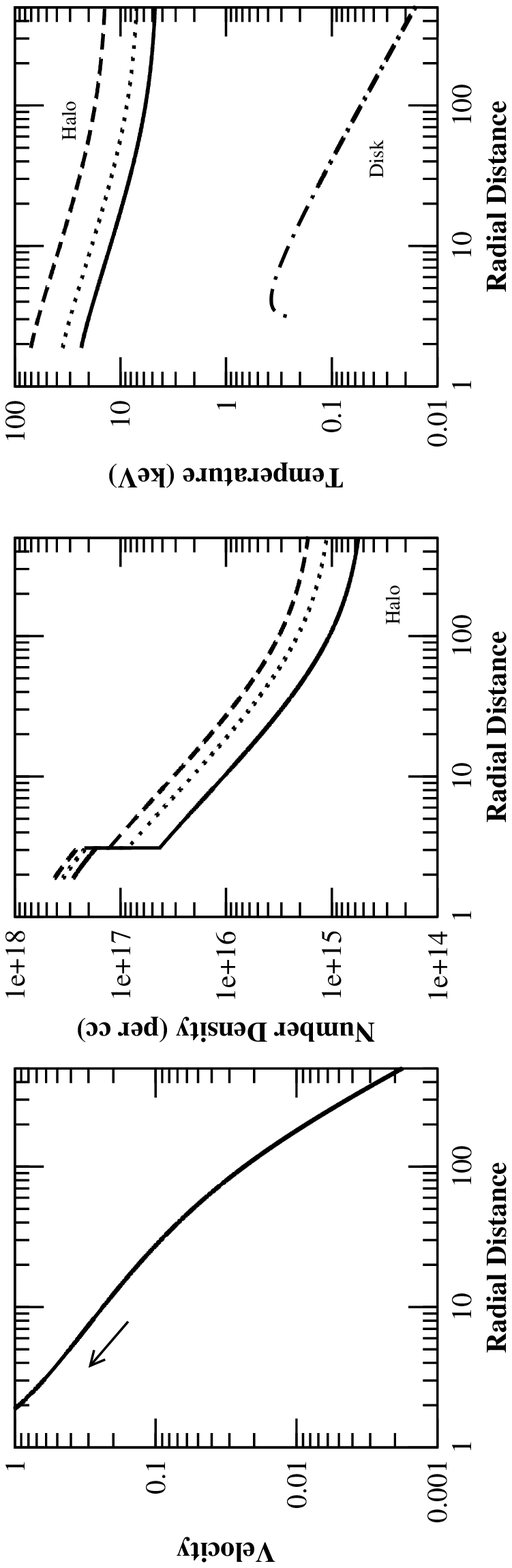}
\vskip 2.0cm
\caption{{\bf (a-c)}: Velocity (left), density (middle) and the temperature 
(right) profiles of Cases 2(a-c) as described in Table 5.2  
are shown with solid ($\dot{m_h} = 0.5$), dotted ($1$) and dashed ($1.5$)
curves. $\dot{m_d} = 1.5$ was used throughout. Velocities are the 
same for all the disk accretion rates (GG10).}
\label{fig.5.4}
\end{center}
\end{figure}

Let us now consider the cases where the jet is absent ($R=1$). 
In Figs. \ref{fig.5.4}(a-c), we show the velocity (left), number density of electrons 
(middle) and temperature (right) profiles of Cases 2(a-c) as described 
in Table 5.2. Here we have fixed $\dot{m_d}=1.5$ and $\dot{m_h}$ is 
varied: ${\dot {m_h}}=\ 0.5$ (solid), $1$ (dotted) and $1.5$ (dashed). 
To study the effects of bulk
motion Comptonization, the temperature of the electron cloud has 
been kept low for these cases. The temperature profile in the 
different cases has been chosen according to the Fig. 3b of CT95. 
In the absence of any shock ($R=1$) the Keplerian disk extends up to the 
marginally stable orbit ($3 r_g$). The temperature profile of the 
Keplerian disk for the above cases has been marked as `Disk'.
\begin{center}
\begin {tabular}[h]{ccccccccc}
\multicolumn{9}{c}{Table 5.2: Summary of the simulation parameters where outflow is absent (GG10).}\\
\hline Case & $\dot{m_h}$, $\dot{m_d}$ & $N_{int}$ & $N_{cs}$ & $N_{ms}$ &  
$N_{subkep}$ & $N_{cap}$ & $p$ & $\alpha_1, \alpha_2$\\
\hline 
2a & 0.5, 1.5 & 1.1E6  & 2.1E8 & 7.4E5 & 3.1E8 & 1.7E5 & 33.3 & -0.09, 0.4\\
2b & 1.0, 1.5 & 1.2E6  & 3.4E8 & 1.0E6 & 6.9E8 & 2.0E5 & 52.7 & -0.13, 0.75\\
2c & 1.5, 1.5 & 1.3E6  & 4.2E8 & 1.3E6 & 1.1E9 & 2.3E5 & 64.9 & -0.13, 1.3\\
\hline
\end{tabular}
\end{center}

In Table 5.2, we summarize the results of simulations where we 
have varied $\dot{m_d}$, for a fixed value of $\dot{m_h}$. 
In all of these cases no jet comes out of the CENBOL 
(i.e., $R=1$). In the last column, we listed two spectral 
slopes $\alpha_1$ (from $10$ to $100$ keV) and $\alpha_2$ 
(due to the bulk motion Comptonization). Here, $N_{ms}$ 
represents the photons that have suffered scattering 
between $r_g=3$ and the horizon of the black hole.

\begin{figure}[]
\begin{center}
\vskip 4.9cm
\includegraphics{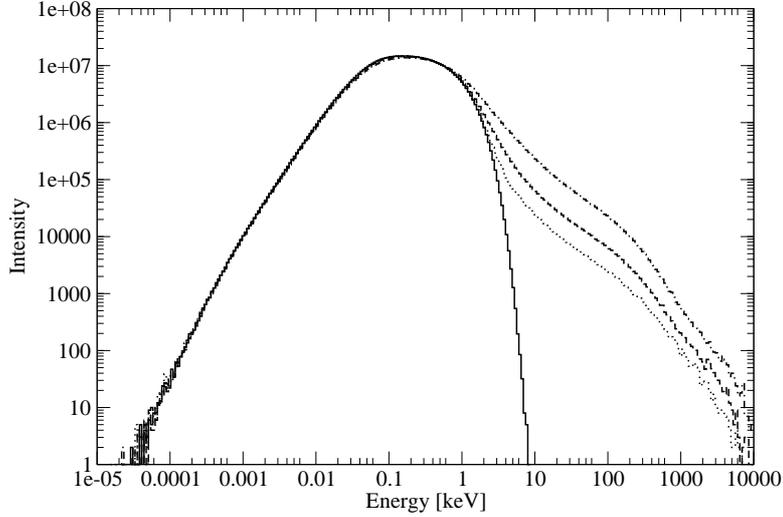}
\vskip 2.3cm
\caption{Bulk motion Comptonization spectrum.
Solid (Injected), dotted ($\dot{M_h} = 0.5$), dashed ($\dot{M_h} = 1$), 
dash-dotted ($\dot{M_h} = 1.5$). $\dot{M_d} = 1.5$ for all the cases.
Keplerian disk extends up to $3.1 r_g$. Table 5.2 summarizes the 
parameters used and the simulation results for these cases.}
\label{fig.5.7}
\end{center}
\end{figure}

In Fig. \ref{fig.5.7}, the emerging spectrum due to the bulk motion 
Comptonization when the halo rate is varied is shown. The solid curve is the injected
spectrum (modified black body). The dotted, dashed, and dash-dotted curves 
are for $\dot{m_h} = 0.5, \ 1$ and $1.5$ respectively. $\dot{m_d} = 1.5$ 
for all the cases. Table 5.2 
summarizes the parameter used and the results of the simulation.
As the halo rate increases, the density of the CENBOL also
increases causing a larger number of scattering. 
From Fig. \ref{fig.5.4}a, we noticed that the bulk velocity variation of 
the electron cloud is the same for all the four cases. Hence, 
the case where the density is maximum, the photons got 
energized to a very high value due to repeated scatterings 
with that high velocity cold matter. As a result, there is 
a hump in the spectrum around 100 keV energy for all the cases.
We find the signature of two power-law regions in the higher energy
part of the spectrum. The spectral indices are  given in Table 5.2. It
is to be  noted that $\alpha_2$ increases with $\dot{m_h}$ and 
becomes softer for high  $\dot{m_h}$. Our geometry here at the 
inner edge is conical which is more realistic, unlike a sphere 
(perhaps nonphysically so) in Laurent \& Titarchuk (2001). 
This may be the reason why our slope is not the same as in 
Laurent \& Titarchuk (2001) where $\alpha_2=1.9$. 

\begin{figure}[]
\begin{center}
\vskip 4.9cm
\includegraphics{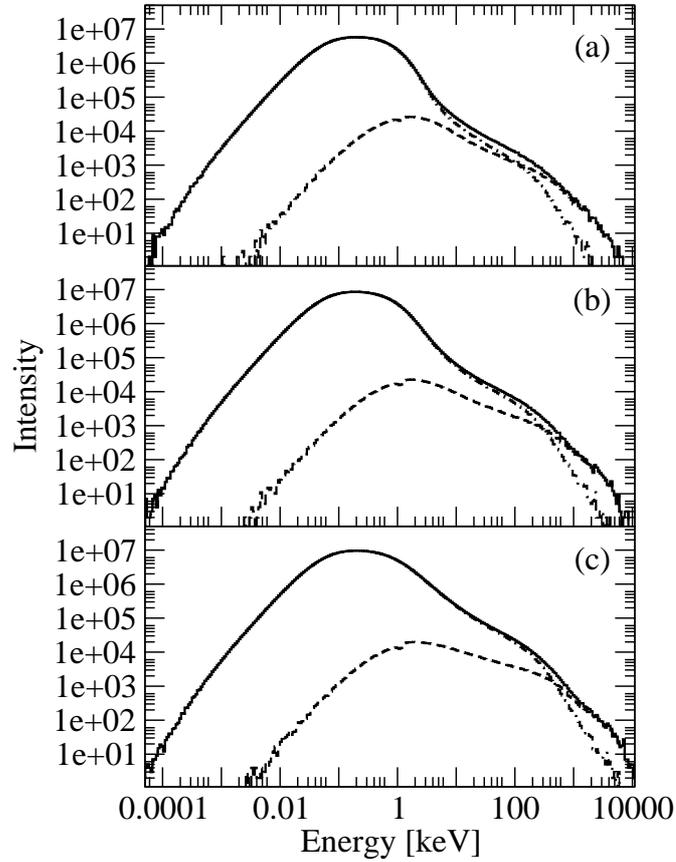}
\vskip 4.0cm
\caption{Components of the emerging spectrum for the Cases 2(a-c).
Solid curves are the spectra of all the photons that have suffered
scattering. The dashed and dash-dotted curves are the spectra
of photons which are emitted from inside and outside of the 
marginally stable orbit ($3 r_g$) respectively. The photons from 
inside the marginally stable radius are Comptonized by the 
bulk motion of the infalling matter. Here the jet is absent.}
\label{fig.5.8}
\end{center}
\end{figure}

In Figs. \ref{fig.5.8}(a-c), we present the components of the emerging spectra.
As in Fig. \ref{fig.5.6}, solid curves are the spectra of all the photons that have suffered
scattering. The dashed and dash-dotted curves are the spectra of 
photons emitted from inside and outside of the 
marginally stable orbit ($3 r_g$) respectively. The photons from 
inside the marginally stable radius are Comptonized by the 
bulk motion of the converging infalling matter and produces the power-law tail 
whose spectral index is given by $\alpha_2$ (Table 5.2).

In the next Chapter we will present the results of a coupled 
Monte Carlo and hydrodynamic simulation. The hydrodynamic 
code self consistently simulates the temperature, density and velocity of the 
matter inside the accretion disk at each time step and the Monte Carlo code 
simulates the spectrum of the photons that are coming out of 
the electron cloud at that particular time. Here, no adhoc 
geometry of the Compton cloud has been assumed. The geometry 
of the flow is also dictated by the hydrodynamic simulation.

\newpage

%% file: CHAPTER6/chap6.tex
\newpage
\markboth{\it Effects of Compton Cooling}
{\it Effects of Compton Cooling}
\chapter{Effects of Compton Cooling on Hydrodynamic and Spectral Properties}



\section{Introduction}

In Chapters 4 and 5 we have considered steady state geometries of the accretion disk 
and simulated the spectrum for them. In realty one should do a time dependent simulation, where the 
hydrodynamics of the flow changes due to the radiative transfer in the accretion disk and 
the time dependence of the output spectra can be observed. Given that the two component flows 
have been found to be useful to understand the spectral and timing properties (Chapter 1), 
it will be important to carry out the numerical simulations of radiative flows around black 
holes which also include shocks. So far, however, only bremsstrahlung  or pseudo-Compton cooling have been added
into the time-dependent flow (MSC96; Chakrabarti et al. 2004). 
In the present Chapter, we present the time dependent simulation results which includes both 
hydrodynamics and radiative transfer (Ghosh et al. 2011). We use the low angular momentum halo 
along with a Keplerian disk. We find how the Comptonization affects the 
temperature distribution of the flow and how this in turn affects the dynamics of the flow as well. 
So far, our solutions have been steady. We obtain the outgoing spectrum of radiation as well. 

In the next Section, we discuss the geometry of the soft photon source and the Compton cloud in our Monte-Carlo simulations. 
The variation of the thermodynamic quantities and other vital parameters inside the Keplerian disk and the Compton cloud which 
are required for the Monte-Carlo simulations are given in \S 6.2. In \S 6.3, we describe the simulation procedure and in \S 6.4, we present the
results of our simulations.

\section{System Description}

\begin{figure}[]
\begin{center}
\vskip 7.5cm
\includegraphics{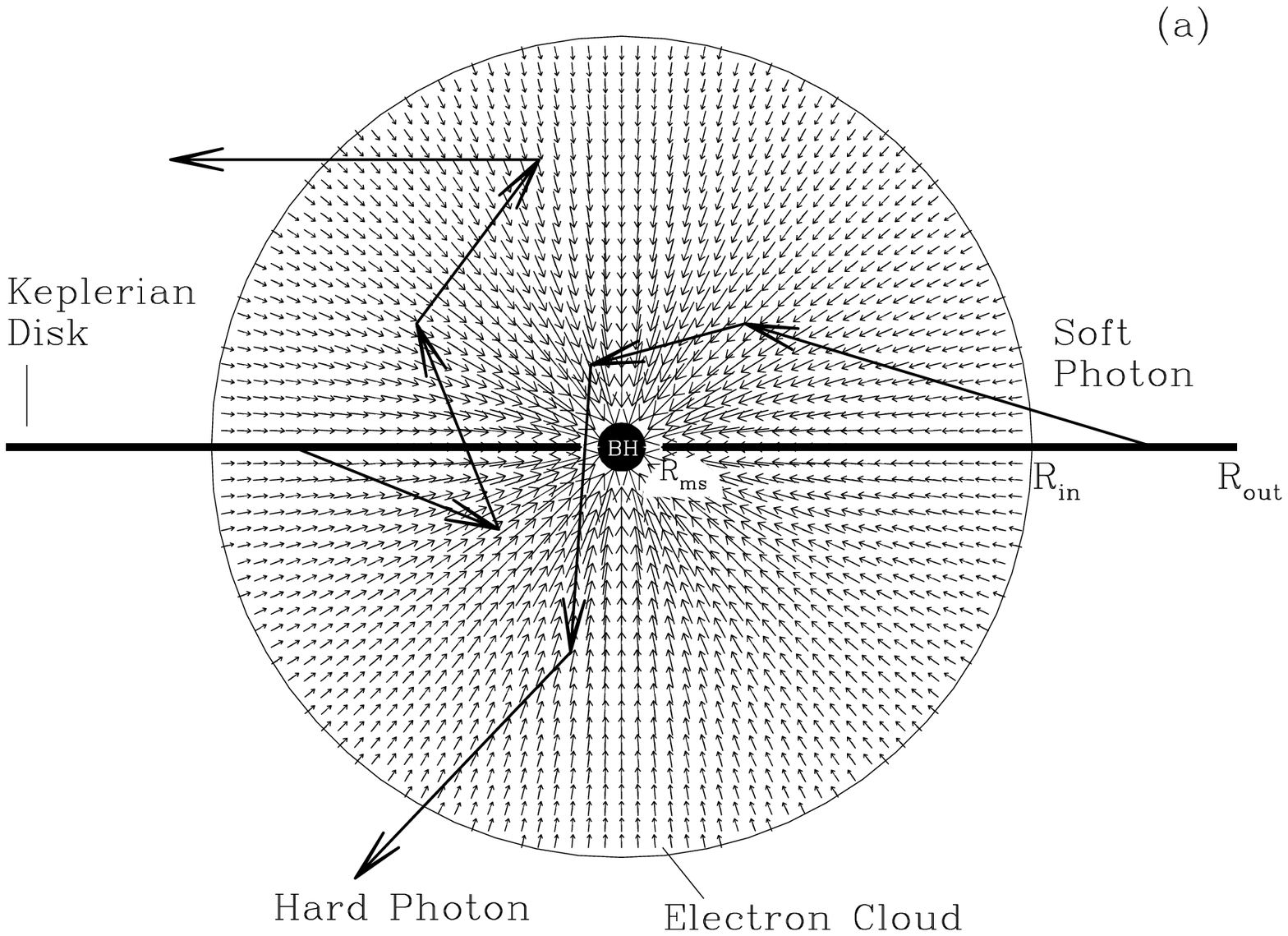}
\includegraphics{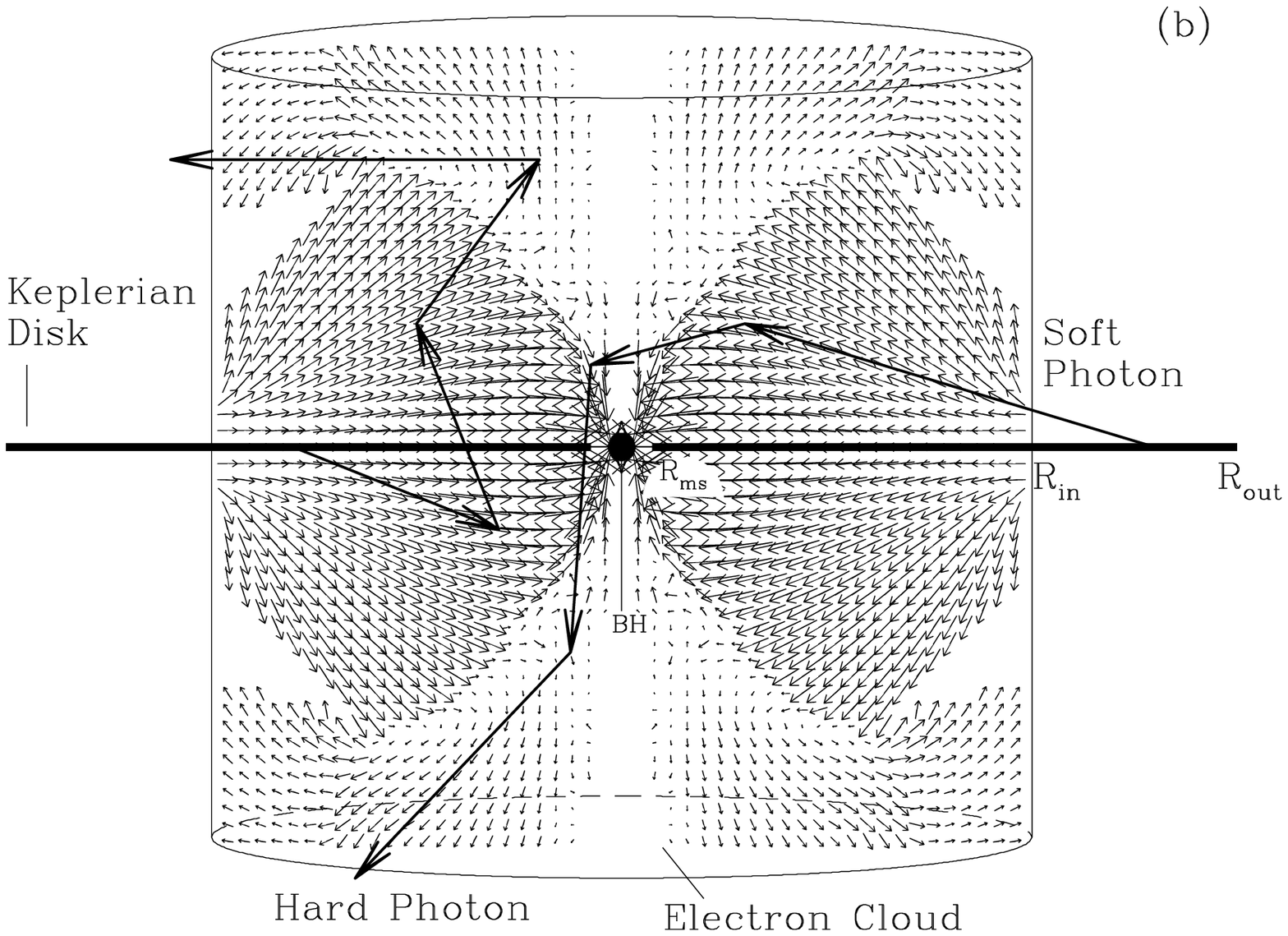}
\vskip 8.2cm
\caption{Schematic diagram of the geometry of our Monte Carlo simulations for (a) $\lambda=0$ 
and for (b) $\lambda=1$. Zigzag trajectories and velocity vectors are typical paths followed by 
the photons and the velocity vectors of the infalling matter inside the cloud (Ghosh et al. 2011).}
\label{fig1ab}
\end{center}
\end{figure}

We present cartoon diagrams of our simulation set up for 
(a) spherical Compton cloud (halo) with zero angular momentum (specific angular momentum $\lambda=0$) 
and (b) rotating Compton cloud (halo) with a specific angular momentum $\lambda=1$ in Figs. \ref{fig1ab}(a-b). 
In the first case (a), we have the electron cloud within a 
sphere of radius $R_{in} = 200 r_g$, the Keplerian disk resides at the equatorial 
plane. The outer edge of this disk is assumed to be at $R_{out} = 300 r_g$ 
and it extends up to the marginally stable orbit $R_{ms} = 3 r_g$. 
At the centre of the sphere, a black hole of mass $10 M_{\odot}$ is located. The spherical matter 
is injected into the sphere from the radius $R_{in}$ from all directions. It intercepts the soft 
photons emerging out of the Keplerian disk and reprocesses them via Compton or inverse    
Compton scattering. 
In the second case (b), due to the 
presence of the angular momentum of the flow, the spherical symmetry of the flow is lost. 
Because of the centrifugal barrier, the matter slows down at some point on its way to 
the central black hole. 
The other parameters of the Keplerian disk and the halo are the same as in (a).

\subsection{Distribution of Temperature and Density inside the Compton Cloud}

\begin{figure}[]
\begin{center}
\vskip 6.0cm
\includegraphics{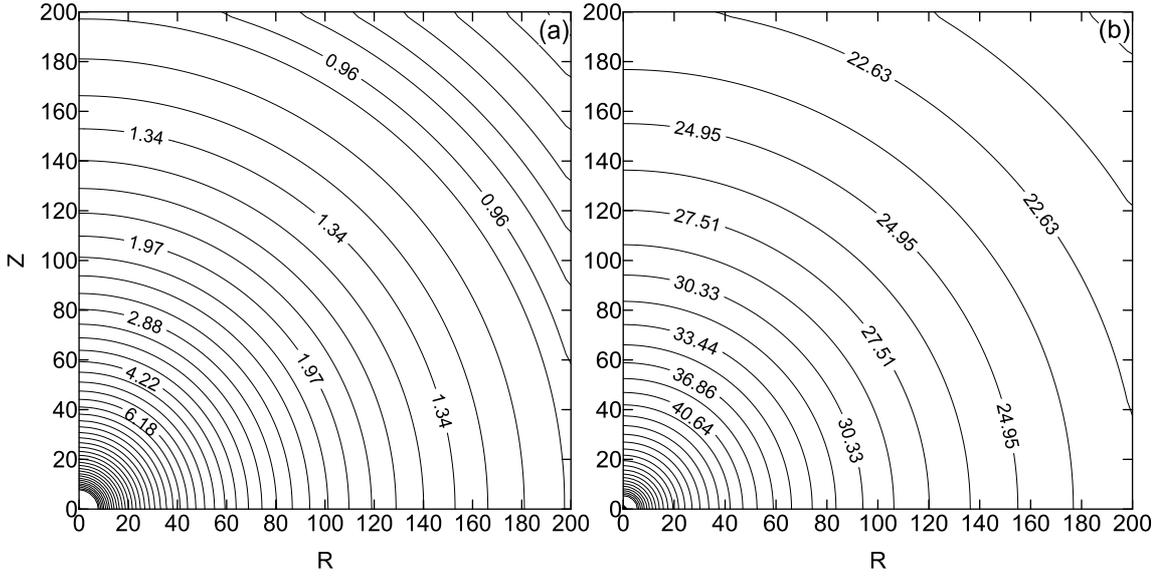}
\vskip 1.2cm
\caption{Density (a) and temperature (b) contours inside a spherical
halo in the absence of Compton cooling. Here, densities are in normalized 
unit and temperatures are in keV. $\lambda = 0$ is chosen. See text for details (Ghosh et al. 2011).}  
\label{fig2ab}
\end{center}
\end{figure}

A realistic accretion disk is expected to be three-dimensional. However, assuming axisymmetry, 
we can reduce one degree of freedom and make it a two dimentional problem. 
We have calculated the flow dynamics in two dimensions using a finite difference method which uses the principle of 
Total Variation Diminishing (TVD) to carry out hydrodynamic simulations (see, Ryu, Chakrabarti \& Molteni, 1997
and references therein; Giri et al. 2010). At each time step, we carry out Monte-Carlo simulation
to obtain the cooling/heating due to Comptonization. We incorporate the cooling/heating of 
each grid while executing the next time step of hydrodynamic simulation.
The numerical calculation for the two-dimensional flow has been carried out with 
$900 \times 900$ cells in a $200 r_g \times 200 r_g$ box. We chose the units in a way that
the outer boundary ($R_{in}$) is chosen to be unity and
the matter density is normalized to become unity. 
We assume the black hole to be non-rotating and we use the pseudo-Newtonian potential $-\frac{1}{2(r-1)}$ 
(PW80) to calculate the flow geometry around a black hole 
(Here, $r$ is in the unit of Schwarzschild radius $r_g=2GM/c^2$). Velocities and angular 
momenta are measured in units of $c$, the velocity of light and $r_g c$ respectively.
In Figs. \ref{fig2ab}(a-b) we show the snapshots of the density 
and temperature (in keV) profiles obtained in a steady state purely from our hydrodynamic
simulation.  The density contour levels
are drawn for 0.65-1.01 (levels increasing by a factor of 1.05) and 1.01-66.93 (successive level
ratio is 1.1). The temperature contour levels are drawn for 16.88-107.8
(successive level ratio is 1.05).

\subsection{Properties of the Keplerian Disk}

The soft photons are produced from a Keplerian disk whose inner edge has been kept 
fixed at the marginally stable orbit $R_{ms}$, while the outer edge is located at $R_{out}$ ($300 r_g$).
The source of soft photons have a multicolor blackbody spectrum coming from a
standard (SS73) disk.
Hydrodynamic properties of the Keplerian disk is the same as described in Sec. 5.2. 
In the Monte-Carlo simulation, we incorporated the directional effects of photons 
coming out of the Keplerian disk with the maximum number of photons emitted in the 
$z$-direction and minimum number of photons are generated along the plane of the disk. 
Thus, in the absence of photon bending effects, the disk is invisible as seen edge on. 
The position of each emerging photon is randomized using the distribution 
function $dN(r) =  4 \pi r \delta r H(r) n_\gamma(r)$, (Eqn. 4.5). We chose the 
mass of black hole $M_{bh} = 10$ in the rest of the Chapter.

\section{Simulation Procedure}

For a simulation, we assume a Keplerian disk rate ($\dot{m}_d$) and a 
sub-Keplerian halo rate ($\dot{m}_h$) to be given. The specific energy ($\epsilon$) of 
the halo provides the hydrodynamic (e.g., number density of the
electrons and the velocity variation) and the thermal properties of matter. Since we
chose the PW80 potential, the radial velocity is not exactly 
unity at $r=1$, the horizon, but it becomes unity just outside. In order not 
to over estimate the effects of bulk motion Comptonization (CT95) 
which is due to the momentum transfer of the moving electrons to the horizon, we kept the 
highest velocity to be 1. We use the absorbing boundary condition at $r=1.5$ ($\lambda=0$ case) 
and $r=2.5$ ($\lambda=1$ case). These simplifying assumptions do not 
affect our conclusions, especially because we are studying inviscid flow and the specific
angular momentum is constant. Photons are generated from the Keplerian disk as mentioned before 
and may be intercepted by the sub-Keplerian halo (sphere in Fig. \ref{fig1ab}a and cylinder in Fig. \ref{fig1ab}b).

The propagation of the photon inside the electron cloud, condition for a photon to 
scatter with an electron, choosing the photon and electron energy and the 
change in photon momentum after the scattering all these processes are implemented 
in the Monte Carlo code by the same procedure as described in Secs. 4.2 \& 5.2. 
 
We take a steady state flow profile from a hydrodynamics code to start the Monte Carlo simulation. 
When a photon interacts with an electron via Compton or inverse-Compton scattering, it 
loses or gains some energy ($\Delta E$). At each grid of the code we compute $\Delta E$.
We modify the energy from the flow by this amount and continue the hydrodynamic code with 
this modified energy. This, in turn, modify the hydrodynamic profile. Thus the Monte Carlo 
code for radiative transport and TVD code for hydrodynamics are coupled. 
In case the final state is steady, the temperature of the cloud would be reduced 
progressively to a steady value from the initial state where no cooling was assumed.

\subsection{Coupling of the Hydrodynamic and Radiative Transfer Codes}

Once a steady state is achieved in the non-radiative hydro-code, we compute the spectrum using the 
Monte Carlo code. This is the spectrum in first approximation. To include cooling in the coupled 
code, we follow these steps: (a) we calculate the velocity, density and 
temperature profiles of the electron cloud from the output of the hydro-code. (b) Using the Monte Carlo code we
calculate the spectrum. (c) Electrons are cooled (heated up) by the
inverse-Compton (Compton) scattering. We calculate the amount of the heat loss
(gain) by the electrons and its new temperature and energy distributions and (d) taking the new
temperature and energy profiles as initial condition, we run the hydro-code for a period of time.
Subsequently, we repeat the steps (a-d). In this way, we get an opportunity to 
see how the spectrum is modified as the iterations proceed. The iterations
stop when two successive steps produce virtually the same temperature profile and the emitted spectrum.

\subsubsection{Calculation of the Compton Cooling using Monte Carlo Code:}

For the Monte Carlo simulation, we divide the Keplerian disk in different annuli of width $D(r)=0.5$. 
Each annulus is characterized by its central temperature $T(r)$. The total number of photons emitted 
from the disk surface of each annulus can be calculated using Eqn. (4.5). 
This total number comes out to be $\sim~10^{39-40}$ for $\dot{m}_d = 1.0$.
In reality, one cannot inject this much number of photons in Monte Carlo simulation
because of the limitation of the computation time. So we replace this large number 
by a low number of bundles, say, $N_{comp}(r)~\sim~10^7$ and calculate a weightage 
factor 
$$
f_W = \frac{dN(r)}{N_{comp}(r)}.
$$ 
Clearly, from each annulus, the number of photons in a bundle will vary. 
This is computed exactly and used to compute the change of energy due 
to Comptonization. When this injected photon is inverse-Comptonized 
(or, Comptonized) by an electron in a volume element of size $dV$, 
we assume that $f_W$ number of photons has suffered similar
scatterings with the electrons inside the volume element $dV$. If the energy
loss (gain) per electron in this scattering is $\Delta E$, we multiply this
amount by $f_W$ and distribute this loss (gain) among all the electrons inside 
that particular volume element. This is continued for all the $10^7$ bundles of photons 
and the revised energy distribution is obtained.

\subsubsection{Computation of the Temperature Distribution after Cooling}

Since the hydrogen plasma considered here is ultra-relativistic ($\gamma=\frac{4}{3}$ throughout 
the hydrodynamic simulation), the thermal energy per particle is $3k_BT$ where $k_B$ is 
Boltzmann constant, $T$ is the temperature of the particle.  
The electrons are cooled by the inverse-Comptonization of the soft photons
emitted from the Keplerian disk. The protons are cooled because of the 
Coulomb coupling with the electrons. Total number of electrons inside 
any box with the centre at location $(ir,iz)$ is given by,
$$
\label{eq:a9} dN_e(ir,iz) = 4\pi rn_e(ir,iz)drdz, 
\eqno{(6.1)}
$$
where, $n_e(ir,iz)$  is the electron number density at $(ir,iz)$ location, and
$dr$ and $dz$ represent the grid size along $r$ and $z$ directions respectively. So, 
the total thermal energy in any box is given by $3k_BT(ir,iz)dN_e(ir,iz) = 12\pi
rk_BT(ir,iz)n_e(ir,iz)drdz,$ where $T(ir,iz)$ is the temperature at $(ir,iz)$
location. We calculate the total energy loss (gain) $\Delta E$ of electrons inside the
box according to what is presented above and subtract that amount to get the
new temperature of the electrons inside that box as 
$$
\label{eq:a10}k_BT_{new}(ir,iz) = k_BT_{old}(ir,iz)-\frac{\Delta E}{3dN_e(ir,iz)}. 
\eqno{(6.2)}
$$

\subsection{Details of the Hydrodynamic Simulation Code}

To model the initial injection of matter, we consider an axisymmetric flow of gas in the Psudo-Newtonian 
gravitational field of a black hole of mass $M_{bh}$ 
located at the centre in the cylindrical coordinates  $[R,\theta,z]$. 
We assume that at infinity, the gas pressure is negligible and the
energy per unit mass vanishes. As mentioned before, the gravitational field 
of the black hole can be described by PW80 potential.
We have assumed a polytropic equation of state for the
accreting (or, outflowing) matter, $P=K \rho^{\gamma}$, where,
$P$ and $\rho$ are the isotropic pressure and the matter density
respectively, $\gamma$ is the adiabatic index (assumed in this 
work to be constant throughout the flow, and is related to the
polytropic index $n$ by $\gamma = 1 + 1/n$) and $K$ is related
to the specific entropy of the flow $s$. The details of the code is 
described in Ryu, Molteni \& Chakrabarti (1997) and in Giri et al. (2010).
When we couple with cooling, $K$ itself will change. However, it remains
constant in between two successive Monte-Carlo coupling of inclusion of the cooling.

Our computational box occupies one quadrant of the R-Z plane with 
$0 \leq R \leq 200$ and $0 \leq z \leq 200$. The incoming gas 
enters the box through the outer boundary, located at $R_{in} = 200$. 
We have chosen the density of the incoming gas ${\rho}_{in} = 1$ 
for convenience since, in the absence of self-gravity and cooling, 
the density is scaled out, rendering the simulation results valid 
for any accretion rate. As we are considering only the energy flows 
while keeping the boundary of the numerical grid at a finite distance, 
we need the  sound speed $a$ (i.e., temperature) of the flow and 
the incoming velocity at the boundary points. For the spherical 
flow with zero angular momentum (Bondi flow), we have taken the 
boundary values from standard pseudo-Bondi solution. We injected 
the matter from both the outerboundary of R and z coordinate.
In order to mimic the horizon of the black hole at the 
Schwarzschild radius, we placed an absorbing inner boundary at 
$r = 1.5 r_g$, inside which all material is completely absorbed 
into the black hole. For the background matter (required to avoid 
division by zero) we used a stationary gas with density 
${\rho}_{bg} = 10^{-6}$ and sound speed (or temperature) the 
same as that of the incoming gas. Hence the incoming material 
has a pressure $10^6$ times larger than that of the background matter. All
the calculations were performed with $900 \times 900$ cells, 
so each grid has a size of $0.22$ in units of the Schwarzschild radius. 
For $\lambda = 1$ case, the matter was injected from the outerboundary of $R$ coordinate.

All the simulations are carried out assuming a stellar mass black hole. 
The procedures remain equally valid for massive/super-massive black holes. 
We carry out the simulations till several thousands of dynamical 
time-scales are passed. In reality, this corresponds to a few seconds in physical units.

\section{Results and Discussions}

In Table 6.1, we summarize all the cases for which the simulations have been presented 
in this Chapter. In Column 1, various cases are marked. Columns 2 gives the angular momentum 
($\lambda$) and the specific energy ($\epsilon$) of the flow. The Keplerian disk rate ($\dot{m}_d$)
and the sub-Keplerian halo rate ($\dot{m}_h$) are listed in Column 3. The number of soft photons, 
injected from the Keplerian disk ($N_{inj}$) for various disk rates can be found in Column 4. 
Columns 5 lists the number of photons ($N_{sc}$) that have suffered at least one scattering 
inside the electron cloud. The number of photons ($N_{unsc}$), escape from the cloud without 
any scattering are listed in Column 6. Columns 7 and 8 give the percentages of injected photons that 
have entered into the black hole ($N_{bh}$) and suffered scattering ($p = \frac{N_{sc}}{N_{inj}}$), 
respectively. The cooling time ($t_0$) of the system is defined as the expected time for the 
system to lose all its thermal energy with the particular flow parameters (namely, $\dot{m}_d$ and $\dot{m}_h$).
We calculate $t_0 = E/{\dot{E}}$ in each time step, where, $E$ is the total energy content of the system and 
${\dot{E}}$ is the energy gain or loss by the system in that particular time step. Column 9 lists the cooling 
time in sec for each case. We present the energy spectral index $\alpha$ $\left[I(E) \sim E^{-\alpha}\right]$ 
obtained from our simulations in the last column.

\subsection{Compton Cloud with Zero Angular Momentum ($\lambda=0$)}
\begin{figure}[]
\begin{center}
\vskip 2.0cm
\includegraphics{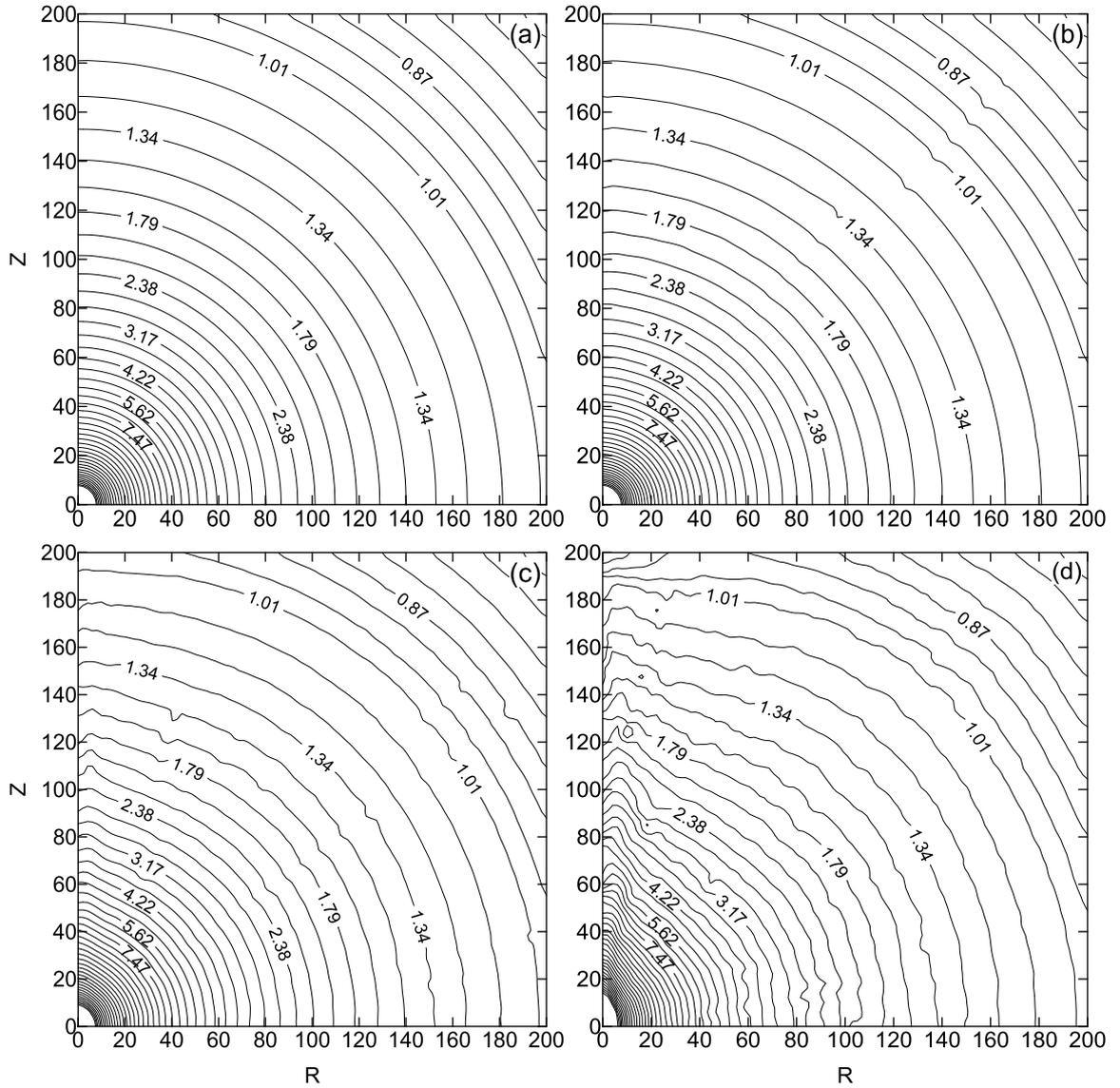}
\vskip 10.0cm
\caption{Changes in the density distribution in presence of cooling. $\lambda = 0$ and 
$\dot{m}_h = 1$ for all the cases. Disk accretion rate $\dot{m}_d$ used 
are (a) 1, (b) 2, (c) 5 and (d) 10 respectively (Cases 1(a-d) of Table 6.1). 
The density contours are drawn using the same contour levels as in Fig. \ref{fig2ab}a (Ghosh et al. 2011).}
\label{fig3abcd}
\end{center}
\end{figure}
\begin{figure}[]
\begin{center}
\vskip 2.0cm
\includegraphics{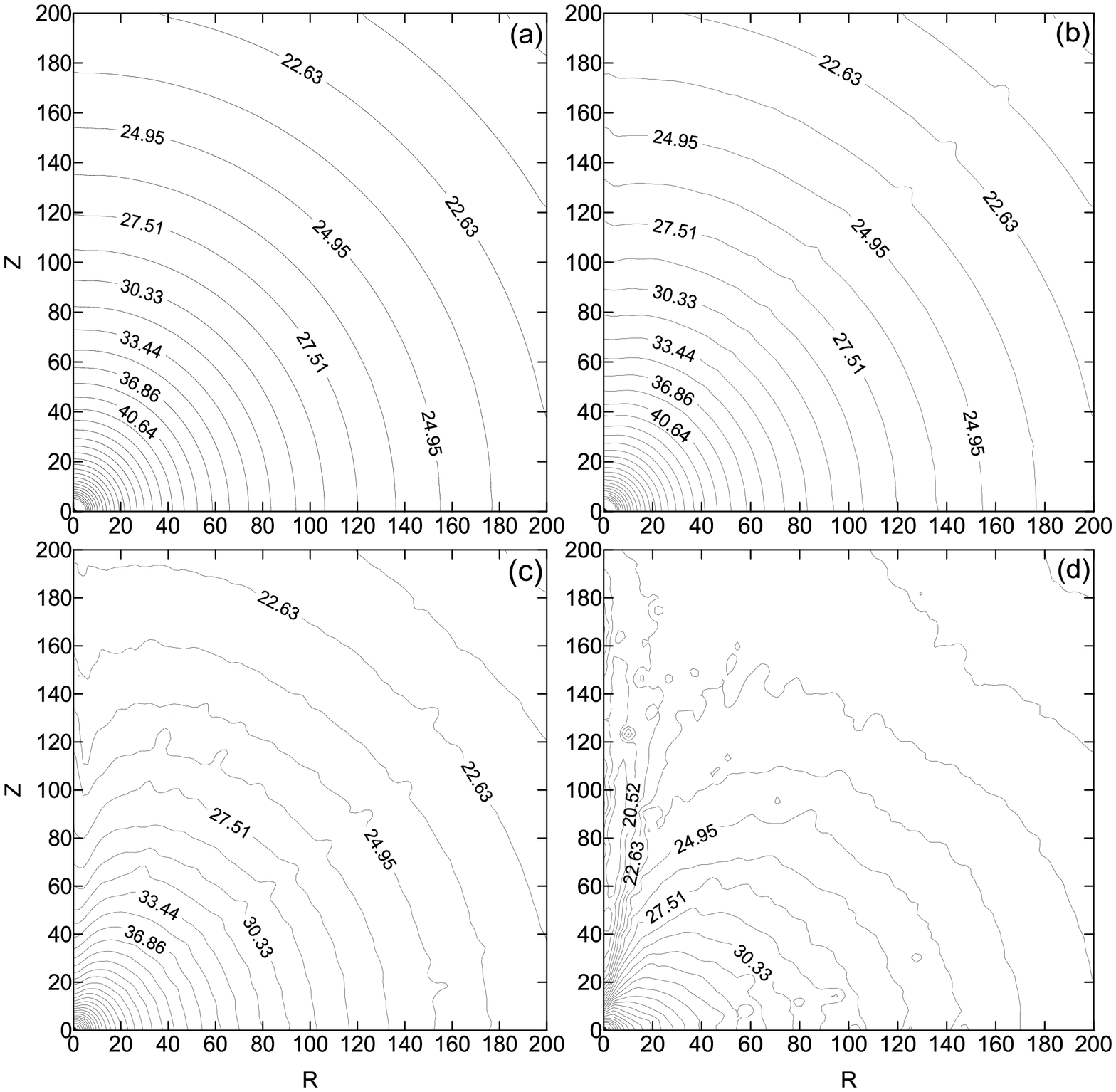}
\vskip 10.0cm
\caption{Change in the temperature distribution in presence of cooling. 
$\lambda = 0$ and $\dot{m}_h = 1$ for all the cases. Disk 
accretion rate $\dot{m}_d$ is (a)1, (b)2, (c)5 and (d)10 
respectively (Cases 1(a-d) of Table 6.1). Contours are drawn using the same levels as 
in Fig. \ref{fig2ab}b (Ghosh et al. 2011).}
\label{fig4abcd}
\end{center}
\end{figure}

First we discuss the results corresponding to the Cases 1(a-d) of Table 6.1. 
In Fig. \ref{fig3abcd}(a-d) we present the changes in density distribution 
as the disk accretion rates are changed: $\dot{m}_d =$ (a) 1, (b) 2, 
(c) 5 and (d) 10 respectively. We notice that as the accretion rate 
of the disk is enhanced, the density distribution losses its spherical symmetry. In particular, the 
density at a given radius is enhanced in a conical region along the axis. This is due to 
the cooling of the matter by Compton scattering. To show this, in Fig. \ref{fig4abcd}(a-d) 
we show the temperature contours of the same four cases. The contours are 
marked with temperatures. We notice that the temperature is reduced along 
the axis (where the optical depth as seen by the soft photons from the 
Keplerian disk is higher) drastically after repeated Compton scattering. 

{\small{
\begin{center}
\begin {tabular}[h]{cccccccccc}
\multicolumn{10}{c}{Table 6.1: Summary of all the simulations presented in \S 6.4.1 and \S 6.4.2 (Ghosh et al. 2011).}\\
\hline 
Case & $\lambda$, $\epsilon$ & $\dot{m}_d$, $\dot{m}_h$ & $N_{inj}$ & 
$N_{sc}$ & $N_{unsc}$ & $N_{bh}$ [$\%$] & $p$ [$\%$] & $t_0$ [sec] & $\alpha$\\
\hline 
1a & 0, 22E-4 & 1,  1   & 4.3E40 & 8.7E39 & 3.5E40 & 0.12 & 20.03 & 228.3 & 1.15, 0.99 \\
1b & 0, 22E-4 & 2,  1   & 1.5E41 & 2.9E40 & 1.2E41 & 0.12 & 20.02 & 63.6  & 1.30, 1.0  \\
1c & 0, 22E-4 & 5,  1   & 7.3E41 & 1.5E41 & 5.9E41 & 0.12 & 19.94 & 12.4  & 1.40, 0.96 \\
1d & 0, 22E-4 & 10, 1   & 2.5E42 & 5.0E41 & 2.0E42 & 0.12 & 19.82 & 4.2   & 1.65, 0.90 \\
1e & 0, 22E-4 & 1,  0.5 & 4.3E40 & 4.7E39 & 3.9E40 & 0.07 & 10.89 & 380.0 & 1.57       \\
1f & 0, 22E-4 & 1,  2   & 4.3E40 & 1.5E40 & 2.8E40 & 0.23 & 34.32 & 118.9 & 1.1        \\
1g & 0, 22E-4 & 1,  5   & 4.3E40 & 2.6E40 & 1.8E40 & 0.50 & 59.01 & 48.0  & 0.7        \\
1h & 0, 22E-4 & 1,  10  & 4.3E40 & 3.3E40 & 1.1E40 & 0.70 & 75.52 & 35.1  & 0.45       \\
\hline
2a & 1, 3E-4 & 1,  1    & 6.3E40 & 1.2E40 & 5.1E40 & 0.29 & 19.20 & 79.7 & 0.88 \\
2b & 1, 3E-4 & 2,  1    & 2.1E41 & 4.1E40 & 1.7E41 & 0.28 & 19.28 & 21.9 & 0.94 \\
2c & 1, 3E-4 & 5,  1    & 1.0E42 & 1.9E41 & 8.1E41 & 0.28 & 19.21 & 4.3  & 1.03 \\
2d & 1, 3E-4 & 10, 1    & 3.6E42 & 6.9E41 & 2.9E42 & 0.29 & 18.94 & 1.4  & 1.17 \\
2e & 1, 3E-4 & 10, 0.5  & 3.6E42 & 3.9E41 & 3.2E42 & 0.19 & 10.77 & 1.9  & 1.37 \\
2f & 1, 3E-4 & 10, 1.5  & 3.6E42 & 9.3E41 & 2.7E42 & 0.37 & 25.49 & 1.1  & 1.01 \\
2g & 1, 3E-4 & 10, 2    & 3.6E42 & 1.1E42 & 2.5E42 & 0.44 & 30.76 & 0.9  & 0.95 \\
2h & 1, 3E-4 & 10, 5    & 3.6E42 & 1.8E42 & 1.7E42 & 0.69 & 51.07 & 0.7  & 0.59 \\
\hline
\end{tabular}
\end{center}
}}
\vskip 0.1cm
\begin{figure}[h]
\begin{center}
\vskip 1.0cm
\includegraphics{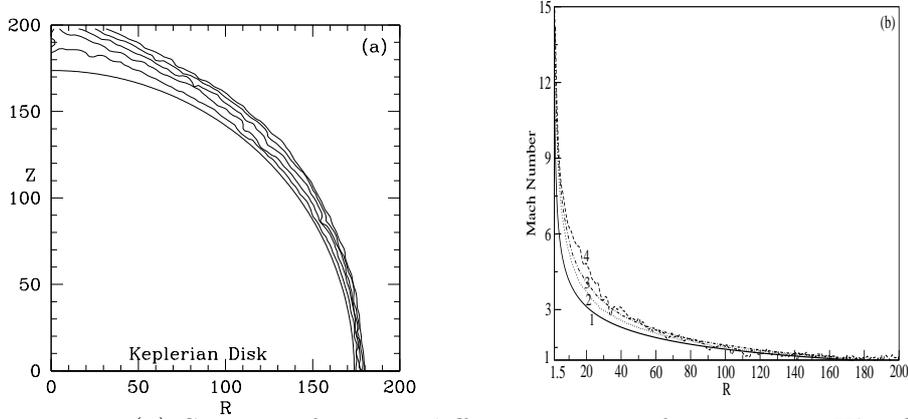}
\includegraphics{CHAPTER6/Fig5b.eps}
\vskip 4.8cm
\caption{(a) Sonic surfaces at different stages of iterations. The final curve
represents the converged solution. The initial spherical sonic surface become 
prolate spheroid due to the presence of the Keplerian disk at the equatorial plane. 
Parameters are for Case 1d (Table 6.1).
(b) Mach number variation as a function of distance after a complete solution of 
the radiative flow is obtained. Plot no. 1 corresponds to the solution from adiabatic Bondi flow. 
Plots 2-4 are the solutions along the equatorial plane, the diagonal and the axis of the disk. 
Parameters are for Case 1d (Table 6.1) (Ghosh et al. 2011).}
\label{fig5abcd}
\end{center}
\end{figure}
\begin{figure}[h]
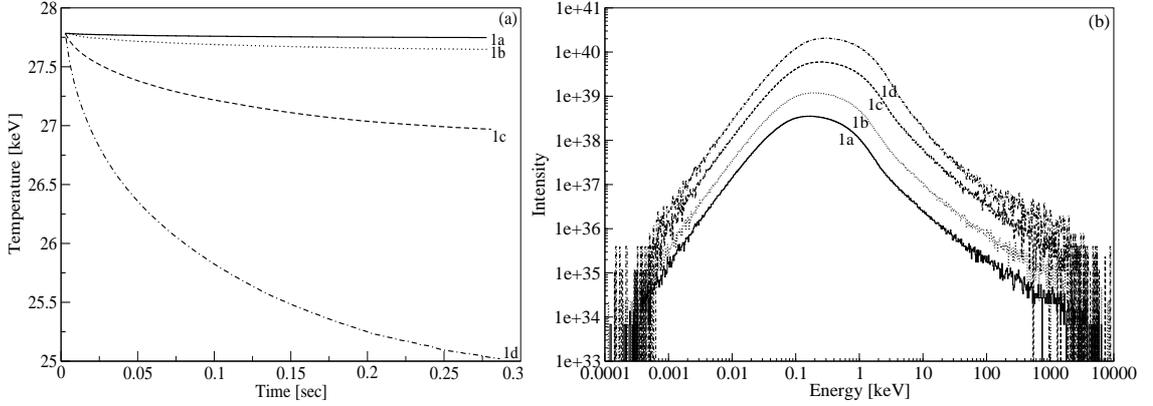

\begin{center}
\vskip -0.3cm
\includegraphics{CHAPTER6/Fig5c.eps}
\includegraphics{CHAPTER6/Fig5d.eps}
\vskip 5.8cm
\caption{(a) Variation of the average temperature of the Compton cloud as the iteration proceeds when 
the disk accretion rate is varied. $\dot{m}_h = 1$. The solid, dotted, dashed and dash-dotted plots are 
for $\dot{m}_d = 1$, $2$, $5$ and $10$ respectively. Case numbers (Table 6.1) are marked. 
With the increase of disk rate, the temperature of the Compton cloud converges to a lower temperature.
(b) Variation of the spectrum with the increase of disk accretion rate. Parameters are the same as in (c).
With the increase in $\dot{m}_d$, the intensity of the spectrum increases due to the increase 
in $N_{inj}$ (see Table 6.1). The spectrum is softer for the higher value of $\dot{m}_d$. 
Spectral slopes for each of these spectra are listed in Table 6.1 (Ghosh et al. 2011).}
\label{5a_b}
\end{center}
\end{figure}

In Fig. \ref{fig5abcd} and Fig. \ref{5a_b}, we show the hydrodynamic and radiative 
properties. In Fig. \ref{fig5abcd}a, we show the sonic surfaces. The lowermost curve corresponds 
to theoretical solution for an adiabatic flow (e.g., C90). Other 
curves from the bottom to top are the iterative solutions for the Case 1d mentioned above. 
As the disk rate is increased, the cooling increases the Mach number 
along the axis at a give distance. Of course, there are other effects: The cooling causes the 
density to go up to remain in pressure equilibrium. In Fig. \ref{fig5abcd}b, 
the Mach number variation is shown. The lower most curve (marked 1) from theoretical 
consideration. Plots 2-4 are the variation of Mach number with radial distance 
along the equatorial plane, along the diagonal and along the vertical axis 
respectively. In Fig. \ref{5a_b}a, the average temperature of the spherical halo is plotted with 
iteration time until almost steady state is reached. The cases are marked on the curves. We note that as the 
injection of soft photons is increased, the average temperature of the halo decreases drastically. 
In Fig. \ref{5a_b}b, we have plotted the energy dependence of the photon intensity.
We find, as we increase the disk rate, keeping the halo rate fixed, number 
of photons coming out of the cloud in a particular energy bin increases and the spectrum becomes softer. 
This is also clear from Table 6.1, $N_{inj}$ increases with $\dot{m}_d$, increasing $\alpha$. We find the 
signature of double slope in these cases. As the disk rate increases, the second slope becomes steeper. 
This second slope is the signature of bulk motion Comptonization. As $\dot{m}_d$ increases, the cloud becomes 
cooler (Fig. \ref{5a_b}a) and the power-law tail due to the bulk motion Comptonization 
(CT95) becomes prominent. 

\begin{figure}[h]
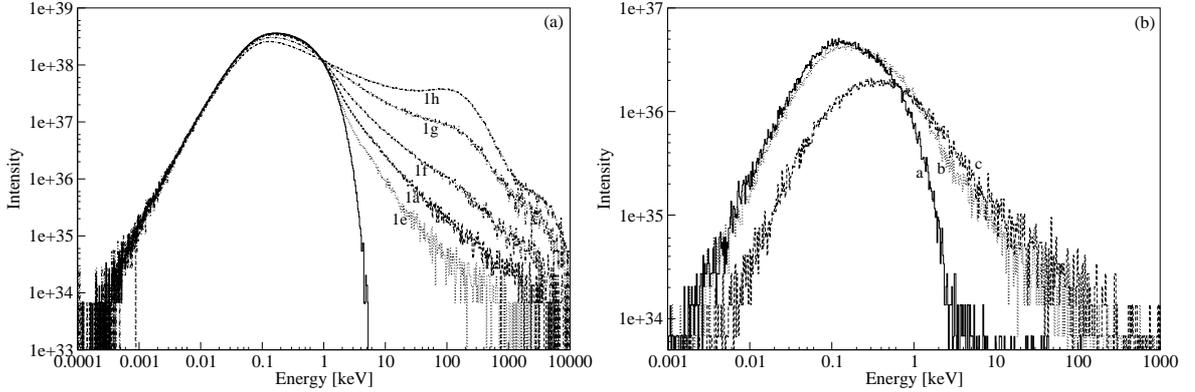

\begin{center}
\vskip 6.1cm
\includegraphics{CHAPTER6/Fig6a.eps}
\includegraphics{CHAPTER6/Fig6b.eps}
\vskip -0.9cm
\caption{(a) Variation of the spectrum with the increase of the halo accretion rate, keeping the 
disk rate ($\dot{m}_d = 1$) and angular momentum of the flow ($\lambda = 0$) fixed. 
The dotted, dashed, dash-dotted, double dot-dashed and double dash-dotted curves show the 
spectra for $\dot{m}_h = 0.5$, $1$, $2$, $5$ and $10$ respectively. 
The injected multicolor blackbody spectrum supplied by the Keplerian disk is shown (solid line). (b) 
Directional variation in the spectrum: $\lambda = 0$, $\dot{m}_h = 2$, $\dot{m}_d = 1$ is 
shown. The solid, dotted and dashed curves are for observing angles $2^{\circ}$, $45^{\circ}$ 
and $90^{\circ}$ respectively. All the angles are measured with respect to the rotation axis ($z$-axis) 
(Ghosh et al. 2011).}
\label{fig6ab}
\end{center}
\end{figure}

In Fig. \ref{fig6ab}a, we show the variation of the energy spectrum with the increase of the halo 
accretion rate, keeping the disk rate ($\dot{m}_d = 1$) and angular momentum of the flow ($\lambda = 0$) fixed. 
The dotted, dashed, dash-dotted, double dot-dashed and double dash-dotted curves show the 
spectra for $\dot{m}_h = 0.5$, $1$, $2$, $5$ and $10$ respectively. The spectrum becomes 
harder for higher values of $\dot{m}_h$ as it becomes difficult to cool. 
The injected multicolor blackbody spectrum supplied by the Keplerian disk is shown (solid line). 
The spectrum becomes harder for higher values of $\dot{m}_h=1$ as it is difficult to cool a higher density 
matter with the same number of injected soft photons. In Fig. \ref{fig6ab}b, we show the 
directional dependence of the spectrum. For $\lambda = 0$, $\dot{m}_h = 2$, 
$\dot{m}_d = 1$ (Case 1f Table 6.1). The solid, dotted and dashed curves are for 
observing angles (a) $2^{\circ}$,  (b) $45^{\circ}$ and (c) $90^{\circ}$ respectively. 
All the angles are measured with respect to the rotation axis ($z$-axis). 
As expected, the photons arriving along the Z-axis would be dominated by the soft photons from the 
Keplerian disk while the power-laws would dominate the spectrum coming edge-on.

\begin{figure}[h]
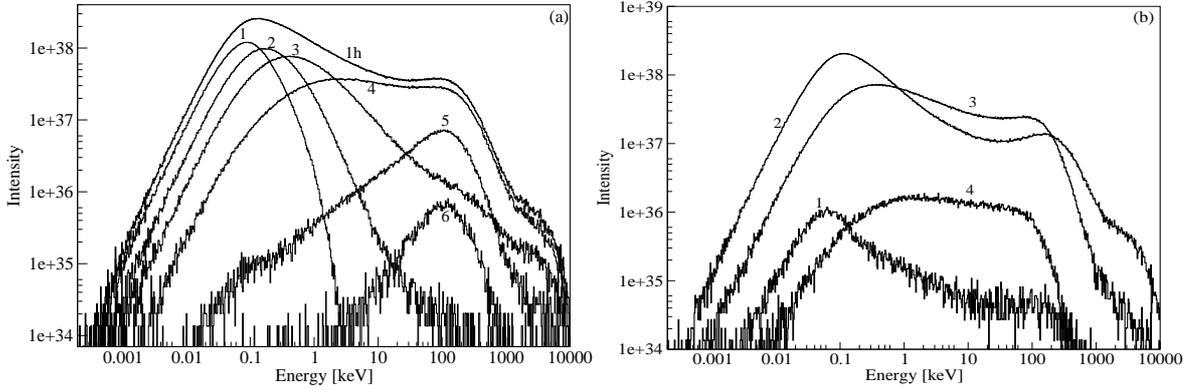

\begin{center}
\vskip 5.0cm
\includegraphics{CHAPTER6/Fig7a.eps}
\includegraphics{CHAPTER6/Fig7b.eps}
\vskip 0.15cm
\caption{Intensity spectra emerging from the cloud (a) after suffering various 
number of scatterings and (b) at four different times immediately after the injection 
of soft photons. Case 1h is assumed. In the left panel (a), the spectra of the photons suffering 0, 1-2, 3-6, 
7-18, 19-28 and more than 29 scatterings are shown by the plots 1, 2, 3, 4, 5 and 6 
respectively, within the cloud. Curve 1h is the net spectrum for which these components are drawn. As 
the number of scattering increases, the photons gain more and more energy from the hot electron cloud 
through inverse Comptonization process. In Panel b, the spectra of the photons spending 
0.01-1, 1-40, 40-100 and more than 100 ms time inside the electron cloud are marked by 1, 2, 3 and 4 
respectively (Ghosh et al. 2011).}
\label{fig7ab}
\end{center}
\end{figure}

We now study the dependence of the spectrum on the time delay between injected and outgoing photons. 
Depending on number of scatterings suffered and length of the path traveled, different 
photons spend different times inside the Compton cloud. The energy gain or loss by any 
photon depends on this time. Fig. \ref{fig7ab}a shows the spectrum of the photons suffering 
different number of scatterings inside the cloud. Here the numbers 1, 2, 3, 4, 5 and 6 show 
the spectrum for 6 different ranges of number of scatterings. Plot 1 shows the spectrum of the 
photons that have escaped from the cloud without suffering any scattering. This spectrum is 
nearly the same as the injected spectrum, only difference is that it is Doppler shifted. As the 
number of scattering increases (spectra 2, 3 and 4), the photons get more and more energies 
via inverse Compton scattering with the hot electron cloud. For 
scatterings more than 19, the high energy photons start loosing energy through Compton scattering 
with the relatively lower energy electrons. Components 5 and 6 show the spectra of the 
photons suffering 19-28 scatterings and the photons 
suffering more than 28 respectively. Here the flow parameters are: 
$\dot{m}_d = 1$, $\dot{m}_h = 10$ and $\lambda = 0$ (Case 1h Table 6.1).

In Fig. \ref{fig7ab}b, we plot the spectrum emerging out of the electron cloud at four 
different time ranges. In the simulation, that the photons take 
$0.01$ to $130$ ms to come out of the system. We divide this time range 
into 4 suitable bins and plot their spectrum. 
Case 1h of Table 6.1 is considered. We observe that the spectral 
slopes and intensities of the four spectra are different. As the photons spend more and 
more time inside the cloud, the spectrum gets harder (plots 1, 2 and 3). However, 
very high energy photons which spend maximum time inside the cloud lose some 
energy to the relatively cooler electrons before escaping from the cloud. 
Thus, the spectrum 4 is actually the spectrum of Comptonized photons.

\subsection{Compton Cloud  with Very Low Angular Momentum ($\lambda=1.0$)}

\begin{figure}[]
\begin{center}
\vskip 6.0cm
\includegraphics{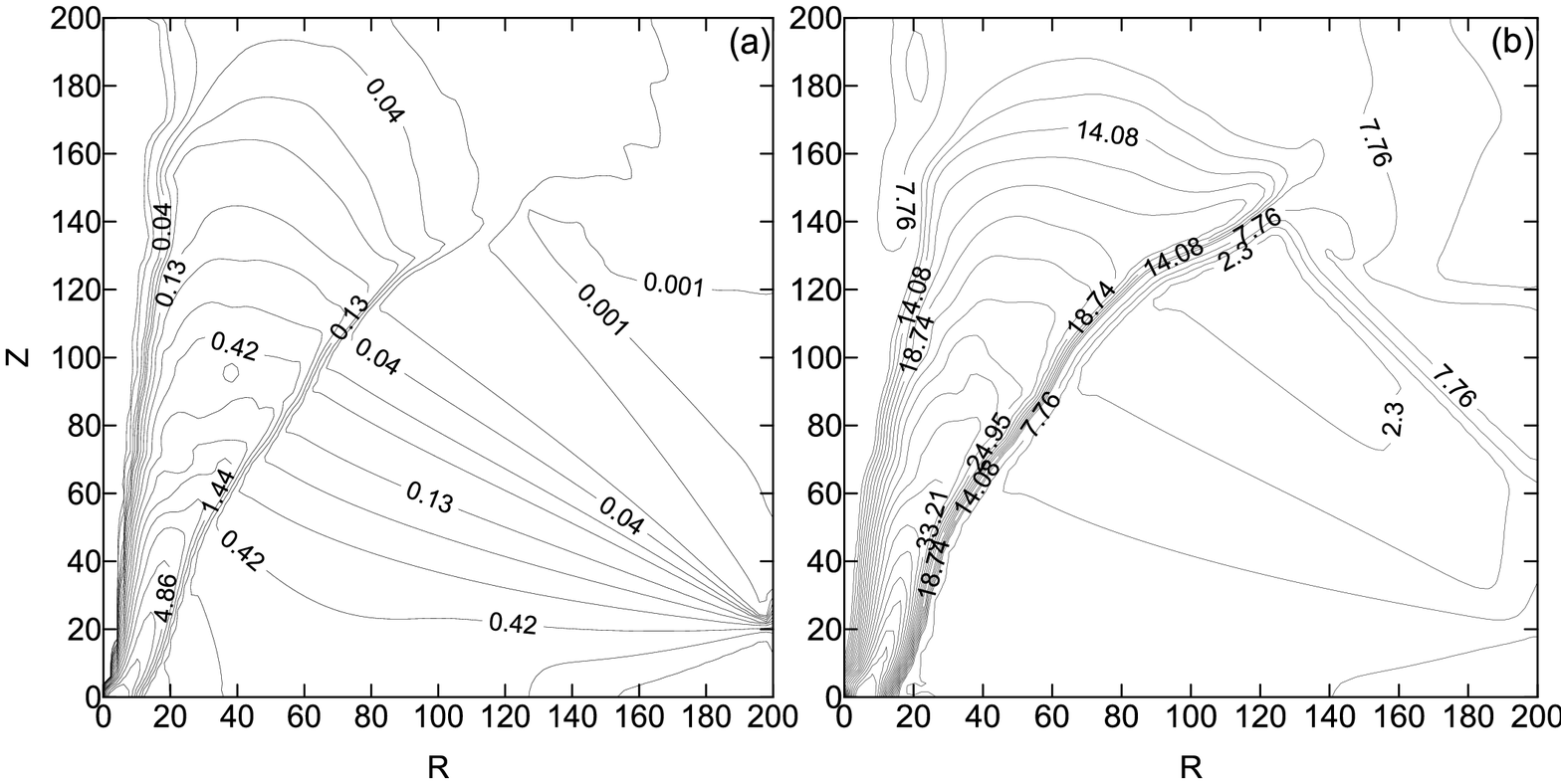}
\vskip 1.5cm
\caption{Density (a) and temperature (b) contours inside the halo ($\lambda=1.0$) in the absence 
of Compton cooling. Densities are in normalized unit and temperatures 
are in keV. See, text for details (Ghosh et al. 2011).}
\label{fig8ab}
\end{center}
\end{figure}

We now turn our attention to the case where the cloud is formed by a low angular momentum flow. In this 
case, the flow is already axisymmetric and due to the centrifugal force, a weak shock wave, or at least 
a pressure wave would be formed. In Fig. \ref{fig8ab}(a-b), we show the contours of constant density (Fig. \ref{fig8ab}a)
and temperature (Fig. \ref{fig8ab}b) when no radiative transfer is included. Here the specific angular momentum 
of $\lambda=1$ was chosen. Density contour levels are drawn from 0.001-55.35 (successive level ratio is 1.5), 
55.35-73.73 (successive level ratio is 1.1). Temperature contour levels are drawn from 
2.3-11.64 (successive level ratio is 1.5), 11.64-64.71 (successive level ratio is  1.1). 
We note that a shock has been formed which bends outwards 
away from the equatorial plane (Ryu, Chakrabarti \& Molteni, 1997; Giri et al., 2010.).
\begin{figure}[]
\begin{center}
\vskip 15.5cm
\includegraphics{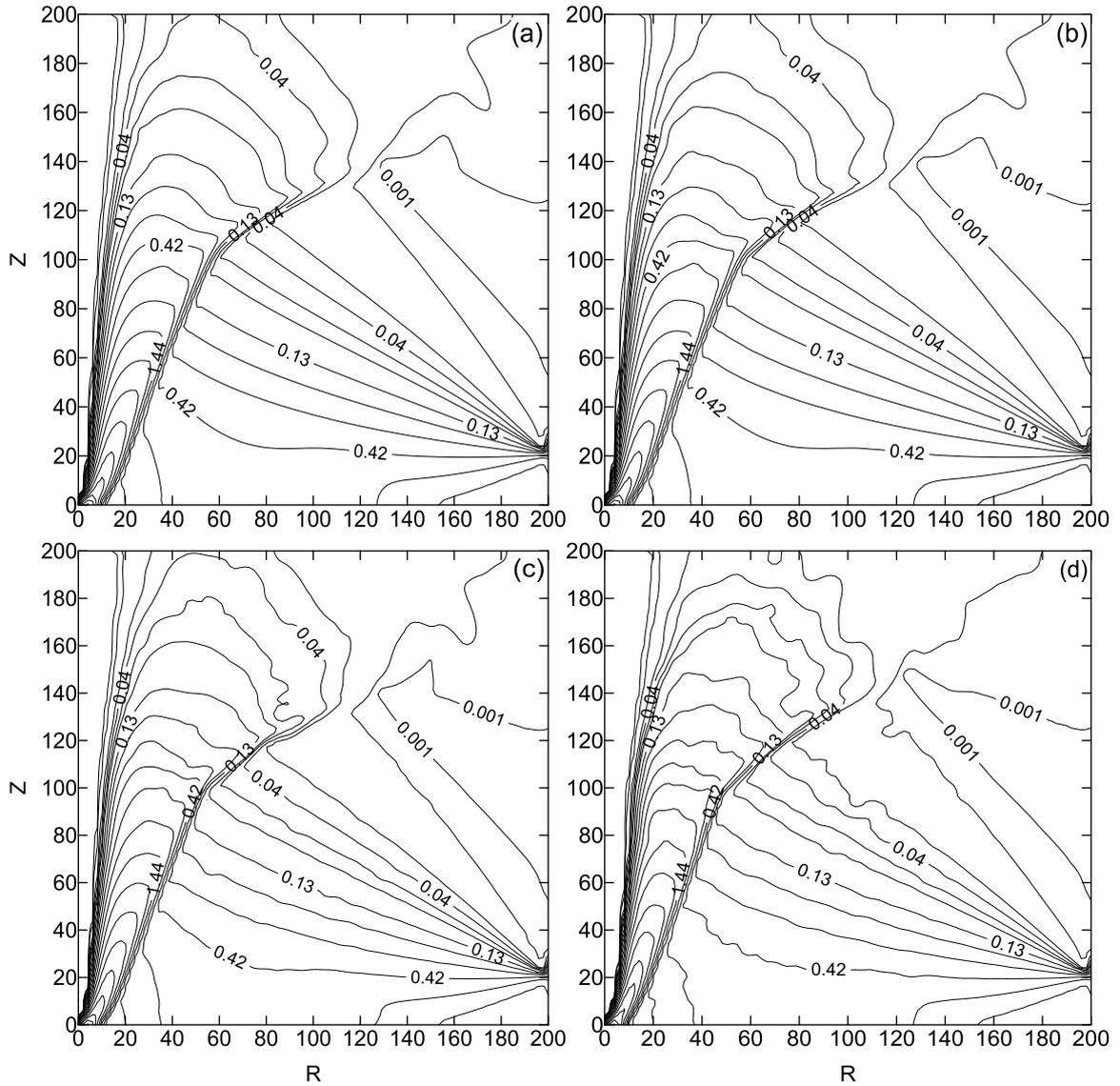}
\vskip 1.5cm
\caption{Change in the density contours in presence of cooling ($\lambda=1$). 
See, text for details. The conical region between the axis and shock wave becomes denser
as the accretion rate of the Keplerian disk is increased (Ghosh et al. 2011).}
\label{fig9abcd}
\end{center}
\end{figure}
\begin{figure}[]
\begin{center}
\vskip 15.5cm
\includegraphics{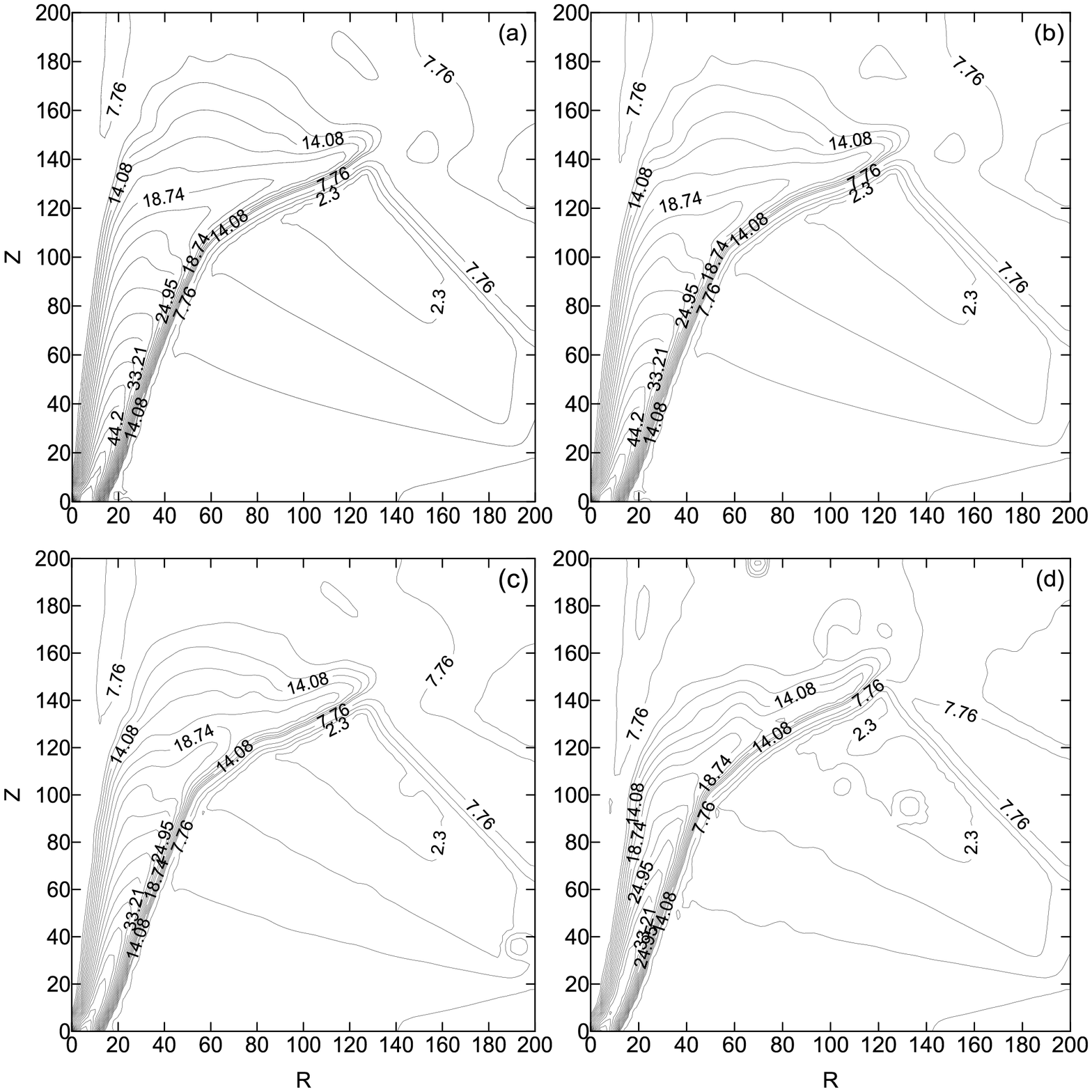}
\vskip 1.5cm
\caption{Change in the temperature contours in presence of cooling. The parameters 
are the same as in Fig. \ref{fig9abcd}(a-d). The temperature values used to draw the contours are 
the same as in Fig. \ref{fig8ab}b. Note that the shock shifts closer to the axis with the increase in
disk accretion rate (Ghosh et al. 2011).}
\label{fig10abcd}
\end{center}
\end{figure}

In Fig. \ref{fig9abcd}(a-d), we show the results of placing a Keplerian disk 
in the equatorial plane. The inner edge is located at $3 r_g$, the marginally 
stable orbit. Here, $\dot{m}_h = 1$ and $\dot{m}_d =$ (a) 1, (b) 2, (c) 5 and (d) 10 
respectively (Cases 2(a-d) of Table 6.1). The densities used to draw the contours are the same as that in 
Fig. \ref{fig8ab}a. As the Keplerian disk rate is increased, the intensity of soft photons interacting with 
the high optical depth (post-shock) region is increased. In Fig. \ref{fig9abcd}d, 
we observe that the conical region around the axis is 
considerably cooler. Thus, the density around the shock is enhanced. However, 
most importantly, with the increase in disk accretion rate, i.e., cooling, the shock location 
moves in closer to the black hole. This result has been already demonstrated in the  
context of the bremsstrahlung cooling (MSC96) inside the sub-Keplerian flow.

In Fig. \ref{fig10abcd}(a-d), we present the corresponding 
temperature distribution. The parameters are the same as in Fig. \ref{fig9abcd}(a-d) and the temperatures 
used to draw the contours are the same as that in Fig. \ref{fig8ab}b. 
The Comptonization in the shocked region cools it down considerably. Otherwise, 
not enough visible changes in the thermodynamic variables are seen. 
To understand the detailed effects of the radiative transfer on the 
dynamics of the flow, we take the differences in the pressure and velocity at each grid 
point of the flow for Cases 1d and 2d of Table 6.1. 
\begin{figure}[h]
\begin{center}
\vskip 5.5cm
\includegraphics{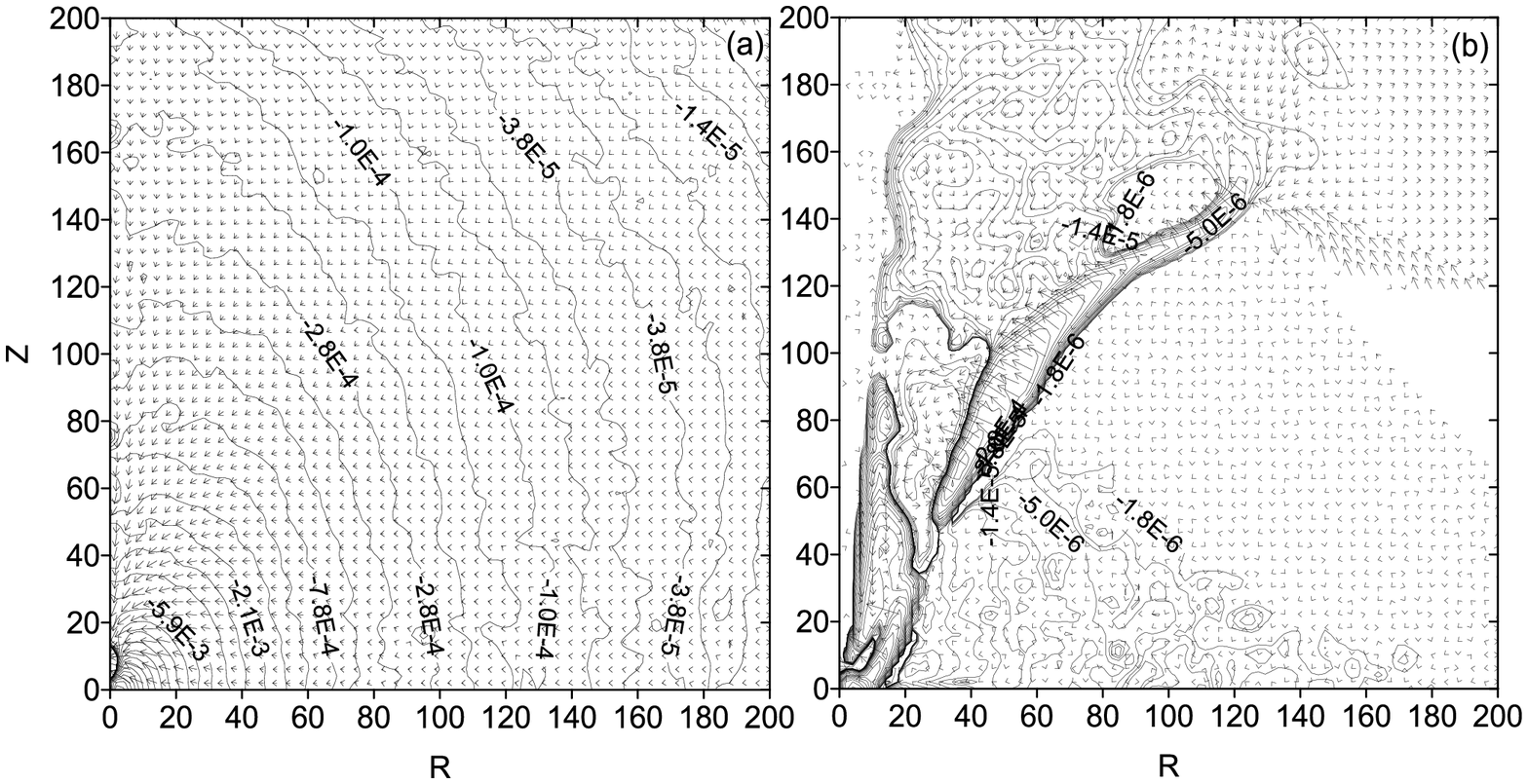}
\vskip 1.5cm
\caption{Difference in pressure and velocities between the flow with Comptonization and 
without Comptonization. Other parameters remain exactly the same. The cases are 
(a) Case 1d and (b) Case 2d of Table 6.1 respectively (Ghosh et al. 2011).} 
\label{fig11ab}
\end{center}
\end{figure}

In Fig. \ref{fig11ab}(a-b), we show the difference between the results of a purely hydrodynamical flow 
and the results by taking the Comptonization into account. Fig. \ref{fig11ab}a is for the flow with no 
angular momentum and Fig. \ref{fig11ab}b is drawn for the specific angular momentum 
$\lambda=1$. The contours are of constant $\Delta P= P_c-P_a$, where $P$ is the pressure
and the subscripts $c$ and $a$ represent the pressure with and without cooling 
respectively. The arrows represent the difference in velocity vectors in each grid. As expected, 
in both the cases the changes are maximum near the axis. The fractional changes in 
pressures and velocities are anywhere between $\sim 0$ (outer edge) and $\sim 25$\% (inner edge
and near the axis). Because of the shifts of the shock location towards the axis, the 
variation of the velocity is also highest in the vicinity of the shock.
Thus we prove that not only the symmetry is lost by the insertion of an axisymmetric soft photon source, 
the cooling process also plays a major role in deciding the dynamics of the flow.

\begin{figure}[]
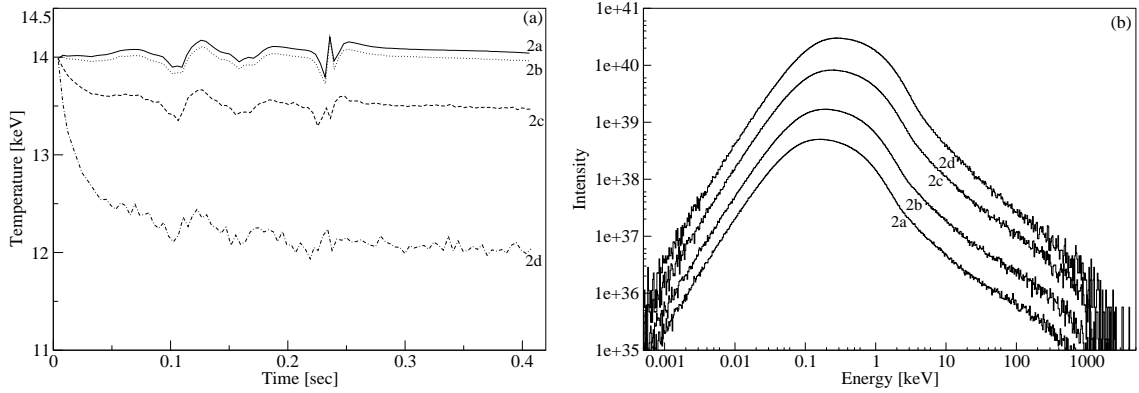

\begin{center}
\vskip 5.2cm
\includegraphics{CHAPTER6/Fig12a.eps}
\includegraphics{CHAPTER6/Fig12b.eps}
\vskip 1.0cm
\caption{Variation of the (a) average temperature of the Compton cloud with iteration time and 
(b) spectrum with the increase of disk accretion rate. $\lambda = 1$ and 
$\dot{m}_h = 1$ are used. The solid, dotted, dashed and dash-dotted plots are for $\dot{m}_d = 1$, 
 $2$, $5$ and $10$ respectively. With the increase in the disk rate, the temperature of the Compton cloud 
saturates at lower temperature. The solid, dotted, dashed and dash-dotted curves show the spectrum for $\dot{m}_d = 1$, 
$2$, $5$ and $10$ respectively. The spectrum is softer for higher value of $\dot{m}_d$ (Ghosh et al. 2011).}
\label{fig12ab}
\end{center}
\end{figure}
We now turn our attention to the dynamical variables and the spectral behavior of the rotating flow. 
In Fig. \ref{fig12ab}a, we show the variation of the average temperature of the 
Compton cloud with iteration time of the coupled code (Cases 2(a-d), Table 6.1). 
With the increase of disk rate, the temperature of the Compton cloud saturates at 
lower temperature. Fig. \ref{fig12ab}b shows the effect of the decrease in cloud temperature 
due to the increase of disk rate over the spectrum. The spectrum becomes softer as we increase 
the disk rate, keeping the halo rate fixed.

\begin{figure}[]
\begin{center}
\vskip 4.9cm
\includegraphics{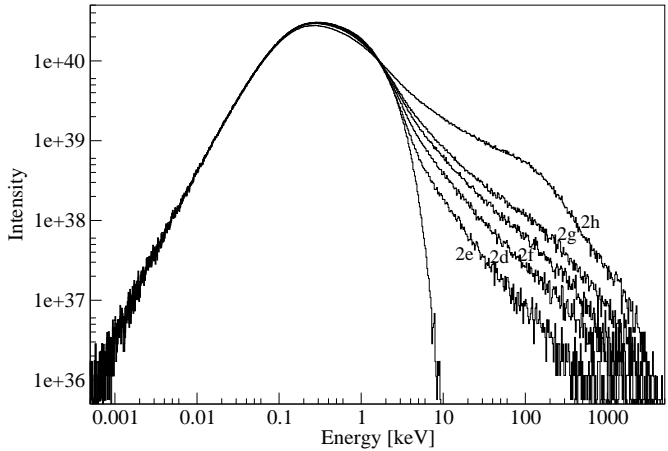}
\vskip 1.0cm
\caption{Variation of the spectrum with the increase of the halo accretion rate, keeping the 
disk rate ($\dot{m}_d = 10$) and angular momentum of the flow ($\lambda = 1$) fixed. 
The case number for which the spectra is drawn is marked in the plot. The unmarked 
plot is the injected spectrum. The spectrum becomes harder for the higher values of $\dot{m}_h$ (Ghosh et al. 2011).}
\label{fig13}
\end{center}
\end{figure}
In Fig. \ref{fig13} we show the effects of increasing the electron number 
density (due to the increase of $\dot{m}_h$) for a fixed disk rate. The spectrum becomes harder as we 
increase the halo rate keeping the number of injected soft the photons the same.

\begin{figure}[h]
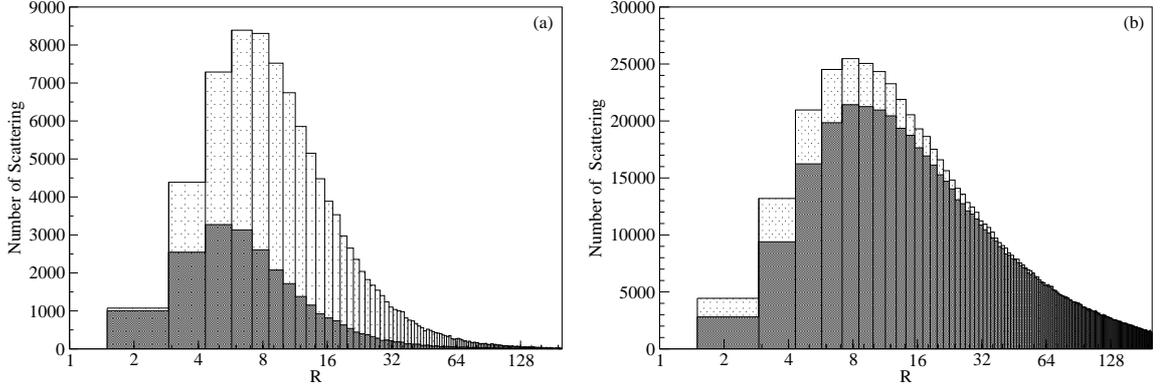

\begin{center}
\vskip 4.5cm
\includegraphics{CHAPTER6/Fig14a.eps}
\includegraphics{CHAPTER6/Fig14b.eps}
\vskip 1.0cm
\caption{Number of scatterings inside the spherical shell between $R$ to $R + \delta R$ ($ \delta R \sim 1.4$). 
The light and dark shaded histograms are for the cloud with and without bulk velocity, respectively. 
(a) Only the photons emerging from the cloud with energies E, where $50$ keV $< E <150$ keV, 
are considered here. (b) All the photons emerging from the cloud are considered here. Parameters used: 
$\dot{m}_d=1$,  $\dot{m}_h=10$ and $\lambda=0$ (Ghosh et al. 2011).}
\label{fig14ab}
\end{center}
\end{figure}
\begin{figure}[h!]
\begin{center}
\vskip 5.3cm
\includegraphics{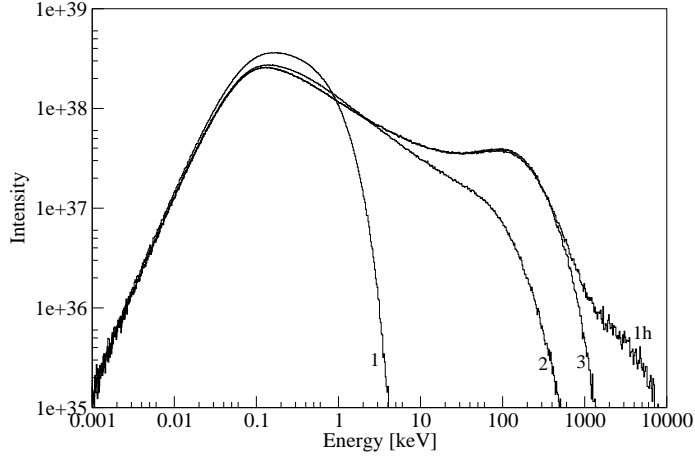}
\vskip 1.0cm
\caption{The spectrum for Case 1h of Table 6.1. The curves marked 2 and 3 
give the spectra when the bulk velocity of the electron
is absent for the whole cloud and for the cloud inside 3 $r_g$, respectively.
The curve marked 1 gives the injected spectrum for the Case 1h of Table 6.1. 
The bulk motion Comptonization of the photons inside the 3 $r_g$ radius creates the hard tail. The 
bump near 100 keV is a combined effect of the temperature and bulk velocity of rest of the cloud (Ghosh et al. 2011).}
\label{fig15}
\end{center}
\end{figure}
We observe that the emerging spectrum has a bump, especially at higher accretion rates of the halo, 
at around $100$ keV (e.g. the spectra marked 1g, 1h in Fig. \ref{fig6ab}a and the 
spectrum marked 2h in Fig. \ref{fig13}). A detailed 
analysis of the emerging photons having energies between 50 to 150 keV is done to see 
where in the Compton cloud they were produced. In Fig. \ref{fig14ab}a, we present the number of scatterings 
inside different spherical shells within the electron cloud suffered by these 
photons ($50 < E < 150$ keV) before leaving the cloud. Parameters used: 
$\dot{m}_d=1$,  $\dot{m}_h=10$ and $\lambda=0$. The light and dark shaded histograms are for the cloud with 
and without bulk velocity components, respectively. We find that the presence of bulk motion 
of the infalling electrons pushes the photons towards the hotter and denser 
[Figs. \ref{fig2ab}(a-b)] inner region of the cloud to suffer more and more scatterings. 
We find that the photons responsible for the bump suffered the maximum number of scatterings around 
8 $r_g$. From the temperature contours, we find that the cloud 
temperature around 8 $r_g$ is $\sim 100$ keV. 
In Fig. \ref{fig14ab}b, we consider all the outgoing photon energies. The inflowing 
bulk velocity of the electrons push the photons to the inner region of the accretion disk. Thus 
the photons that come out of the electron cloud suffer more scatterings in presence 
of the bulk velocity. As the effects of bulk velocity become more dominant, 
the difference between the two histograms also increases. 
\begin{figure}[h]
\begin{center}
\vskip 5.1cm
\includegraphics{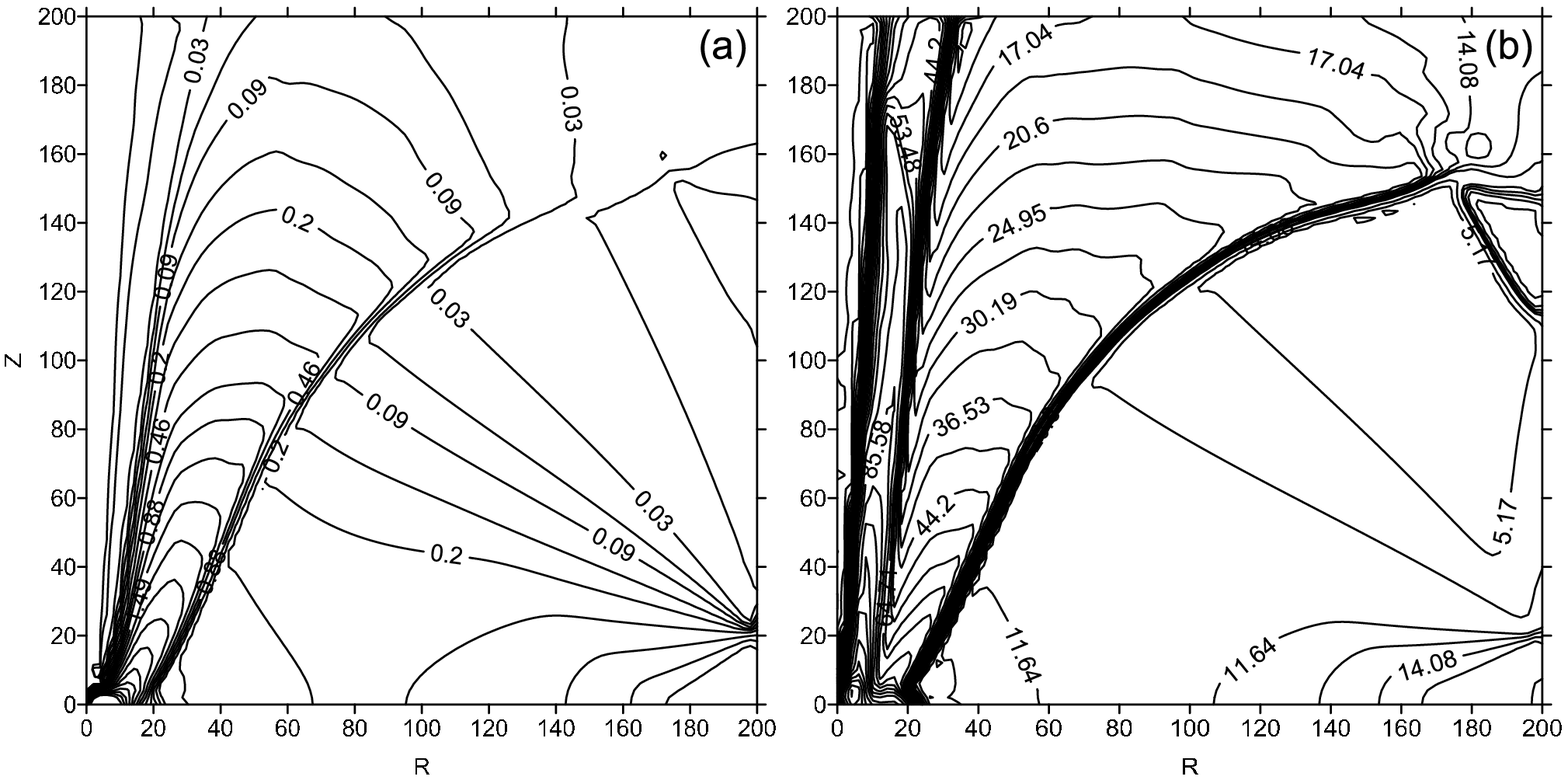}
\vskip 1.5cm
\caption{Same as Fig. \ref{fig8ab}(a-b) but for ($\lambda$, $\epsilon$) = (1.5, 0.0007). 
Densities are in normalized unit and the temperatures are in keV. See, text for the 
values of the contour levels.
}
\label{fig16l1pt5}
\end{center}
\end{figure}

In Fig. \ref{fig15}, we explicitly show the effects of the bulk velocity on the spectrum. 
We note that the bump disappears when the bulk velocity of the electron cloud 
becomes zero (Curve marked 2). This fact shows that the region around 8 $r_g$ behaves more like 
a black body emitter, which creates the bump. Since the photons are 
suffering large number of scatterings near this region (8 $r_g$), most of them
emerge from the cloud with the characteristic temperature of the region. The 
effect of bulk velocity in this region is to force the photons to suffer larger 
number of scatterings. This bump vanishes for lower density cloud (low $\dot{m}_h$) 
as the photons suffer lesser number of scatterings. 
The photons which are scattered close to the black hole horizon and escape
without any further scattering, produce the high energy tail in the output spectrum. 
Curve 3 of Fig. \ref{fig15} shows the intensity spectrum of Case 1h (Table 6.1), 
when there are zero bulk velocity inside 3 $r_g$. We find that in the 
absence of bulk velocity inside 3 $r_g$, the high 
energy tail in the Curve 1h vanishes. This is the clear signature of the presence of bulk 
motion Comptonization near the black hole horizon.

\subsection{Compton Cloud  with $\lambda=1.5$}

Let us now increase the specific angular momentum of the flow to $\lambda = 1.5$. In the case of non-dissipative 
flows in a vertical equilibrium, a flow is not supposed to form a shock wave (C90). However, when cooling
is present we see that a shock has been formed due to the centrifugal barrier. In Fig. \ref{fig16l1pt5}(a-b), 
we have shown the contours of constant density (Fig. \ref{fig16l1pt5}a) and temperature 
(Fig. \ref{fig16l1pt5}b) in absence of Compton cooling. Density contour levels 
are drawn from $0.001-0.03$ (successive level ratio is $5$), $0.03-0.06$ 
(successive level ratio is $2$), $0.06-0.68$ (successive level ratio is $1.5$), 
$0.68-20.46$(successive level ratio is $1.3$). Temperature contour levels are 
drawn from $2.3-11.64$ (successive level ratio is $1.5$), $11.64-64.71$ 
(successive level ratio is $1.1$), $64.71-346.21$ (successive level ratio is $1.15$). 
We note that a shock stronger than the case $\lambda=1$, has been formed 
which bends outwards away from the equatorial plane. The specific 
energy $\epsilon$ for the simulations presented in this Section is $0.0007$. 
\begin{figure}[]
\begin{center}
\vskip 4.9cm
\includegraphics{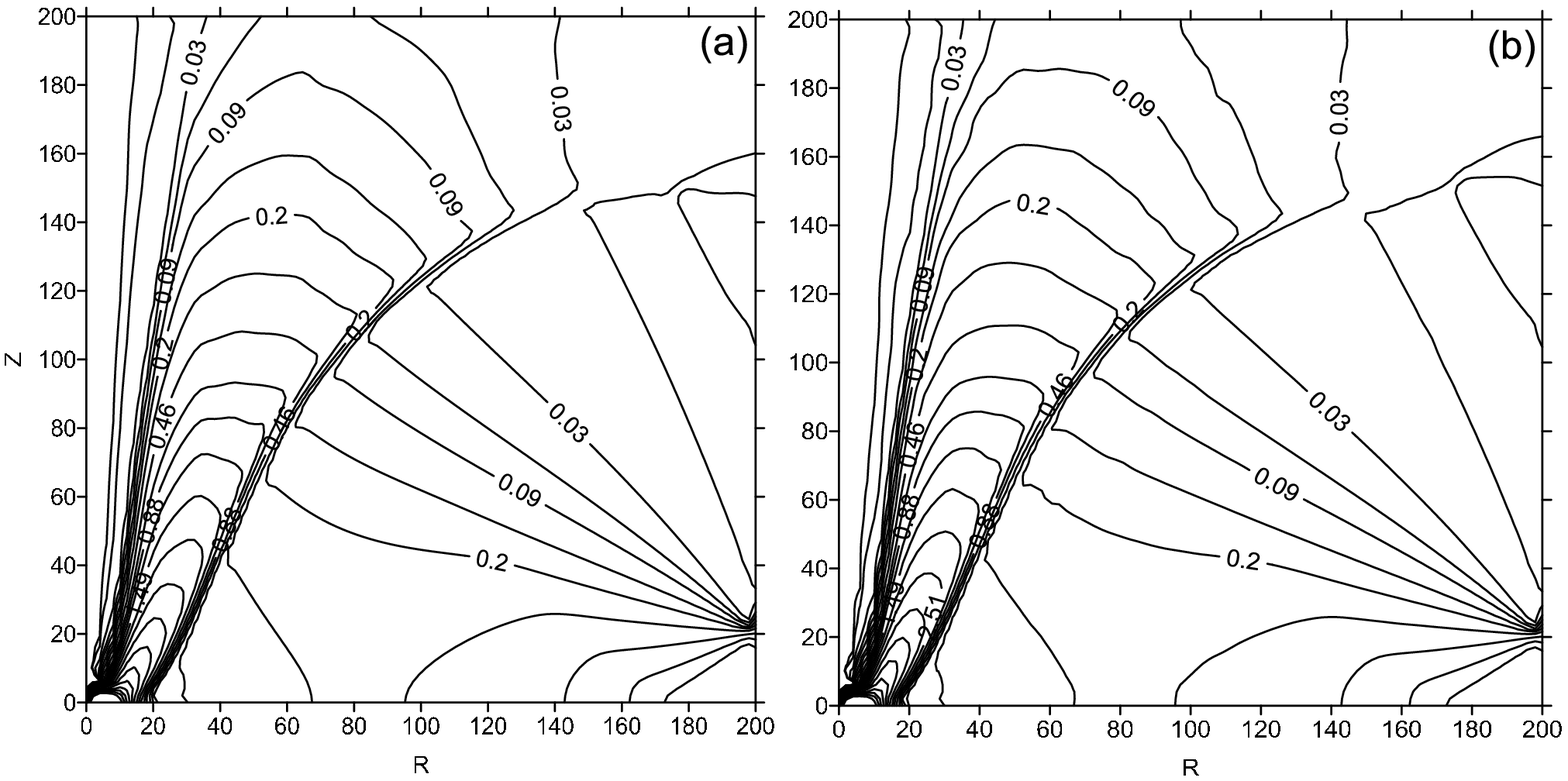}
\vskip 1.5cm
\caption{Change in the density contours in presence of cooling ($\lambda = 1.5$). 
The accretion rate of the halo is fixed $\dot{m}_h=1$. (a) $\dot{m}_d=1$ (b) $\dot{m}_d=10$. 
Due to the increase of disk rate ($N_{inj}$), the size of the post shock Compton cloud 
decreases. The density values used to draw the contours are same as in Fig. \ref{fig16l1pt5}a.}
\label{fig17l1pt5}
\end{center}
\end{figure}

\begin{figure}[]
\begin{center}
\vskip 5.0cm
\includegraphics{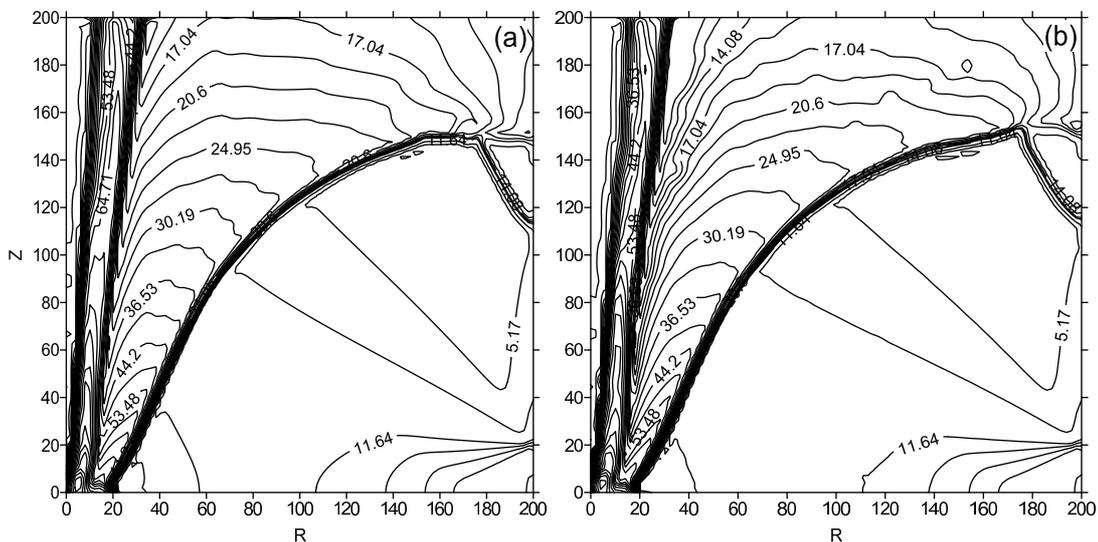}
\vskip 1.5cm
\caption{Change in the temperature contours in presence of cooling. The parameters are the same 
as in Fig. \ref{fig17l1pt5}(a-b).}
\label{fig18l1pt5}
\end{center}
\end{figure}
\begin{figure}[]
\begin{center}
\vskip 5.2cm
\includegraphics{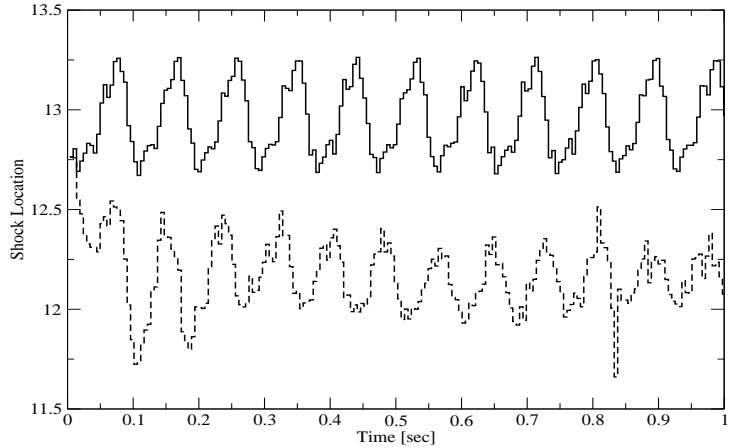}
\vskip 1.0cm
\caption{Variation of the shock locations with iteration time. Parameters used: $\dot{m}_d=1$ 
(solid curve), $\dot{m}_d=10$ (dotted curve). $\dot{m}_h=1$ for both the cases.}
\label{fig19l1pt5}
\end{center}
\end{figure}
\begin{figure}[b]
\begin{center}
\vskip 6.2cm
\includegraphics{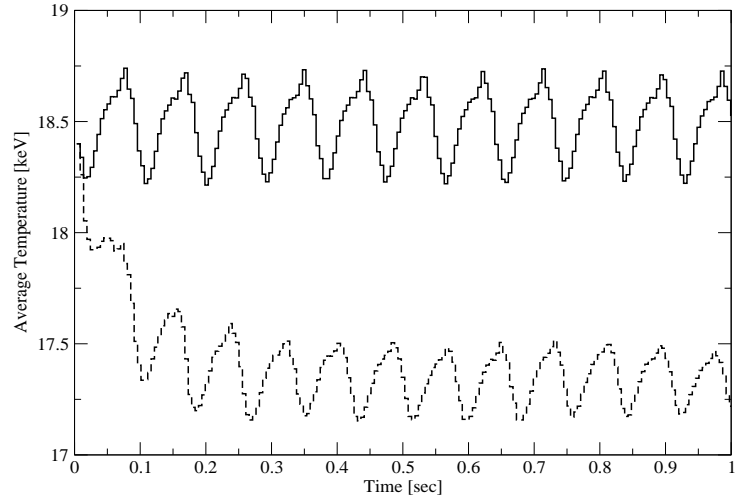}
\vskip 1.0cm
\caption{Same as Fig. \ref{fig12ab}(a) but for ($\lambda$, $\epsilon$) = (1.5, 0.0007). 
With the increase in disk rate the average temperature of the cloud decreases.}
\label{fig20l1pt5}
\end{center}
\end{figure}

In Figs. \ref{fig17l1pt5}(a-b) and \ref{fig18l1pt5}(a-b) we show the effects of 
the inclusion of a Keplerian disk at the equatorial plane. We find that the shock location 
is shifted inwards due to the Compton cooling. In Fig. \ref{fig19l1pt5} we have shown 
the variation of the location of the shock at the equatorial plane with the 
iteration time. We notice that the shock has been formed at a lower radius 
for $\dot{m}_d=10$ than the $\dot{m}_d=1$ case. This happens because for 
a higher disk rate the Compton cloud becomes cooler and the size of the Compton cloud 
becomes smaller. In Fig. \ref{fig20l1pt5} the variation of the average temperature of the cloud 
as the time goes, is shown. The cloud becomes cooler as the disk rate increases.

\begin{figure}[]
\begin{center}
\vskip 4.9cm
\includegraphics{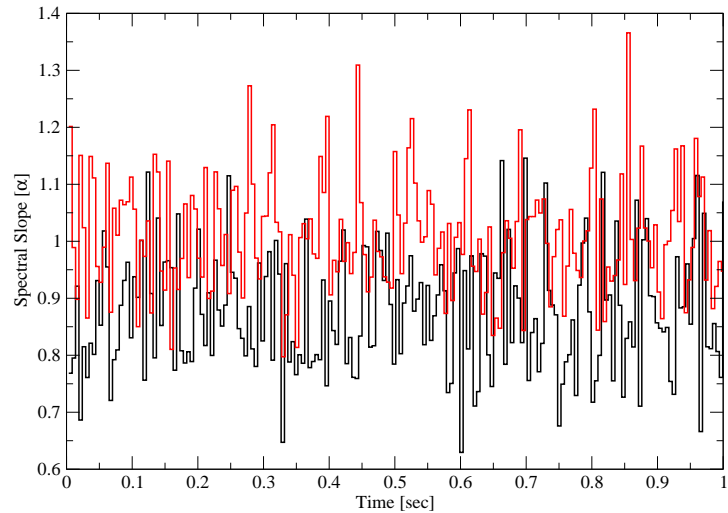}
\vskip 1.0cm
\caption{Variation of the spectral slope $\alpha$ with iteration time. Parameters used: $\dot{m}_d=1$ 
(black curve), $\dot{m}_d=10$ (red curve). $\dot{m}_h=1$ for both the cases. As $\dot{m}_d$ increases 
the spectrum becomes softer.}
\label{fig21l1pt5}
\end{center}
\end{figure}

In Fig. \ref{fig21l1pt5} we plot the variation the spectral slope $\alpha$ with time 
for $\dot{m}_d=1$ and $\dot{m}_d=10$ keeping the halo rate $\dot{m}_h=1$ fixed. We 
see that the spectrum becomes softer as the disk rate increases. 

\begin{figure}[]
\begin{center}
\vskip 6.25cm
\includegraphics{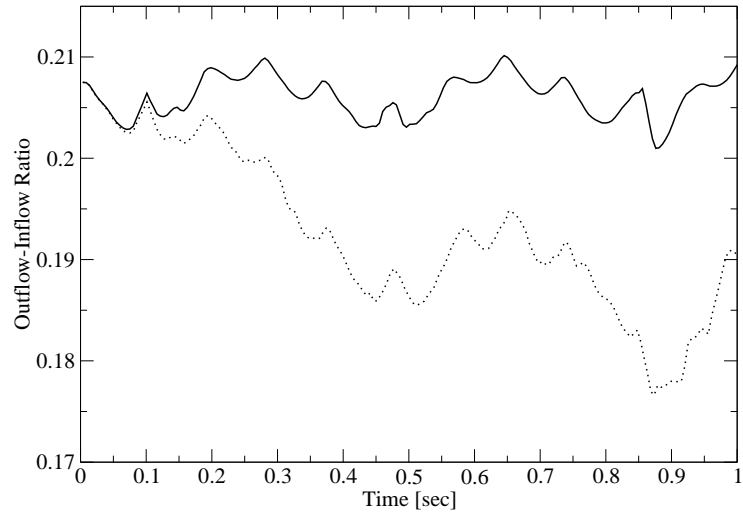}
\vskip 1.0cm
\caption{Variation of the ratio between outflow and inflow rate with iteration time. 
Parameters used: $\dot{m}_d=1$ (solid curve), $\dot{m}_d=10$ (dotted curve). $\dot{m}_h=1$ 
for both the cases. For the same halo profile, as the number of soft photons increases 
(due to the increase in disk rate) the outflow rate decreases.}
\label{fig22l1pt5}
\end{center}
\end{figure}

In Fig. \ref{fig22l1pt5}, the variation of ratio of outflowing matter to 
injected matter with time when $\dot{m}_d$ is increased keeping $\dot{m}_h$ fixed 
is shown. The outflow rate decreases due to the increase in disk rate. We have 
also verified that the outflow rate increases with the increase in the 
specific angular momentum of the flow (Giri et al., 2010).

\begin{figure}[]
\begin{center}
\vskip 6.0cm
\includegraphics{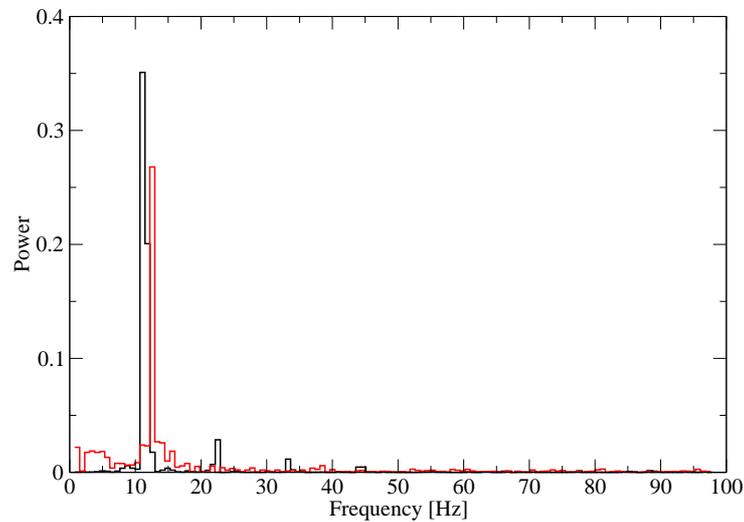}
\vskip 1.0cm
\caption{Power density spectra of the time variation of the shock locations are shown. 
Parameters used: $\dot{m}_d=1$ (black curve), $\dot{m}_d=10$ (red curve). 
$\dot{m}_h=1$ for both the cases. As the disk rate increases, the post shock region 
oscillates at a slightly higher frequency but with a lower power.}
\label{fig23l1pt5}
\end{center}
\end{figure}

We now turn our attention to the variation of the shock location (Fig. \ref{fig19l1pt5}). 
We find that the shock is oscillating in both the cases. In Fig. \ref{fig23l1pt5} we 
have plotted the power density spectrum (PDS) for these two cases. The PDS gives the 
frequency of oscillation for $\dot{m}_d = 1$ to be $\nu_{QPO}=10$ Hz. The oscillation 
frequency slightly increases for the $\dot{m}_d = 10$ case. We also note that 
for the higher disk rate, the amplitude of the QPO has a lower value. 

\section{Fate of the Jet and the High Angular Momentum Flows in Presence of Cooling}

So far, we have concentrated on the spectral properties of disk-jet system with passive 
velocity distribution. We also studied low angular momentum systems which do not produce 
very strong jets. However, in  a realistic system, the angular momentum close to the black hole
could be around the marginally stable and bound values ($\sim 1.83-2$ in our units). 

Physically, when the angular momentum is increased, the shocks can form at a larger
radii. With the increase of cooling (Keplerian disk rate) the shock moves closer
to the black hole. The shock oscillations occur as before (previous Section), but the 
oscillations damp out as the cooling is increased. The post-shock region also collapses
with the increase in cooling (Garain, Ghosh and Chakrabarti, 2012). Most interestingly, since the 
formation of the jet is from CENBOL in this model, our result clearly shows that the outflow rate is greatly reduced
as the cooling rate is increased. This quenching of jet the phenomenon directly shows that correlation
between the spectral states and the outflow rates (Garain, Ghosh and Chakrabarti, 2012). 

\clearpage
\newpage

%% file: conclusion.tex
\newpage
\markboth{\it Conclusions and Future Plans}
{\it Conclusions and Future Plans}
\chapter{Conclusions and Future Plans}

So far, we described the developement of a radiative transfer Monte Carlo code 
and presented its applications to various astrophysical problems. 
We started with a brief introduction of `compact objects' and 
also presented a simple classifications of such objects based 
on their mass limit. We mentioned the importance of accretion 
process around a compact object and briefly described the 
history of the observational evidences from black holes. In Section 1.4 
we have discussed the spectral properties of galactic and extragalactic 
black holes in X-rays and their possible origin. There are quite a few 
accretion disk models present in the literature. We have listed some of 
the basic accretion disk models in Sections 1.5 and 1.6, starting from 
Bondi flow to TCAF. In this Thesis, 
we have mainly focused on the spectral properties of the radiations coming 
from an accretion disk. The processes responsible for the emission 
of radiation from an accretion flow has been listed in Section 1.7. 
A black hole system is general relativistic 
in nature. In our work we have used a pseudo-Newtonian 
potential which mimics the space time of a Schwarzschild black hole.

The mechanisms responsible for the emission of radiations from an accretion 
disk are discussed in details in Chapter 2. We have started with the 
black body radiation and calculated the total number of photons 
generated from a black body at temperatute $T$. 
Next we discuss in detail the Compton scattering process between an electron 
and a photon. We calculate the probability of a photon to be scattered and 
the energy exchange between an electron and a photon due to the scattering.

In Chapter 3, we discuss the Monte Carlo techniques through which we have 
incorporated the radiative transfer into our code. We have shown the results 
of some simple Monte Carlo simulations at the end of Chapter 3.

In Chapter 4, we have presented several results of Monte Carlo simulation 
of Comptonization by hot electron clouds which surround the black hole 
in the form of a toroidal shaped centrifugal pressure dominated boundary 
layer. The soft photons are supplied by a Keplerian disk which reside 
just outside this cloud. 
We verify several of the previously reported conclusions obtained by 
theoretical methods. We find that for a given supply of the injected 
soft photons, the spectrum can become harder (i.e., spectral index 
$\alpha$ can go down) when either the optical depth is increased (electron
number density goes up) and/or the electron temperature is increased. 
Furthermore, we found how the spectral shape changes when the Compton 
cloud expands and shrinks. We compute exactly what fraction of the 
photons are intercepted and processes and compute the percentage of 
scattered photons as functions of the flow variables. These results 
would be valuable to interpret the observational results from black 
hole candidates, especially when the spectral index is found to be changed.

In Chapter 5, we have extended the results of our previous work 
on Monte-Carlo simulations. We include the outflow in 
conjunction with the inflow. The outflow rate is self-consistently 
computed from the inflow rate using the well-known considerations 
present in the literature (Das et al. 2001 and references therein). 
We compute the effects of the thermal and the bulk motion Comptonization 
on the soft photons emitted from a Keplerian disk around a black 
hole by the post-shock region of a sub-Keplerian flow which 
surrounds the Keplerian disk. A shock in the inflow increases the 
CENBOL temperature, increases the electron number density and 
reduces the bulk velocity. Thermal Comptonization and bulk 
motion Comptonization inside the CENBOL increases photon energy. 
However, the CENBOL also generates the outflow of matter which 
down-scatters the photons to lower energy. We show that the thermal 
Comptonization and the bulk motion Comptonization are possible 
in both the accretion and the outflows. While the converging flow
up-scatters the radiation, the outflow down-scatters. However, 
the net effect is not simple. The outflow parameters are strongly 
coupled to the inflow parameters and thus for a given inflow 
and outflow geometry, the strength of the shock can also determine 
whether the net scattering by the jets would be significant or not.
Sometimes the spectrum may become very complex with two 
power-law indices, one from thermal and the other from the 
bulk motion Comptonization. Since the volume of the jet may be 
larger than that of the CENBOL, sometimes the number of 
scatterings suffered by softer photons from the electrons in 
the jet may be high. However, whether the CENBOL or the 
jet emerging from it will dominate in shaping the spectrum 
strongly depends on the geometry of the flow and the strength 
of the shock. We also found that the halo can Comptonize and 
harden the spectrum even without the CENBOL.

Chapter 4 and Chapter 5 contained the results of the simulations 
where a particular geometry of the electron cloud is assumed 
(toroidal electron cloud in Chapter 4 and spherical inflow in 
conjunction with a conical jet in Chapter 5). In Chapter 6, 
we have shown the results of the Monte Carlo simulations, 
where no such geometry for the Compton cloud is assumed. The 
geometry and the hydrodynamics of the flow is now simulated at each 
iteration time step by a Total Variation Diminishing (TVD) code 
and is used as the initial conditions in the radiative transfer code. 
We have included the effects of Compton cooling into the Monte Carlo 
code. The coupled Monte Carlo-TVD code simulates the spectra and 
hydrodynamics of the accretion disk around a stellar mass black hole.
In Section 6.3, we describe the procedure to couple the hydrodynamic 
and the radiative transfer code. Our major conclusions in this Chapter 
are: first, in the presence of an axisymmetric disk which supplies soft
photons to the Compton cloud, even an originally spherically symmetric
accreting Compton cloud becomes axisymmetric. This is because, due 
to the higher optical depth, there is a significant cooling
near the axis of the intervening accreting halo between the disk
and the axis. Second, due to the cooling effects close to the axis, 
the pressure drops significantly, which may change the flow velocity up to
25 per cent. Third, this effect becomes more for low angular momentum 
flows which produce shock waves close to the axis. The post-shock 
region cools down and the outflow falls back to the disk. 
This shows that the Chakrabarti \& Manickam (2000) mechanism of 
the effects of Comptonization of outflows does take place. Fourth, 
the emitted spectrum is direction-dependent. The spectrum 
along the axis shows a large soft bump, while the spectrum along
the equatorial plane is harder. Fifth, we also find that 
photons which spend more time 
(up to 100 ms in the case considered) inside the Compton 
cloud produce harder spectrum as they scatter several times
However, if they spend too much (above 100 ms) time, 
they transfer their energies back to the cooler electrons 
while escaping. These results would be valuable for 
interpreting the timing properties of the radiation 
from black hole candidates. 

At the end of Chapter 6, we have shown some simulation results 
which we obtained using the coupled Monte Carlo-TVD simulation for 
an accretion disk having a higher angular momentum than the 
previously presented cases. Here a shock has been generated 
because of the centrifugal barrier. The shock location 
oscillates with time and the oscillation gives rise to the 
Quasi Periodic Oscillation. We find the spectral slope also 
changing during the oscillation of the shock.


In this Thesis, we have worked with zero or very low 
angular momentum accretion flows only. In future we will 
extend our simulations for those cases where strong 
shocks are present. In presence of cooling an oscillating 
shock may stop oscillation or a steady shock may start 
oscillating. We like to study the spectral properties 
for those systems. In future we will include synchrotron 
and bremsstrahlung radiation in our code. Stochastic magnetic fields would produce
synchrotron radiations everywhere and thus the soft photon source would be 
distributed. We will also carry out satellite data analysis and fit our 
simulated spectra, time lag/lead, angle dependent spectral properties etc.
with observed results.

The entire work of this Thesis has been done in 
the Schwarzschild black hole geometry using pseudo-Newtonian 
potential. However, since black holes form out of rotating collapsing stars
and sometimes collapse is induced by the rotating accretion matter, it is not unlikely that 
many, if not all, of the black holes are actually Kerr type. In future, we shall continue
our work in Kerr geometry and look for tell tale signatures of the rotation parameter of the black hole.

\newpage

%% file: appendx.tex
\newpage
